\documentclass{iopart}
\usepackage{bbm,iopams}

\usepackage{times}

\usepackage[all]{xy}
\usepackage{eepic,psfrag}
\xyoption{ps}
\xyoption{dvips}

\usepackage{graphicx}
\usepackage[usenames]{color}

\usepackage{amssymb}
\usepackage{bbm}
\usepackage{bm}

\usepackage{url}
\usepackage{cite}

\usepackage[scriptsize]{subfigure}

\newcommand{\av}[1]{\overline{#1}} 
\newcommand{\avV}{\laser{V}}

\newcommand{\efield}{\mathcal{E}}
\newcommand{\Corr}{\mathcal{P}}
\newcommand{\Power}{\mathcal{P}}
\newcommand{\Bessel}{J}

\newcommand{\calC}{\mathcal{C}}
\newcommand{\diko}{\mathcal{D}}
\newcommand{\Ker}{\Phi_0} 
\newcommand{\Cur}{\bPhi_1}

\newcommand{\dos}{\mathcal{N}_0}
\newcommand{\avdos}{\mathcal{N}}

\newcommand{\ells}{\ell_{s}}					
\newcommand{\elltr}{\ell}						
\newcommand{\tautr}{\tau}                         
\newcommand{\taus}{\tau_{s}}					
\newcommand{\elltrb}{\boltz{\ell}}				
\newcommand{\elltrwl}{\ell^*}
\newcommand{\scat}[1]{#1_{s}}
\newcommand{\boltz}[1]{#1_\mathrm{B}}
\newcommand{\wl}[1]{#1^*}
\newcommand{\cut}[1]{#1_{c}}					
\newcommand{\inel}[1]{#1_\mathrm{i}}
\newcommand{\atom}[1]{#1_A}
\newcommand{\laser}[1]{#1_\mathrm{L}}
\newcommand{\light}[1]{#1}
\newcommand{\sat}[1]{#1_\mathrm{s}}

\newcommand{\recoil}[1]{#1_R}
\newcommand{\broglie}[1]{#1_\mathrm{dB}}
\newcommand{\mobil}[1]{#1_\Delta}
\newcommand{\mobby}[1]{#1_\mathrm{c}}
\newcommand{\loc}[1]{#1_\mathrm{loc}}
\newcommand{\maxi}[1]{#1_\mathrm{m}}
\newcommand{\usphere}{\ensuremath{S_d}}
\newcommand{\ie}{i.\,e.~}
\newcommand{\hikami}[2]{#1_{#2}}
\newcommand{\fuchs}{f}

\newcommand{\fluct}{\delta V}
\renewcommand{\Re}{\mathrm{Re}}
\renewcommand{\Im}{\mathrm{Im}}
\newcommand{\dd}{\rmd^d} 
\newcommand{\dkpi}[1]{\frac{\dd #1}{(2\pi)^d}}
\newcommand{\bra}[1]{\left\langle\smash{#1}\right|}
\newcommand{\ket}[1]{\left|\smash{#1}\right\rangle}
\newcommand{\tfrac}[2]{\frac{#1}{#2}}
\newcommand{\sfrac}[2]{#1/#2}
\newcommand{\abs}[1]{|{#1}|}

\newcommand{\nref}[1]{(\ref{#1})}

\newcommand{\operator}[1]{\ensuremath{#1}}

\newlength{\up}
\newlength{\shrink}
\newlength{\hoch}
\newlength{\breit}
\newlength{\breiter}
\newlength{\lang}
\newlength{\ick}
\newlength{\inn}
\newlength{\iki}
\newlength{\uku}
\newlength{\aka}
\newlength{\oko}

\newcommand{\st}[1]{\ensuremath{~{#1}~}}

\newcommand{\diagram}[1]{\ensuremath{
\setlength{\up}{.7\up}
\setlength{\shrink}{.7\shrink}
\setlength{\hoch}{.7\hoch}
\setlength{\breit}{.7\breit}
\setlength{\breiter}{.7\breiter}
\setlength{\lang}{.7\lang}
\setlength{\ick}{.7\ick}
\setlength{\inn}{.7\inn}
\setlength{\unitlength}{.7\unitlength}
\def\objectstyle{\scriptstyle}
\def\labelstyle{\scriptstyle}
{#1}}}

\newcommand{\scinq}{\xymatrix@1@=\shrink{
*{\otimes}\ar@{-}@*{[butt][|<\inn>]}[r]\ar@{.}@/^{1.8\up}/^{~}[rrrr]&
*{\bullet}\ar@{-}@*{[butt][|<\inn>]}[r]\ar@{.}@/^{1.4\up}/^{~}[rr]\ar@{.}@/^{0.8\up}/[r]&
*{\bullet}\ar@{-}@*{[butt][|<\inn>]}[r]\ar@{.}@/^{0.8\up}/[r]&
*{\bullet}\ar@{-}@*{[butt][|<\inn>]}[r]& 
*{\otimes}}}

\newcommand{\scinqb}{\xymatrix@1@=\shrink{
*{\bullet}\ar@{-}@*{[butt][|<\inn>]}[r]\ar@{.}@/^{1.8\up}/^{~}[rrrr]\ar@{.}@/^{0.6\up}/^{~}[r]&
*{\bullet}\ar@{-}@*{[butt][|<\inn>]}[r]\ar@{.}@/^{0.6\up}/[r]&
*{\bullet}\ar@{-}@*{[butt][|<\inn>]}[r]\ar@{.}@/^{0.6\up}/[r]&
*{\bullet}\ar@{-}@*{[butt][|<\inn>]}[r]\ar@{.}@/^{0.6\up}/^{~}[r]& 
*{\bullet}}}

\newcommand{\vgv}{\xymatrix@1@=\shrink{
*{\otimes}\ar@{.}@/^\up/[r]&
*{\otimes}\ar@{-}@*{[butt][|<\inn>]}[l]}}

\newcommand{\ffgff}{
	\xymatrix@1@=\shrink{
	*{\bullet}\ar@{:}@/^\up/[r]&
	*{\bullet}\ar@{-}@*{[butt][|<\inn>]}[l]}
}
\newcommand{\ffleerff}{
	\xymatrix@1@=\shrink{
	*{\bullet}\ar@{:}@/^\up/[r]&
	*{\bullet}}
}

\newcommand{\vleerv}{\xymatrix@1@=\shrink{
*{\otimes}\ar@{.}@/^\up/[r]&
*{\otimes}}}

\newcommand{\oleero}{\xymatrix@1@=\shrink{
*{\bullet}\ar@{.}@/^\up/[r]&
*{\bullet}}}

\newcommand{\vggv}{\xymatrix@1@=\shrink{
*{\otimes}\ar@{.}@/^\up/[r]&
*{\otimes}\ar@{-}@*{[butt][|<\ick>]}[l]}}

\newcommand{\vgvgv}{\xymatrix@1@=\shrink{
*{\otimes}\ar@{-}@*{[butt][|<\inn>]}[r]\ar@{.}[r]+<0pt,{1.2\up}>&
*{\otimes}\ar@{-}@*{[butt][|<\inn>]}[r]\ar@{.}[]+<0pt,{1.2\up}>&
*{\otimes}\ar@{.}[l]+<0pt,{1.2\up}>}}

\newcommand{\croggo}{\xymatrix@1@=\shrink{
*{\otimes}\ar@{-}@*{[butt][|<\ick>]}[r]&
*{\otimes}\ar@{-}@*{[butt][|<\ick>]}[r]\ar@{.}@/^\up/[rr]&
*{\otimes}\ar@{-}@*{[butt][|<\ick>]}[r]\ar@{.}@/_\up/[ll]&
*{\otimes}}}

\newcommand{\crogg}{\xymatrix@1@=\shrink{
*{\otimes}\ar@{-}@*{[butt][|<\ick>]}[r]&
*{\otimes}\ar@{-}@*{[butt][|<\ick>]}[r]\ar@{.}@/^\up/^{~}[rr]&
*{\otimes}\ar@{-}@*{[butt][|<\ick>]}[r]\ar@{.}@/_\up/_{~}[ll]&
*{\otimes}}}

\newcommand{\fvgvgv}{\xymatrix@1@=1.25\shrink{
*{\bullet}\ar@{-}@*{[butt][|<\inn>]}[r]\ar@{.}@/^{1.4\up}/^{~}[rr]\ar@{.}@/^{0.8\up}/[r]&
*{\bullet}\ar@{-}@*{[butt][|<\inn>]}[r]\ar@{.}@/^{0.8\up}/[r]&
*{\bullet}}}

\newcommand{\fggs}{\xymatrix@1@=1.25\shrink{
*{\bullet}\ar@{-}@*{[butt][|<\ick>]}[r]\ar@{.}@/^{1\up}/[rr]\ar@{.}@/^{0.6\up}/[r]&
*{\bullet}\ar@{-}@*{[butt][|<\ick>]}[r]\ar@{.}@/^{0.6\up}/[r]&
*{\bullet}}}

\newcommand{\pggs}{\xymatrix@1@=\shrink{
*{\otimes}\ar@{-}@*{[butt][|<\ick>]}[r]\ar@{.}[r]+<0pt,{1\up}>&
*{\otimes}\ar@{-}@*{[butt][|<\ick>]}[r]\ar@{.}[]+<0pt,{1.\up}>&
*{\otimes}\ar@{.}[l]+<0pt,{1\up}>}}

\newcommand{\doppel}{\xymatrix@1@=\shrink{
*{\otimes}\ar@{-}@*{[butt][|<\inn>]}[r]&
*{\otimes}\ar@{-}@*{[butt][|<\inn>]}[r]\ar@{.}@/_\up/^{~}[l]&
*{\otimes}\ar@{-}@*{[butt][|<\inn>]}[r]\ar@{.}@/^\up/^{~}[r]&
*{\otimes}}}

\newcommand{\cross}{\xymatrix@1@=\shrink{
*{\otimes}\ar@{-}@*{[butt][|<\inn>]}[r]&
*{\otimes}\ar@{-}@*{[butt][|<\inn>]}[r]\ar@{.}@/^\up/^{~}[rr]&
*{\otimes}\ar@{-}@*{[butt][|<\inn>]}[r]\ar@{.}@/_\up/^{~}[ll]&
*{\otimes}}}

\newcommand{\crosso}{\xymatrix@1@=\shrink{
*{\otimes}\ar@{-}@*{[butt][|<\inn>]}[r]&
*{\otimes}\ar@{-}@*{[butt][|<\inn>]}[r]\ar@{.}@/^\up/[rr]&
*{\otimes}\ar@{-}@*{[butt][|<\inn>]}[r]\ar@{.}@/_\up/[ll]&
*{\otimes}}}

\newcommand{\bowso}{\xymatrix@1@=\shrink{
*{\otimes}\ar@{-}@*{[butt][|<\inn>]}[r]\ar@{.}@/^{1\up}/[rrr]&
*{\otimes}\ar@{-}@*{[butt][|<\inn>]}[r]\ar@{.}@/^{.8\up}/[r]&
*{\otimes}\ar@{-}@*{[butt][|<\inn>]}[r]&
*{\otimes}}}

\newcommand{\bows}{\xymatrix@1@=\shrink{
*{\otimes}\ar@{-}@*{[butt][|<\inn>]}[r]\ar@{.}@/^{1.2\up}/^{~}[rrr]&
*{\otimes}\ar@{-}@*{[butt][|<\inn>]}[r]\ar@{.}@/^{.8\up}/[r]&
*{\otimes}\ar@{-}@*{[butt][|<\inn>]}[r]&
*{\otimes}}}

\newcommand{\vier}{\xymatrix@1@=\shrink{
*{\otimes}\ar@{}[rrr]|{}="m"\ar@{-}@*{[butt][|<\inn>]}[r]\ar@{.}[];"m"+<0pt,{1.2\up}>&
*{\otimes}\ar@{-}@*{[butt][|<\inn>]}[r]\ar@{.}[];"m"+<0pt,{1.2\up}>&
*{\otimes}\ar@{-}@*{[butt][|<\inn>]}[r]\ar@{.}[];"m"+<0pt,{1.2\up}>&
*{\otimes}\ar@{.}[];"m"+<0pt,{1.2\up}>}}

\newcommand{\fvieri}{\xymatrix@1@=1.25\shrink{
*{\bullet}\ar@{-}@*{[butt][|<\inn>]}[r]\ar@{.}@/^{1.6\up}/^{~}[rrr]\ar@{.}@/^{.7\up}/[r]&
*{\bullet}\ar@{-}@*{[butt][|<\inn>]}[r]\ar@{.}\ar@{.}@/^{.7\up}/[r]&
*{\bullet}\ar@{-}@*{[butt][|<\inn>]}[r]\ar@{.}\ar@{.}@/^{.7\up}/[r]&
*{\bullet}}}

\newcommand{\fvierii}{\xymatrix@1@=1.25\shrink{
*{\bullet}\ar@{-}@*{[butt][|<\inn>]}[r]\ar@{.}@/^{1.6\up}/^{~}[rrr]\ar@{.}@/^{.8\up}/[rr]&
*{\bullet}\ar@{-}@*{[butt][|<\inn>]}[r]\ar@{.}\ar@{.}@/^{.8\up}/[rr]\ar@{.}@/^{1.2\up}/[r]&
*{\bullet}\ar@{-}@*{[butt][|<\inn>]}[r]\ar@{.}&
*{\bullet}}}

\newcommand{\fvieriii}{\xymatrix@1@=1.25\shrink{
*{\bullet}\ar@{-}@*{[butt][|<\inn>]}[r]\ar@{.}@/^{1.6\up}/^{~}[rr]\ar@{.}@/^{.8\up}/[r]&
*{\bullet}\ar@{-}@*{[butt][|<\inn>]}_{~}[r]\ar@{.}\ar@{.}@/^{1.6\up}/^{~}[rr]&
*{\bullet}\ar@{-}@*{[butt][|<\inn>]}[r]\ar@{.}\ar@{.}@/^{.8\up}/[r]&
*{\bullet}}}

\newcommand{\govgvgv}{\xymatrix@1@=\shrink{
*{}\ar@{}[r]|{}="m" \ar@{}[];"m"|{}="n"
&*{\otimes}\ar@{-}@*{[butt][|<\inn>]}[];"n"
\ar@{-}@*{[butt][|<\inn>]}[r]\ar@{.}[r]+<0pt,{1.2\up}>
&*{\otimes}\ar@{-}@*{[butt][|<\inn>]}[r]
\ar@{.}[]+<0pt,{1.2\up}>
&*{\otimes}\ar@{}[r]|{}="o" \ar@{}"o";[r]|{}="p"
\ar@{.}[l]+<0pt,{1.2\up}>
\ar@{-}@*{[butt][|<\inn>]}[];"p" &*{} }}

\newcommand{\govgv}{\xymatrix@1@=\shrink{
*{}\ar@{}[r]|{}="m" \ar@{}[];"m"|{}="n"
&*{\otimes}\ar@{-}@*{[butt][|<\inn>]}[];"n"
\ar@{.}@/^\up/[r]\ar@{-}@*{[butt][|<\inn>]}[r]
&*{\otimes}\ar@{}[r]|{}="o" \ar@{}"o";[r]|{}="p"
\ar@{-}@*{[butt][|<\inn>]}[];"p" &*{} }}

\newcommand{\gov}{\xymatrix@1@=\shrink{
*{}\ar@{}[r]|{}="m" \ar@{}[];"m"|{}="n"
&*{\otimes}\ar@{-}@*{[butt][|<\inn>]}[];"n"
\ar@{}[r]|{}="o" \ar@{}"o";[r]|{}="p"
\ar@{-}@*{[butt][|<\inn>]}[];"p"&*{} }}

\newcommand{\go}{\xymatrix@1@=1.2\shrink{
*{}\ar@{-}@*{[butt][|<\inn>]}[r]&*{} }}

\newcommand{\gm}{\xymatrix@1@=1.2\shrink{
*{}\ar@{-}@*{[butt][|<\ick>]}^{\phantom{x}}_{\phantom{x}}[r]&*{} }}

\newcommand{\eny}{\xymatrix@1@=\shrink{
*=0{\circle*{6}}}}

\newcommand{\teny}{\xymatrix@1@=\shrink{
*=0{\circle*{5.5}}}}

\newcommand{\tey}{\xymatrix@1@=\shrink{
*=0{\blacklozenge}}}

\newcommand{\stab}{
\vcenter{\xymatrix@1@=\shrink{
*{\otimes}\ar@{.}[d]\\
*{\otimes}}}}

\newcommand{\herz}{
\vcenter{
\xymatrix@1@=\shrink{
*{\bullet}\ar@{-}@*{[butt][|<\ick>]}[r]\ar@{.}@/^{\up}/[r]
&*{\bullet}\\
*{}\ar@{}[r]|*{\bullet}="m"&*{}\ar@{.}[ul];"m"\ar@{.}[u];"m"
}}}

\newcommand{\pik}{\vcenter{\xymatrix@1@=\shrink{
*{}\ar@{}[r]|*{\bullet}="m"&*{}\\
*{\bullet}\ar@{-}@*{[butt][|<\ick>]}[r]\ar@{.}@/_{\up}/[r]\ar@{.}[];"m"&
*{\bullet}\ar@{.}[];"m"
}}}

\newcommand{\baum}{\vcenter{\xymatrix@1@=\shrink{
*{\otimes}\ar@{-}@*{[butt][|<\ick>]}[r]&
*{\otimes}\ar@{-}@*{[butt][|<\ick>]}[r]\ar@{.}[d]&
*{\otimes}\ar@{.}@/_{1.2\up}/
[ll]\\
*{}&*{\otimes}&*{} }}}

\newcommand{\blatt}{\vcenter{\xymatrix@1@=\shrink{
*{\bullet}\ar@{-}@*{[butt][|<\ick>]}[r]\ar@{.}@/^{\up}/[r]\ar@{.}[dr]&
*{\bullet}\ar@{-}@*{[butt][|<\ick>]}[r]\ar@{.}@/^{\up}/[r]&
*{\bullet}\ar@{.}[dl]\\
*{}&*{\bullet}&*{} }}}

\newcommand{\blume}{\vcenter{\xymatrix@1@=\shrink{
*{}&*{\otimes}&*{}\\
*{\otimes}\ar@{-}@*{[butt][|<\ick>]}[r]&
*{\otimes}\ar@{-}@*{[butt][|<\ick>]}[r]\ar@{.}[u]&
*{\otimes}\ar@{.}@/^{1.2\up}/^{~}[ll] }}}

\newcommand{\spade}{\vcenter{\xymatrix@1@=\shrink{
*{}&*{\bullet}\ar@{.}[dr]\ar@{.}[dl]&*{}\\
*{\bullet}\ar@{-}@*{[butt][|<\ick>]}[r]\ar@{.}@/_{\up}/[r]&
*{\bullet}\ar@{-}@*{[butt][|<\ick>]}[r]\ar@{.}@/_{\up}/[r]&
*{\bullet}}}}

\newcommand{\lad}{\vcenter{\xymatrix@1@=\shrink{
*{\otimes}\ar@{.}[d]\ar@{-}@*{[butt][|<\ick>]}[r]&*{\otimes}\ar@{.}[d]\\
*{\otimes}\ar@{-}@*{[butt][|<\ick>]}[r]&*{\otimes}}}}

\newcommand{\ladder}{\vcenter{\xymatrix@1@=\shrink{
*{\otimes}\ar@{.}[d]\ar@{-}@*{[butt][|<\ick>]}[r]&
*{\otimes}\ar@{.}[d]\ar@{-}@*{[butt][|<\ick>]}[r]&
*{\otimes}\ar@{.}[d]\\
*{\otimes}\ar@{-}@*{[butt][|<\ick>]}[r]&
*{\otimes}\ar@{-}@*{[butt][|<\ick>]}[r]&
*{\otimes} }}}

\newcommand{\icks}{\vcenter{\xymatrix@1@=\shrink{
*{\otimes}\ar@{.}[dr]\ar@{-}@*{[butt][|<\ick>]}[r]&*{\otimes}\\
*{\otimes}\ar@{.}[ur]\ar@{-}@*{[butt][|<\ick>]}[r]&*{\otimes}}}}

\newcommand{\icksix}{\vcenter{\xymatrix@1@=\shrink{
*{\otimes}\ar@{}[r]|{}="b"\ar@{}[];"b"|{}="a"\ar@{}"b";[r]|{}="c"
\ar@{.}[dr]\ar@{-}@*{[butt][|<\ick>]}[r]&*{\otimes}\\
*{\otimes}\ar@{}[r]|{}="e"\ar@{}[];"e"|{}="d"\ar@{}"e";[r]|{}="f"
\ar@{.}[ur]\ar@{-}@*{[butt][|<\ick>]}[r]&*{\otimes}
\ar@{.}"a";"f"\ar@{.}"b";"e"\ar@{.}"c";"d"}}}

\newcommand{\icksplode}{\vcenter{\xymatrix@1@=\shrink{
*{\otimes}\ar@{.}@/^{\up}/
[r]\ar@{}[r]|{}="b"\ar@{}[];"b"|{}="a"\ar@{}"b";[r]|{}="c"
\ar@{.}[dr]\ar@{-}@*{[butt][|<\ick>]}[r]&*{\otimes}\\
*{\otimes}\ar@{}[r]|{}="e"\ar@{}[];"e"|{}="d"\ar@{}"e";[r]|{}="f"
\ar@{.}[ur]\ar@{-}@*{[butt][|<\ick>]}[r]&*{\otimes}
\ar@{.}@/^{\up}/
[l]\ar@{.}"a";"f"\ar@{.}"b";"e"\ar@{.}"c";"d"}}}

\newcommand{\laddad}{\vcenter{\xymatrix@1@=\shrink{
*{\otimes}\ar@{}[r]|{}="b"\ar@{}[];"b"|{}="a"\ar@{}"b";[r]|{}="c"
\ar@{.}[d]\ar@{-}@*{[butt][|<\ick>]}[r]&*{\otimes}\\
*{\otimes}\ar@{}[r]|{}="e"\ar@{}[];"e"|{}="d"\ar@{}"e";[r]|{}="f"
\ar@{-}@*{[butt][|<\ick>]}[r]&*{\otimes}\ar@{.}[u]
\ar@{.}"a";"d"\ar@{.}"b";"e"\ar@{.}"c";"f"}}}

\newcommand{\lupo}{\vcenter{\xymatrix@1@=\shrink{
*{\otimes}\ar@{}[r]|{}="b"\ar@{}[];"b"|{}="a"\ar@{}"b";[r]|{}="c"
\ar@{.}[d]\ar@{-}@*{[butt][|<\ick>]}[r]&*{\otimes}
\ar@{.}[d]\ar@{}[r]|{}="v"\ar@{}[];"v"|{}="u"\ar@{}"v";[r]|{}="w"
\ar@{.}[dr]\ar@{-}@*{[butt][|<\ick>]}[r]&*{\otimes}
\ar@{.}@/_{\up}/
[l]
\ar@{}[r]|{}="j"\ar@{}[];"j"|{}="i"\ar@{}"j";[r]|{}="k"
\ar@{.}[d]\ar@{-}@*{[butt][|<\ick>]}[r]&*{\otimes}\\
*{\otimes}\ar@{}[r]|{}="e"\ar@{}[];"e"|{}="d"\ar@{}"e";[r]|{}="f"
\ar@{-}@*{[butt][|<\ick>]}[r]&*{\otimes}
\ar@{}[r]|{}="y"\ar@{}[];"y"|{}="x"\ar@{}"y";[r]|{}="z"
\ar@{.}[ur]\ar@{-}@*{[butt][|<\ick>]}[r]&*{\otimes}
\ar@{.}@/^{\up}/
[l]
\ar@{}[r]|{}="m"\ar@{}[];"m"|{}="l"\ar@{}"m";[r]|{}="n"
\ar@{-}@*{[butt][|<\ick>]}[r]&*{\otimes}
\ar@{.}"i";"l"\ar@{.}"j";"m"\ar@{.}"k";"n"
\ar@{.}"u";"z"\ar@{.}"v";"y"\ar@{.}"w";"x"
\ar@{.}"a";"d"\ar@{.}"b";"e"\ar@{.}"c";"f"}}}

\newcommand{\opul}{\vcenter{\xymatrix@1@=\shrink{
*{\otimes}\ar@{}[r]|{}="b"\ar@{}[];"b"|{}="a"\ar@{}"b";[r]|{}="c"
\ar@{.}[drrr]\ar@{-}@*{[butt][|<\ick>]}[r]&*{\otimes}
\ar@{}[r]|{}="v"\ar@{}[];"v"|{}="u"\ar@{}"v";[r]|{}="w"
\ar@{.}[d]\ar@{-}@*{[butt][|<\ick>]}[r]&*{\otimes}
\ar@{.}@/_{\up}/
[l]
\ar@{}[r]|{}="j"\ar@{}[];"j"|{}="i"\ar@{}"j";[r]|{}="k"
\ar@{.}[d]\ar@{-}@*{[butt][|<\ick>]}[r]&*{\otimes}\\
*{\otimes}\ar@{.}[urrr]\ar@{}[r]|{}="e"\ar@{}[];"e"|{}="d"\ar@{}"e";[r]|{}="f"
\ar@{-}@*{[butt][|<\ick>]}[r]&*{\otimes}
\ar@{}[r]|{}="y"\ar@{}[];"y"|{}="x"\ar@{}"y";[r]|{}="z"
\ar@{-}@*{[butt][|<\ick>]}[r]&*{\otimes}
\ar@{.}@/^{\up}/
[l]
\ar@{}[r]|{}="m"\ar@{}[];"m"|{}="l"\ar@{}"m";[r]|{}="n"
\ar@{-}@*{[butt][|<\ick>]}[r]&*{\otimes}
\ar@{.}"i";"f"\ar@{.}"j";"e"\ar@{.}"k";"d"
\ar@{.}"u";"x"\ar@{.}"v";"y"\ar@{.}"w";"z"
\ar@{.}"a";"n"\ar@{.}"b";"m"\ar@{.}"c";"l"}}}

\newcommand{\baumix}{\vcenter{\xymatrix@1@=\shrink{
*{\otimes}\ar@{-}@*{[butt][|<\ick>]}[r]&
*{\otimes}\ar@{}[r]|{}="b"\ar@{}[];"b"|{}="a"\ar@{}"b";[r]|{}="c"
\ar@{.}[dr]\ar@{-}@*{[butt][|<\ick>]}[r]&
*{\otimes}\ar@{-}@*{[butt][|<\ick>]}[r]&
*{\otimes}\ar@{.}@/_{1.2\up}/
[lll]\\
*{}&
*{\otimes}\ar@{}[r]|{}="e"\ar@{}[];"e"|{}="d"\ar@{}"e";[r]|{}="f"
\ar@{.}[ur]\ar@{-}@*{[butt][|<\ick>]}[r]&*{\otimes}
\ar@{.}"a";"f"\ar@{.}"b";"e"\ar@{.}"c";"d"
&*{} }}}

\newcommand{\blumix}{\vcenter{\xymatrix@1@=\shrink{
*{}&
*{\otimes}\ar@{}[r]|{}="e"\ar@{}[];"e"|{}="d"\ar@{}"e";[r]|{}="f"
\ar@{.}[dr]\ar@{-}@*{[butt][|<\ick>]}[r]&*{\otimes}
&*{}\\
*{\otimes}\ar@{-}@*{[butt][|<\ick>]}[r]&
*{\otimes}\ar@{}[r]|{}="b"\ar@{}[];"b"|{}="a"\ar@{}"b";[r]|{}="c"
\ar@{.}"a";"f"\ar@{.}"b";"e"\ar@{.}"c";"d"
\ar@{.}[ur]\ar@{-}@*{[butt][|<\ick>]}[r]&
*{\otimes}\ar@{-}@*{[butt][|<\ick>]}[r]&
*{\otimes}\ar@{.}@/^{1.2\up}/
[lll]
}}}

\newcommand{\ickser}{\vcenter{\xymatrix@1@=\shrink{
*{\otimes}\ar@{.}[drr]\ar@{-}@*{[butt][|<\ick>]}[r]&
*{\otimes}\ar@{.}[d]\ar@{-}@*{[butt][|<\ick>]}[r]&
*{\otimes}\\
*{\otimes}\ar@{.}[urr]\ar@{-}@*{[butt][|<\ick>]}[r]&
*{\otimes}\ar@{-}@*{[butt][|<\ick>]}[r]&
*{\otimes}}}}

\newcommand{\U}{\vcenter{\xymatrix@1@=\shrink{
*+<\uku,\iki>{\phantom{\bullet}}&*+<\uku,\iki>{\phantom{\bullet}}\\
*+<\uku,\iki>{\phantom{\bullet}}&*+<\uku,\iki>{\phantom{\bullet}}
\ar@{}[ul]|*++{\operator{U}}
\save[].[ul]*[F]\frm{}\restore
}}}

\newcommand{\R}{\vcenter{\xymatrix@1@=\shrink{
*+<\uku,\iki>{\phantom{\bullet}}&*+<\uku,\iki>{\phantom{\bullet}}\\
*+<\uku,\iki>{\phantom{\bullet}}&*+<\uku,\iki>{\phantom{\bullet}}
\ar@{}[ul]|*++{\operator{R}}
\save[].[ul]*[F]\frm{}\restore
}}}

\renewcommand{\L}{\vcenter{\xymatrix@1@=\shrink{
*+<\uku,\iki>{\phantom{\bullet}}&*+<\uku,\iki>{\phantom{\bullet}}\\
*+<\uku,\iki>{\phantom{\bullet}}&*+<\uku,\iki>{\phantom{\bullet}}
\ar@{}[ul]|*++{\operator{L}}
\save[].[ul]*[F]\frm{}\restore
}}}

\newcommand{\C}{\vcenter{\xymatrix@1@=\shrink{
*+<\uku,\iki>{\phantom{\bullet}}&*+<\uku,\iki>{\phantom{\bullet}}\\
*+<\uku,\iki>{\phantom{\bullet}}&*+<\uku,\iki>{\phantom{\bullet}}
\ar@{}[ul]|*++{\operator{C}}
\save[].[ul]*[F]\frm{}\restore
}}}

\newcommand{\CC}[1]{\vcenter{\xymatrix@1@=\shrink{
*+<\uku,\iki>{\phantom{\bullet}}&*+<\uku,\iki>{\phantom{\bullet}}\\
*+<\uku,\iki>{\phantom{\bullet}}&*+<\uku,\iki>{\phantom{\bullet}}
\ar@{}[ul]|*++{\operator{C}_{#1}}
\save[].[ul]*[F]\frm{}\restore
}}}

\newcommand{\PP}{\vcenter{\xymatrix@1@=\shrink{
*+<\uku,\iki>{\phantom{\bullet}}&*+<\uku,\iki>{\phantom{\bullet}}\\
*+<\uku,\iki>{\phantom{\bullet}}&*+<\uku,\iki>{\phantom{\bullet}}
\ar@{}[ul]|*++{\operator{\Phi}}
\save[].[ul]*[F]\frm{}\restore
}}}

\newcommand{\Ux}{\vcenter{\xymatrix@1@=\shrink{
*{}\ar@{}[r]|{}="m" \ar@{}[];"m"|{}="n"&
*+<0pt,\iki>{}\ar@{<-}_>{\vecc{\kappa}_1}@[];"n"&
*+<0pt,\iki>{}\ar@{}[r]|{}="o" \ar@{}"o";[r]|{}="p"\ar@{->}^>{\vecc{\kappa}_2}@[];"p"&*{}\\
*{}\ar@{}[r]|{}="q" \ar@{}[];"q"|{}="r"&
*+<0pt,\iki>{}\ar@{->}^>{\vecc{\kappa}_4}@[];"r"&
*+<0pt,\iki>{}\ar@{}[ul]|*++{\operator{U}}
\save[].[ul]*[F]\frm{}\restore
\ar@{}[r]|{}="s" \ar@{}"s";[r]|{}="t"\ar@{<-}_>{\vecc{\kappa}_3}@[];"t"&*{}
}}}

\newcommand{\Uxx}{\vcenter{\xymatrix@1@=\shrink{
*{}\ar@{}[r]|{}="m" \ar@{}[];"m"|{}="n"&
*+<0pt,\iki>{}\ar@{<-}_>{\vecc{\kappa}+\vec{q}/2}@[];"n"&
*+<0pt,\iki>{}\ar@{}[r]|{}="o" \ar@{}"o";[r]|{}="p"\ar@{->}^>{\vecc{\kappa}'+\vec{q}/2}@[];"p"&*{}\\
*{}\ar@{}[r]|{}="q" \ar@{}[];"q"|{}="r"&
*+<0pt,\iki>{}\ar@{->}^>{\vecc{\kappa}-\vec{q}/2}@[];"r"&
*+<0pt,\iki>{}\ar@{}[ul]|*++{\operator{U}}
\save[].[ul]*[F]\frm{}\restore
\ar@{}[r]|{}="s" \ar@{}"s";[r]|{}="t"\ar@{<-}_>{\vecc{\kappa}'-\vec{q}/2}@[];"t"&*{}
}}}

\newcommand{\laddadxi}{\vcenter{\xymatrix@1@=\shrink{
*{}\ar@{}[r]|{}="m" \ar@{}[];"m"|{}="n"&
*{\otimes}\ar@{<-}_>{\vecc{\kappa}+\vec{q}/2}@[];"n"
\ar@{}[r]|{}="b"\ar@{}[];"b"|{}="a"\ar@{}"b";[r]|{}="c"
\ar@{.}[d]\ar@{-}@*{[butt][|<\ick>]}[r]&
*{\otimes}\ar@{}[r]|{}="o" \ar@{}"o";[r]|{}="p"\ar@{->}^>{~~\vecc{\kappa}'+\vec{q}/2}@[];"p"&*{}\\
*{}\ar@{}[r]|{}="q" \ar@{}[];"q"|{}="r"&
*{\otimes}\ar@{->}^>{-\vecc{\kappa}'+\vec{q}/2~~}@[];"r"
\ar@{}[r]|{}="e"\ar@{}[];"e"|{}="d"\ar@{}"e";[r]|{}="f"
\ar@{-}@*{[butt][|<\ick>]}[r]&*{\otimes}\ar@{.}[u]
\ar@{}[r]|{}="s" \ar@{}"s";[r]|{}="t"\ar@{<-}_>{-\vecc{\kappa}+\vec{q}/2}@[];"t"&
*{}\ar@{.}"a";"d"\ar@{.}"b";"e"\ar@{.}"c";"f"}}}

\newcommand{\icksixxi}{\vcenter{\xymatrix@1@=\shrink{
*{}\ar@{}[r]|{}="m" \ar@{}[];"m"|{}="n"&
*{\otimes}\ar@{<-}_>{\vecc{\kappa}+\vec{q}/2}@[];"n"
\ar@{}[r]|{}="b"\ar@{}[];"b"|{}="a"\ar@{}"b";[r]|{}="c"
\ar@{.}[dr]\ar@{-}@*{[butt][|<\ick>]}[r]&
*{\otimes}\ar@{}[r]|{}="o" \ar@{}"o";[r]|{}="p"\ar@{->}^>{~~\vecc{\kappa}'+\vec{q}/2}@[];"p"&*{}\\
*{}\ar@{}[r]|{}="q" \ar@{}[];"q"|{}="r"&
*{\otimes}\ar@{->}^>{-\vecc{\kappa}'+\vec{q}/2~~}@[];"r"
\ar@{}[r]|{}="e"\ar@{}[];"e"|{}="d"\ar@{}"e";[r]|{}="f"
\ar@{.}[ur]\ar@{-}@*{[butt][|<\ick>]}[r]&*{\otimes}
\ar@{}[r]|{}="s" \ar@{}"s";[r]|{}="t"\ar@{<-}_>{-\vecc{\kappa}+\vec{q}/2}@[];"t"&
*{}\ar@{.}"a";"f"\ar@{.}"b";"e"\ar@{.}"c";"d"}}}

\newcommand{\laddadxt}{\vcenter{\xymatrix@1@=\shrink{
*{}\ar@{}[r]|{}="m" \ar@{}[];"m"|{}="n"&
*{\otimes}\ar@{<-}_>{\vec{K}+\vec{Q}/2~}@[];"n"
\ar@{}[r]|{}="b"\ar@{}[];"b"|{}="a"\ar@{}"b";[r]|{}="c"
\ar@{.}[d]\ar@{-}@*{[butt][|<\ick>]}[r]&
*{\otimes}\ar@{}[r]|{}="o" \ar@{}"o";[r]|{}="p"\ar@{->}^>{~~~\vec{K}'+\vec{Q}/2}@[];"p"&*{}\\
*{}\ar@{}[r]|{}="q" \ar@{}[];"q"|{}="r"&
*{\otimes}\ar@{->}^>{\vec{K}-\vec{Q}/2~}@[];"r"
\ar@{}[r]|{}="e"\ar@{}[];"e"|{}="d"\ar@{}"e";[r]|{}="f"
\ar@{-}@*{[butt][|<\ick>]}[r]&*{\otimes}\ar@{.}[u]
\ar@{}[r]|{}="s" \ar@{}"s";[r]|{}="t"\ar@{<-}_>{~~~\vec{K}'-\vec{Q}/2}@[];"t"&
*{}\ar@{.}"a";"d"\ar@{.}"b";"e"\ar@{.}"c";"f"}}}

\newcommand{\laddadxx}{\vcenter{\xymatrix@1@=\shrink{
*{}\ar@{}[r]|{}="m" \ar@{}[];"m"|{}="n"&
*{\otimes}\ar@{<-}_>{\vecc{\kappa}+\vec{q}/2}@[];"n"
\ar@{}[r]|{}="b"\ar@{}[];"b"|{}="a"\ar@{}"b";[r]|{}="c"
\ar@{.}[d]\ar@{-}@*{[butt][|<\ick>]}[r]&
*{\otimes}\ar@{}[r]|{}="o" \ar@{}"o";[r]|{}="p"\ar@{->}^>{~~\vecc{\kappa}'+\vec{q}/2}@[];"p"&*{}\\
*{}\ar@{}[r]|{}="q" \ar@{}[];"q"|{}="r"&
*{\otimes}\ar@{->}^>{\vecc{\kappa}-\vec{q}/2}@[];"r"
\ar@{}[r]|{}="e"\ar@{}[];"e"|{}="d"\ar@{}"e";[r]|{}="f"
\ar@{-}@*{[butt][|<\ick>]}[r]&*{\otimes}\ar@{.}[u]
\ar@{}[r]|{}="s" \ar@{}"s";[r]|{}="t"\ar@{<-}_>{~~\vecc{\kappa}'-\vec{q}/2}@[];"t"&
*{}\ar@{.}"a";"d"\ar@{.}"b";"e"\ar@{.}"c";"f"}}}

\newcommand{\icksixxx}{\vcenter{\xymatrix@1@=\shrink{
*{}\ar@{}[r]|{}="m" \ar@{}[];"m"|{}="n"&
*{\otimes}\ar@{<-}_>{\vecc{\kappa}+\vec{q}/2}@[];"n"
\ar@{}[r]|{}="b"\ar@{}[];"b"|{}="a"\ar@{}"b";[r]|{}="c"
\ar@{.}[dr]\ar@{-}@*{[butt][|<\ick>]}[r]&
*{\otimes}\ar@{}[r]|{}="o" \ar@{}"o";[r]|{}="p"\ar@{->}^>{~~\vecc{\kappa}'+\vec{q}/2}@[];"p"&*{}\\
*{}\ar@{}[r]|{}="q" \ar@{}[];"q"|{}="r"&
*{\otimes}\ar@{->}^>{\vecc{\kappa}-\vec{q}/2}@[];"r"
\ar@{}[r]|{}="e"\ar@{}[];"e"|{}="d"\ar@{}"e";[r]|{}="f"
\ar@{.}[ur]\ar@{-}@*{[butt][|<\ick>]}[r]&*{\otimes}
\ar@{}[r]|{}="s" \ar@{}"s";[r]|{}="t"\ar@{<-}_>{~~\vecc{\kappa}'-\vec{q}/2}@[];"t"&
*{}\ar@{.}"a";"f"\ar@{.}"b";"e"\ar@{.}"c";"d"}}}

\newcommand{\Lxx}{\vcenter{\xymatrix@1@=\shrink{
*{}\ar@{}[r]|{}="m" \ar@{}[];"m"|{}="n"&
*{\otimes}\ar@{.}[d]\ar@{-}@*{[butt][|<\ick>]}[r]
\ar@{<-}_>{\vecc{\kappa}+\vec{q}/2}@[];"n"&
*{\otimes}\ar@{.}[d]
\ar@{}[r]|{}="o" \ar@{}"o";[r]|{}="p"\ar@{->}^>{~~\vecc{\kappa}'+\vec{q}/2}@[];"p"&
*{}\\
*{}\ar@{}[r]|{}="q" \ar@{}[];"q"|{}="r"&
*{\otimes}\ar@{-}@*{[butt][|<\ick>]}[r]
\ar@{->}^>{\vecc{\kappa}-\vec{q}/2}@[];"r"&
*{\otimes}\ar@{}[r]|{}="s" \ar@{}"s";[r]|{}="t"\ar@{<-}_>{~~\vecc{\kappa}'-\vec{q}/2}@[];"t"&
*{}
}}}

\newcommand{\Lxi}{\vcenter{\xymatrix@1@=\shrink{
*{}\ar@{}[r]|{}="m" \ar@{}[];"m"|{}="n"&
*{\otimes}\ar@{.}[d]\ar@{-}@*{[butt][|<\ick>]}[r]\ar@{<-}_>{\vecc{\kappa}+\vec{q}/2}@[];"n"&
*{\otimes}\ar@{.}[d]\ar@{}[r]|{}="o" \ar@{}"o";[r]|{}="p"\ar@{->}^>{~~\vecc{\kappa}'+\vec{q}/2}@[];"p"&
*{}\\
*{}\ar@{}[r]|{}="q" \ar@{}[];"q"|{}="r"&
*{\otimes}\ar@{-}@*{[butt][|<\ick>]}[r]\ar@{->}^>{-\vecc{\kappa}'+\vec{q}/2~~}@[];"r"&
*{\otimes}\ar@{}[r]|{}="s" \ar@{}"s";[r]|{}="t"\ar@{<-}_>{-\vecc{\kappa}+\vec{q}/2}@[];"t"&
*{}
}}}

\newcommand{\Lxt}{\vcenter{\xymatrix@1@=\shrink{
*{}\ar@{}[r]|{}="m" \ar@{}[];"m"|{}="n"&
*{\otimes}\ar@{.}[d]\ar@{-}@*{[butt][|<\ick>]}[r]\ar@{<-}_>{\vec{K}+\vec{Q}/2}@[];"n"&
*{\otimes}\ar@{.}[d]\ar@{}[r]|{}="o" \ar@{}"o";[r]|{}="p"\ar@{->}^>{~~\vec{K}'+\vec{Q}/2}@[];"p"&
*{}\\
*{}\ar@{}[r]|{}="q" \ar@{}[];"q"|{}="r"&
*{\otimes}\ar@{-}@*{[butt][|<\ick>]}[r]\ar@{->}^>{\vec{K}-\vec{Q}/2}@[];"r"&
*{\otimes}\ar@{}[r]|{}="s" \ar@{}"s";[r]|{}="t"\ar@{<-}_>{~~\vec{K}'-\vec{Q}/2}@[];"t"&
*{}
}}}

\newcommand{\Cxx}{\vcenter{\xymatrix@1@=\shrink{
*{}\ar@{}[r]|{}="m" \ar@{}[];"m"|{}="n"&
*{\otimes}\ar@{.}[dr]\ar@{-}@*{[butt][|<\ick>]}[r]\ar@{<-}_>{\vecc{\kappa}+\vec{q}/2}@[];"n"&
*{\otimes}\ar@{.}[dl]\ar@{}[r]|{}="o" \ar@{}"o";[r]|{}="p"\ar@{->}^>{~~\vecc{\kappa}'+\vec{q}/2}@[];"p"&
*{}\\
*{}\ar@{}[r]|{}="q" \ar@{}[];"q"|{}="r"&
*{\otimes}\ar@{-}@*{[butt][|<\ick>]}[r]\ar@{->}^>{\vecc{\kappa}-\vec{q}/2}@[];"r"&
*{\otimes}\ar@{}[r]|{}="s" \ar@{}"s";[r]|{}="t"\ar@{<-}_>{~~\vecc{\kappa}'-\vec{q}/2}@[];"t"&
*{}
}}}

\newcommand{\legs}{\vcenter{\xymatrix@1@=\shrink{
*+<-\aka,\oko>{\phantom{\bullet}} \ar@{-}@*{[butt][|<\ick>]}[r]&*+<-\aka,\oko>{\phantom{\bullet}}\\
*+<-\aka,\oko>{\phantom{\bullet}} \ar@{-}@*{[butt][|<\ick>]}[r]&*+<-\aka,\oko>{\phantom{\bullet}}
}}}

\newcommand{\leg}{\vcenter{\xymatrix@1@R=\hoch@C=\breit{
*+<0pt,\iki>{} \ar@{-}@<-1.5pt>@*{[butt][|<\ick>]}[r]&*=0{}\\
*+<0pt,\iki>{} \ar@{-}@<1.5pt>@*{[butt][|<\ick>]}[r]&*=0{}
}}}

\newcommand{\kreiseins}{\vcenter{\xymatrix@1@=\shrink{
&*+++[o][F-]{1}}}
}

\newcommand{\kreiszwei}{\vcenter{\xymatrix@1@=\shrink{
&*+++[o][F-]{2}}}
}
\begin{document}

\title[]{Coherent Matter Wave Transport in Speckle Potentials}

\author{R C Kuhn$^{1,2}$, O Sigwarth$^1$, C Miniatura$^{2,3}$, D Delande$^4$
and C A M{\"u}ller$^1$}

\address{$^1$ Physikalisches Institut, Universit\"at Bayreuth,
D-95440 Bayreuth} 
\address{$^2$ Institut Non Lin\'eaire de Nice, UNSA, CNRS, 1361 route des Lucioles, F-06560 Valbonne}
\address{$^3$ Department of Physics, Faculty of Science, National University of Singapore, Singapore 117542}
\address{$^4$ Laboratoire Kastler Brossel, Universit\'e Pierre et
Marie Curie-Paris 6, 4 Place Jussieu, F-75005 Paris}

\ead{cord.mueller@uni-bayreuth.de}


\begin{abstract}
This article studies multiple scattering of matter waves by a disordered
optical potential in two and in three dimensions. We
calculate fundamental transport quantities such as the scattering
mean free path $\ells$, the Boltzmann transport mean free path
$\elltrb$, and the Boltzmann diffusion constant $\boltz{D}$, using a
diagrammatic Green functions approach in the weak scattering regime. 
Coherent multiple scattering
induces interference corrections known as weak
localization which entail a reduced diffusion constant. We derive the corresponding expressions for matter
wave transport in an correlated speckle potential and provide the
relevant parameter values for a possible experimental study of this
coherent transport regime, including the critical crossover to the
regime of strong or Anderson localization.  
\end{abstract}


\section{Introduction}
 
The observation  of the first gaseous Bose-Einstein condensates in 1995
and, a few years later, of the first gaseous ultra-cold fermion gases constitutes a major achievement
in the field of atomic physics.
Loading ultra-cold degenerate gases, be it fermions, bosons, or
fermion-boson mixtures into optical lattices has opened fascinating
new perspectives for the study of condensed-matter quantum physics
\cite{lewenstein, dalfovo, coherent, duine}.
A particularly promising line of research is the
experimental and theoretical study of quantum phase transitions
\cite{sachdev}. 
Indeed, the high degree of control and precision achieved in these
experiments has allowed systematic studies of physical phenomena
that are observed until now mostly in solid-state systems.

A natural evolution in this domain is the investigation of the
influence of \textit{disorder}
\cite{ahufinger}  
which can induce a Bose-glass phase for strongly interacting bosons 
\cite{fisher}, and a Lifshitz glass
\cite{Lugan06} or an Anderson glass \cite{damski} for weaker interaction. 
Experimentally, a major milestone has been reached with the realization
of quasi-1D condensates evolving in a speckle light
field \cite{florence, aspect, fort, schulte, bouyer}.
Such disordered optical potentials can be easily generated and their
statistical properties are well known \cite{goodman}. 
In two dimensions, atomic diffusion has been
studied in optical quasi-crystals with five-fold symmetry for atomic clouds
in the dissipative regime \cite{verkerk} and Bose-Einstein condensates
\cite{santos} covering the intermediate regime between ordered and
completely disordered systems. 
Most recently the experimental observation of 
the onset of the Bose glass phase for
ultra-cold atoms in a bichromatic optical lattice has been reported \cite{bose}.

Disorder has long proven to be a crucial ingredient to understand coherent transport properties.
The most prominent example is the
\textit{weak localization}
phenomenon in mesoscopic physics \cite{mesoscopic, akker, rammer},
which has been studied extensively for electrons
\cite{bergmann, lee} and for classical waves \cite{rossum, resonant}.
Weak localization
arises from interference between 
multiply scattered waves in a random medium. 
This interference survives
the configuration average over many realizations of
disorder and reduces the conductivity that enters the Drude model
for electron transport \cite{ashcroft} and the diffusion constant of classical radiative transport theory
\cite{ishimaru, chandra}.
Disorder can even induce a
metal-insulator transition, known as the strong (or Anderson)
localization transition \cite{anderson, gang4, bvtloc, mott, kramer},
that has been studied extensively in the framework of the tight binding model \cite{kroha,kopp}.
In the localized regime the quantum states cease to extend over the
whole system and become spatially localized with the consequence that
quantum transport through the system is exponentially suppressed. 

From a theoretical point of view, atom transport shares many
similarities with the radiative transfer theory \cite{henkelreport} and even
more with electron transport theory \cite{akker, rammer}.
Unfortunately, the unambiguous observation of wave localization phenomena for matter
waves is difficult if the quantum
evolution is interaction-dominated. 
But there are several ways to achieve the interaction-free
regime: for example, one can let the ultra-cold gas first expand, so
as to decrease its spatial density, before switching on the speckle
potential. 
A more elaborate way would be to tune the two-body interactions by using a Feshbach
resonance \cite{duine,timmermans}.
In this case, by scanning a magnetic field, one can continuously go
from the independent-particle regime to the strongly 
interacting regime. This offers the possibility to study in a
controlled way how localization is affected by interactions.

In this paper we focus on the 
impact of disorder in the independent-particle regime and discard both atom-atom interactions and quantum
statistical effects. Such a situation can be reached with ultra-cold atoms produced from a Bose-Einstein
condensate or a Fermi degenerate gas by opening the trap and decreasing the atomic density
far below the quantum degenerate regime.
Our results therefore apply to non-interacting ultra-cold gases at low densities.
Our paper is inspired by Vollhardt and W\"olfle's seminal work on electron transport theory \cite{vollhardt}
and is an extension of our previously published results on
localization of matter waves in two-dimensional speckle potentials
\cite{letter}. 
We calculate experimentally relevant transport quantities, such as the
scattering mean free path, the transport mean free path, 
the diffusion constant and the weak localization corrections, treating
the two- and three-dimensional cases in parallel.  
Contrary to the case of 1D transport studied recently by
Sanchez-Palencia et al.\ \cite{LSP06b} to which this theory
can also be applied, the 2D and 3D cases are
of special interest because 3D allows for the Anderson localization
transition at a finite disorder strength, whereas 2D is the marginal
case (the lower critical
dimension) where analytical results are particularly important. 
We show that it should be possible, with the
current experimental state of the art, to observe weak and
strong localization effects provided that phase-breaking mechanisms
are under control.
 
The paper is organized as follows: in \sref{speckle} we
discuss the optical potential and its statistical properties and
we derive the dimensionless Schr\"odinger equation which
governs the atomic motion in the speckle potential.
Sections \ref{AvPro} and \ref{diff.trans.sec} (and corresponding appendices containing
the technical details)
are intended to give an introduction to the
diagrammatic perturbation theory. 
We obtain analytic expressions for the
configuration-averaged propagator and the scattering mean free
path as well as for the intensity relaxation kernel and the transport mean
free path. Section \ref{coscat} (and corresponding appendices) are devoted to the calculation
of quantum corrections to classical transport. We derive the reduced diffusion constant
in 2 and 3D, and study the strong localization onset. 

\section{Atomic Hamiltonian Dynamics}
\label{speckle}

In the present section, we formulate the general description of the
single-particle dynamics for noninteracting cold atoms in disordered speckle
potentials. 

\subsection{Light shifts}

\begin{figure}
\centering
\psfrag{I}{$I(\bi{r})$}
\psfrag{x}{$x/\zeta$}
\psfrag{y}{$y/\zeta$}
\psfrag{R}{$R$}
\psfrag{z}{$z$}
\psfrag{L}{$L$}
\includegraphics[width=0.7\linewidth]{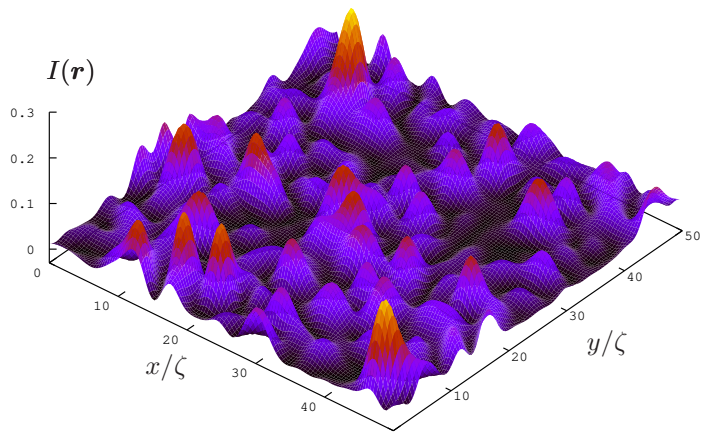}
\caption{
Intensity plot of a 2D speckle pattern, numerically generated as described by Horak et al. \cite{horak}.
The positions $x$ and $y$ are given in units of the speckle correlation length $\zeta$.
}
\label{fig:diffusion}
\end{figure}

When an atom is exposed to electromagnetic radiation, it is polarized,
and its energy levels are shifted. 
In the case of interaction with a laser light field, these energy
shifts are called light shifts \cite{cohen}. In the dipolar
approximation, the light shifts are proportional to the field
intensity evaluated at the center-of-mass of the atom. If
the field intensity is space dependent, so are the light shifts, 
and a moving atom experiences dipolar forces altering its trajectory.
By conveniently tailoring the space and time 
dependence of the light field, one can produce a great
variety of potentials for guiding the atomic motion.

In the present paper, we consider the interaction of a
two-level atom (mass $m$, internal electronic ground state
$|g\rangle$, energy separation $\hbar \atom{\omega}$ to the excited state $|e\rangle$ with natural energy
linewidth $\hbar \Gamma$, electric dipole moment $d_e$) 
with a monochromatic electromagnetic laser
field $\light{\efield}(\bi{r})$ (wave number $\laser{k}$,
wavelength $\laser{\lambda} = 2\pi/\laser{k}$, angular
frequency $\laser{\omega} = c \laser{k}$). The two-level description
is appropriate for atoms like
strontium \cite{yannick} with a nondegenerate electronic ground state
or atoms like rubidium whose ground-state degeneracy
is lifted by a strong magnetic field \cite{olivier}.

The coupling strength between the atom and the electromagnetic laser
field is described by the angular Rabi frequency 
$\hbar\light{\Omega}(\bi{r})= -d_e \light{\efield}(\bi{r})$.
The detuning
$\laser{\delta} = \laser{\omega}-\atom{\omega}$   from an optical
resonance is generally small compared to
the transition frequency, 
$\laser{\delta} \ll \laser{\omega}$. In this case, 
anti-resonant interaction terms can be ignored (rotating-wave
approximation) \cite{eberly}.
Hereafter, we assume that the atom is initially prepared in
its internal ground state, and that the laser detuning satisfies
$\laser{\delta} \gg \Gamma, \laser{\Omega}$, where $\laser{\Omega}$ is
the mean value of the Rabi frequency. The 
transition amplitude to the excited state is then small, 
and the ground-state light shift at position $\bi{r}$ 
caused by the laser intensity $\light{I}(\bi{r})=(\frac{1}{2}\varepsilon_0 c)|\light{\efield}(\bi{r})|^2$ 
is 
\begin{equation}\label{eq:pot}
   V(\bi{r}) \approx \frac{\hbar|\light{\Omega}(\bi{r})|^2}{4\delta} = 
   \frac{\hbar \Gamma}{8} \, \frac{\Gamma}{\laser{\delta}} \,
\frac{\light{I}(\bi{r})}{\sat{I}} .
\end{equation}
The saturation intensity $\sat{I}$ is a characteristic of the atom under
consideration.
The atomic motion is then governed by the effective Hamiltonian
$H=\bi{p}^2/2m+V(\bi{r})$. 

The atomic evolution is purely Hamiltonian only up to
some time $\inel{\tau}$: because of the coupling to the photon vacuum fluctuations, the
light-atom interaction also contains a dissipative term 
which limits the temporal coherence of the atomic wave
function. 
The average inelastic
scattering rate or inverse inelastic time can be calculated using the optical Bloch
equations \cite{cohen} and reads 
\begin{equation}\label{eq:inel}
\inel{\gamma} = \frac{1}{\inel{\tau}}\approx \frac{\Gamma}{\laser{\delta}} \frac{\laser{V}}{\hbar}.
\end{equation}
Here $\laser{V}=\av{V(\bi{r})}$ is the average light shift, given 
in terms of the average intensity
$\laser{I}=\av{\light{I}(\bi{r})}$:  
\begin{equation}
\laser{V}=\frac{\hbar\Gamma}{8}\,\frac{\Gamma}{\laser{\delta}}\,
\frac{\laser{I}}{\sat{I}} . 
\end{equation}
Coherent interference of multiple scattering amplitudes can
affect the atomic dynamics if 
this dissipation is controlled at an arbitrarily low rate by using a
sufficiently far-detuned speckle field $\laser{\delta}\gg\Gamma$ at 
constant average potential height $\laser{V}$.

\subsection{Speckle potential}
\label{stat}

A monochromatic speckle pattern with random 
intensity and phase variations defines a disordered potential
$V(\bi{r})$ as displayed in \fref{fig:diffusion}. 
The corresponding electric field
$\light{\efield}(\bi{r})$ is a superposition
of many complex field amplitudes with zero mean. Then, as
stated by the central limit theorem, the real and imaginary parts
of the field amplitude are uncorrelated Gaussian random
variables \cite{goodman}. Thanks to the Gaussian moment
theorem \cite{mandelwolf}, all correlation functions
of the electric field can be expressed in terms of the pair correlation function
$\Corr_{\efield}(\bi{r},\bi{r}')=
\av{\light{\efield}^*(\bi{r})\light{\efield}(\bi{r}')}$. 
The ensemble average typically restores translational invariance,
$\Corr_{\efield}(\bi{r},\bi{r}')= \Corr_{\efield}(\bi{r}-\bi{r}')$, and 
one defines the dimensionless field correlation function 
or so-called complex degree of coherence \cite{bornwolf}: 
\begin{equation}
 \gamma_d(\bi{r}) =
\frac{\av{\light{\efield^*}(\bi{r}+\bi{r}')\light{\efield}(\bi{r}')}}
{\av{|\light{\efield}(\bi{r})|^2}}. 
\label{correlE}
\end{equation}
The index $d=2,3$ indicates that the correlation function depends on
the choice of a two- or three-dimensional geometry as discussed below. 
Since the light-shift potential \nref{eq:pot} is proportional to the \textit{square} of the field, 
the speckle potential itself
does \emph{not} have a Gaussian distribution. Nevertheless, because of the Gaussian
character of the underlying field, all potential correlation
functions can be split down to sums of products of the field correlation
function (\ref{correlE}). 
The potential pair
correlation function $\av{V(\bi{r}')V(\bi{r}'+\bi{r})}$ is
proportional to the fourth-order field correlation $\av{\light{\efield^*}(\bi{r}')\light{\efield}(\bi{r}')
\light{\efield^*}(\bi{r}'+\bi{r})\light{\efield}(\bi{r}'+\bi{r})}$.
It is given by
\begin{equation}\label{eq:specs}
\av{V(\bi{r}')V(\bi{r}'+\bi{r})} = \laser{V}^2 [1+|\gamma_d(\bi{r})|^2].
\end{equation}
Obviously, the potential dispersion is proportional to the 
average potential value $\avV$ itself.
We rewrite the light-shift potential as
$V(\bi{r})=\avV[1+\fluct(\bi{r})]$ such that 
the dimensionless fluctuations $\fluct(\bi{r})$ have the reduced two-point correlation function 
\begin{equation}\label{correl}
\Corr_d(\bi{r}) = \av{\fluct(\bi{r}') \fluct(\bi{r}' + \bi{r})} =
|\gamma_d (\bi{r})|^2. 
\end{equation}
$\Corr_d(\bi{r})$ is normalized such
that $\Corr_d(0)=1$ and decays to zero over a characteristic
spatial scale $\zeta$ known as the correlation length of the speckle potential.
As shown below, $\zeta$ is a crucial parameter for the
dynamics of the cold atoms in the speckle potential.

A two-dimensional ($d=2$) speckle as displayed in \fref{fig:diffusion} can be produced by reflection of a laser 
from a rough surface or by transmission through a phase mask
\cite{boiron}.
In the far-field $z \gg R$ from the phase mask,
where $z$ denotes the distance from the phase mask to the observation
plane and $R$ denotes the radius of the phase mask, the speckle
interference pattern can be regarded as 
quasi two-dimensional since the speckle grains are
very elongated in the $z$-direction, orthogonal to the illuminated
surface.
 Confining atoms in the plane transverse to $z$ in
 the far-field thus realizes a 2D situation.
A uniformly illuminated circular phase mask with radius $R$ yields
a complex degree of coherence that takes the
following form for small relative distances $r \ll z$ \cite{goodman}
\begin{equation}
\label{goodman}
 \gamma_2(\bi{r})= 2\,\frac{\Bessel_1(\alpha \laser{k}
r)}{\alpha\laser{k} r}. 
\end{equation}
$\Bessel_1(x)$ denotes the first-order Bessel function. The
geometrical factor $\alpha \approx R/z \ll 1$ is the numerical aperture of the imaging device; 
a typical
order of magnitude in recent experiments is $\alpha\approx 0.1$
\cite{florence} or better, $\alpha \approx 0.45$  \cite{Clement06}. 
 The speckle fluctuations originate from the coherent superposition
of purely monochromatic wave vectors. This implies a diffraction limit:
two points in the speckle field are correlated if
sufficiently close to each other, $\lim_{r\to 0}\gamma_2(\bi{r}) =1$. 
In the 2D case, the correlation function $ \gamma_2(\bi{r})$ decreases to zero over a
characteristic length scale $\zeta=1/\alpha\laser{k} 
= \laser{\lambda}/2\pi\alpha$ such that
the potential fluctuations are uncorrelated only for distances larger
than $\zeta$. 

To produce a three-dimensional ($d=3$) disordered configuration, the
speckle grains should be obtained as the interference
pattern of many wave vectors spanning the largest possible angular
aperture. 
Ideally, this situation corresponds to the interference
pattern obtained inside a closed optical cavity, e.g., an
integrating sphere. The complex degree of coherence is then given
by \cite{berry}
\begin{equation}
\label{berry} 
\gamma_3(\bi{r})= \frac{\sin (\laser{k} r)}{\laser{k} r},
\end{equation}
where the correlation length is now
$\zeta=1/\laser{k}$, corresponding to a numerical aperture $\alpha \to
1$.

Another possibly interesting configuration is provided by the
speckle field at the proximity of a rough interface illuminated by
monochromatic coherent light \cite{apostol}. 
Due to the contributions of
\emph{evanescent} components, the average intensity then decreases
exponentially with the distance $z$ to the surface. 
If the surface is rough on short scales, 
the near-field speckle correlation length $\zeta$ is 
smaller than $\laser{\lambda}$, which is impossible for far-field
speckle patterns.
Experimentally, one would have to bring the atoms sufficiently close to the surface in a
controllable way and restrict the dynamics to a plane parallel to the
surface \cite{perrin}. This situation deserves a special study and will not be considered in
the following.

A popular model for disordered potentials in the quantum transport
literature is a spatially $\delta$-correlated potential with a Gaussian distribution
of potential strength. The previous considerations show that a monochromatic speckle
potential is neither $\delta$-correlated nor Gaussian. 
However, 
only the potential fluctuations as seen by the moving
atoms are relevant for the atomic dynamics. 
We will see in the following that 
speckle fluctuations appear effectively $\delta$-correlated only when the atomic de Broglie wavelength is much larger than the correlation length $\zeta$.
Attaining this quantum regime requires sub-recoil cooling techniques.

\subsection{Dimensionless Schr\"odinger Equation}

The correlation length $\zeta$ of the speckle fluctuations defines a
natural physical length scale. In turn, it also defines natural scales for momentum,
wave number, velocity, energy, angular frequency, and time:
\begin{equation} 
p_{\zeta} = \hbar k_\zeta = m v_{\zeta} = \frac{\hbar}{\zeta},
\qquad 
E_\zeta = \hbar \omega_{\zeta} = \frac{\hbar}{\tau_{\zeta}} =
\frac{\hbar^2}{m\zeta^2}.
\end{equation}
In the 3D-speckle case (\ref{berry}), one has $\zeta = \laser{k}^{-1}$
such that 
$v_{\zeta}= \hbar \laser{k} /m = \recoil{v}$ is the familiar atomic
recoil velocity, whereas $E_\zeta = 2 \recoil{E}$ is twice
the atomic recoil energy $\recoil{E}=\frac{1}{2}{m\recoil{v}^2}$. In the
2D-speckle case (\ref{goodman}), $\zeta = (\alpha \laser{k})^{-1}$
and $v_{\zeta}= \alpha \recoil{v} \ll \recoil{v}$, whereas $E_\zeta = 2
\alpha^2 \recoil{E} \ll \recoil{E}$.
Scaling all dynamical variables with these units ($\bi{k}=\bi{p}/\hbar$ is the wave vector
of the atom),
\begin{equation}\label{dimensionless.eq} 
\brho = \bi{r}/\zeta, \quad
\tau =t/\tau_{\zeta}, \quad 
\bkappa =\bi{k}/k_{\zeta}, \quad
\varepsilon =E/E_\zeta,
\end{equation}
the Schr\"odinger equation reappears in the dimensionless form
$i\partial_\tau|\psi\rangle = H|\psi\rangle$ 
with the Hamiltonian
\begin{equation}
H = \frac{\bkappa^2}{2} + \eta + \eta \,\fluct(\brho).
\label{eq:dimH}
\end{equation}
In the rescaled units, the canonical commutator reads $[\brho,\bkappa] =i$. The parameter $\eta$ is the strength of the potential fluctuations
in units of the correlation energy, 
\begin{equation}\label{eq:dim}
\eta = \frac{\avV}{E_\zeta} =
\frac{\hbar\Gamma}{8E_\zeta}\;
\frac{\Gamma}{\laser{\delta}}\;\frac{\laser{I}}{\sat{I}}. 
\end{equation}
The constant term $\eta$ in the Hamiltonian (\ref{eq:dimH}) can be
reabsorbed by fixing the origin of energies at $\avV$. 
In spatial representation and rescaled units, the stationary Schr\"odinger
equation at energy $\varepsilon$ is now:
\begin{equation}\label{eq:schdimf}
\big[\frac{1}{2} \nabla_{\brho}^2 + \varepsilon - \eta\, \fluct(\brho)\big]\,\psi(\brho) = 0.
\end{equation}
The 2-point correlation functions of the potential fluctuations in our rescaled
units are
\numparts
\begin{eqnarray}
\Corr_2(\brho) = \Big[2\,\frac{\Bessel_1(\rho)}{\rho}\Big]^2, \label{eq:corl2d}
\\
\Corr_3(\brho) = \Big[\frac{\sin\,
(\rho)}{\rho}\Big]^2 .\label{eq:corl3d}
\end{eqnarray}
\endnumparts

The equation \nref{eq:schdimf} differs from the Helmholtz equation
obtained for classical electromagnetic waves propagating in random dielectric media.
In our case the potential fluctuations
do not
depend on energy. Hence the treatment for matter waves (be it atoms or
electrons) is much simpler than for classical waves, where this energy
dependence implies significant corrections to dynamical quantities
such as the transport speed of light 
\cite{resonant}.

\section{Effective Medium}
\label{AvPro}

The states of a wave scattered by a disordered potential
are different from one realization of disorder
to another. As a consequence, only expectation values obtained by averaging over
many such realizations provide a useful characterization of
the transport processes at work. 
In the language of standard quantum
transport, tracing out the speckle impurities leads to a 
dispersion relation with finite spectral width $\gamma_{\bkappa}$
 for the plane-wave states $\ket{\bkappa}$. 
Equivalently, but put into the language of atom optics, averaging over
speckle realizations introduces an effective medium that is
characterized by a complex wave vector $\bkappa(\varepsilon)$ for the
propagating matter wave at energy $\varepsilon$.
In this section, we calculate the
dispersion relation in the weakly disordered regime where diagrammatic
perturbation theory \cite{akker, rammer, pingsheng}
proves particularly powerful. 
This theory is well known for
the description of classical waves in a
medium with a fluctuating index of refraction or the transport of
electrons in a disordered metal. However, in contrast to
point-like impurities encountered for the scattering of electrons,
the fluctuations of the optical potential exhibit spatial
correlations leading to anisotropic scattering.
In this sense the scattering of atomic matter waves in a speckle
potential shares many similarities with the scattering of light in
nematic liquid crystals where correlations between thermal
fluctuations of the nematic director play an important
role \cite{bartlq}. 

\subsection{Retarded Green operator}
\label{SEOP}

The retarded Green operator or resolvent $G(\varepsilon)$ for the 
stationary Schr\"odinger equation (\ref{eq:schdimf}) at reduced energy
$\varepsilon$ is the Fourier
transform of the forward time evolution operator 
$\Theta(\tau) \,U(\tau) = \frac{i}{2\pi} \int \rmd\varepsilon \; G(\varepsilon) \,
e^{-i \varepsilon \tau}$,
with the Heaviside function $\Theta(\tau)$. The Green operator
satisfies the equation
\begin{equation}\label{eq:LSop}
G(\varepsilon)=G_0(\varepsilon) + G_0(\varepsilon) \,\eta\,\fluct
\,G(\varepsilon). 
\end{equation}
The free retarded Green operator
$G_0(\varepsilon)=[\varepsilon-H_0+i 0^+]^{-1}$ 
is diagonal in momentum space with propagator matrix elements 
$\bra{\bkappa'}G_0(\varepsilon)\ket{\bkappa} = (2\pi)^d
\delta(\bkappa-\bkappa')G_0(\kappa,\varepsilon)$,
where
$\delta(\bkappa)=\delta(\kappa_1)\cdots\delta(\kappa_d)$,
and 
\begin{equation}
\label{G0.eq}
G_0(\kappa,\varepsilon)= [\varepsilon- \kappa^2/2+i0^+]^{-1}.
\end{equation} 
Iteration of
\nref{eq:LSop} yields the usual Born series (suppressing energy arguments
for brevity) 
\begin{equation}
 G = G_0 + \eta \,G_0 \fluct G_0 + \eta^2\, G_0 \fluct G_0
\fluct G_0 + \ldots 
\end{equation} 
Taking the configuration average of the series, one obtains 
\begin{equation}
\av{G} = G_0
+ \eta^2\, G_0 \,\av{\fluct G_0 \fluct}\, G_0
+ \eta^3\, G_0 \,\av{\fluct G_0 \fluct G_0 \fluct}\, G_0
+ \ldots ,
\label{eq:dborn}
\end{equation}
where the linear term in $\eta$ vanishes because $\av{\fluct}=0$. 
It appears that knowing the average resolvent requires to calculate all higher-order
correlation functions of the potential fluctuations $\delta
V$. However, one is interested in determining the effect of the
disorder on the energy levels $\varepsilon(\bkappa)$. To this end, the
Born series is recast into the following form, known as the Dyson equation:
\begin{equation}
\av{G} = G_0 + G_0 \Sigma \, G_0 + G_0 \Sigma \,G_0 \Sigma\,G_0 + \ldots = G_0 + G_0 \Sigma\,\av{G}.
\label{eq:dysong}
\end{equation}
The retarded self-energy operator
$\Sigma(\varepsilon)$ contains all irreducible correlation
functions, \ie correlations that cannot be split into products of independent
factors by suppressing a single propagator $G_0$ \cite{frisch}. 
Recognizing a geometric series, one can formally solve the Dyson 
equation as 
$\av{G}(\varepsilon)=[G_0(\varepsilon)^{-1}-\Sigma(\varepsilon)]^{-1}$.

The average over many realizations of disorder restores translational invariance. Consequently,
$\av{G}(\varepsilon)$ is diagonal in momentum space just like the
free-space propagator $G_0(\varepsilon)$. 
Since the disordered potential also preserves space isotropy on
average, as exemplified by the scalar correlation functions (\ref{goodman}) and (\ref{berry}),
the propagator matrix elements $\av{G}(\kappa,\varepsilon)$ in momentum
space can only depend on the momentum modulus
$\kappa=|\bkappa|$.
The same conclusion holds for the self-energy
$\Sigma(\kappa,\varepsilon)$ such that
\begin{equation}
\label{eq:ave}
\av{G}(\kappa,\varepsilon)=[\varepsilon-\kappa^2/2-\Sigma(\kappa,\varepsilon)]^{-1}.
\end{equation}

\subsection{Disorder-broadened dispersion relation}

Information about the relative weight, energy, and life time of 
excitations dressed by the disordered medium
is contained in the spectral function \cite{Mahan}
\begin{equation} 
\fl A(\kappa,\varepsilon)= -2\,\Im \av{G}(\kappa,\varepsilon) 
= \frac{ -2\,\Im \Sigma (\kappa,\varepsilon) }{(\varepsilon - \kappa^2/2
-\Re\Sigma (\kappa,\varepsilon))^2 + ( \Im \Sigma (\kappa,\varepsilon))^2}. 
\label{spectral_function.eq}
\end{equation}
The spectral function is positive, $A(\kappa,\varepsilon)\ge 0$
(because the retarded self-energy has a negative imaginary part), 
and normalized to unity
$\int(\rmd\varepsilon/2\pi)A(\kappa,\varepsilon) = 1$. It can thus be
seen to represent the probability density for excitations with 
wave vector
$\kappa$ to have an energy $\varepsilon$.
Its trace over momentum states yields the average density of states
per unit volume, 
\begin{equation}\label{avdos.def}
 \int \frac{\dd\kappa}{(2\pi)^{d}} A(\kappa,\varepsilon) = 2\pi
\avdos(\varepsilon) . 
\end{equation} 
The spectral function of a free particle, 
$A_0(\kappa,\varepsilon)=2\pi\delta(\varepsilon-\kappa^2/2)$, projects onto the
energy shell $\kappa^2= 2\varepsilon$, such that the free density of states reads as usual 
\begin{equation}
\dos(\varepsilon) 
=  \int \frac{\dd\kappa}{(2\pi)^d} \;\delta (\varepsilon-\kappa^2/2)
=\frac{\usphere}{(2\pi)^d} 
	(2\varepsilon)^{d/2-1}		   .
\end{equation}
Here $\usphere=\int\rmd \Omega_d = 2\pi^{d/2}/\Gamma(d/2)$ denotes
the surface of the unit sphere, with the Euler gamma function $\Gamma(x)$.

Whenever the corrections due to disorder are finite but small,
$A(\kappa,\varepsilon)$ as a function of $\varepsilon$ at fixed $\kappa$ is
strongly peaked around the energy $\varepsilon_\kappa$ defined as a zero
of the real dispersion relation 
$\varepsilon_\kappa-\kappa^2/2-\Re\Sigma(\kappa,\varepsilon_\kappa)=0$. 
A Taylor expansion to lowest order around this point, 
$\varepsilon - \kappa^2/2
-\Re\Sigma (\kappa,\varepsilon) = (\varepsilon-\varepsilon_\kappa)
Z(\kappa)^{-1}$, 
defines the so-called renormalization constant
$
Z(\kappa)^{-1} = 1- \left. \partial_\varepsilon
\Re\Sigma(\kappa,\varepsilon)\right |_{\varepsilon=\varepsilon_\kappa}$. 
This brings the spectral function into Lorentzian form, 
\begin{equation} 
A(\kappa,\varepsilon)= Z(\kappa) 
 \frac{ \scat{\gamma} }{(\varepsilon - \varepsilon_\kappa)^2
+ \scat{\gamma}^2/4}.
\end{equation}
The mode $\ket{\bkappa}$ now represents an excitation with spectral weight
$Z(\kappa)$ at energy $\varepsilon_\kappa$, and with a finite spectral
width or elastic scattering rate 
\begin{equation}
\scat{\gamma} = -2 Z(\kappa) \Im
\Sigma(\kappa,\varepsilon_\kappa). 
\label{gamma_kappa}
\end{equation}
This mode describes an atom that is dressed
by the speckle fluctuations and scattered into a different mode
$\ket{\bkappa'}$ on average after a time
$\scat{\tau}=1/\scat{\gamma}$.

Equivalently, one can determine the wave vector $\kappa(\varepsilon)$ 
corresponding to a given energy $\varepsilon$ as the
solution to the complex dispersion relation $\varepsilon -
\kappa_\varepsilon^2/2 - \Sigma(\kappa_\varepsilon,\varepsilon)=0$.
This is the standard approach in optics \cite{bornwolf} and atom
optics \cite{adams}.
Determining the effective dispersion 
relation of matter waves in disordered speckle potentials is thus reduced to
calculating the self-energy.

\subsection{The weak scattering approximation}
\label{weak}

The self-energy defined by the Dyson equation (\ref{eq:dysong}) can be 
expanded as a power series
\begin{equation}\label{self}
\Sigma = \sum_{n \ge 2} \Sigma_n,
\end{equation}
where each term $\Sigma_n$ is the sum of all those irreducible diagrams
that contain products of $n$ field correlation functions
\nref{correlE}.  

For a weakly disordered system, the self-energy can be calculated
analytically because only the first contribution $\Sigma_2$ needs to
be computed. Indeed, the detailed analysis in appendix \ref{app:self-energy} shows that 
the ratio of two consecutive terms $\Sigma_n$ and
$\Sigma_{n+1}$ is proportional to the effective scattering parameter 
$g=\eta/{\kappa}$.
If the effective scattering is small, $g \ll 1$,
the series expansion for $\Sigma$ can be truncated
after the first diagram because the effective medium deviates only
slightly from free space. Therefore
\begin{equation}
\Sigma \approx \Sigma_2,
\label{eq:bornapprox}
\end{equation}
which is known as the weak scattering or Born 
approximation.
In terms of the atomic kinetic energy, the weak scattering condition
reads 
$\varepsilon= E/E_\zeta= \sfrac{{\kappa}^2}{2} \gg \eta^2$ or 
\begin{equation}
\Delta \equiv \frac{\eta^2}{\varepsilon}= \frac{\laser{V}^2}{E E_\zeta}
\ll 1. 
\label{Delta.def.eq} 
\end{equation}
The weak scattering condition $\Delta \ll 1$ determines the range of validity for the
diagrammatic perturbation theory. It can be given a simple
physical interpretation by using a semiclassical approximation. The propagation of an incoming plane atomic wave
in a speckle potential can in principle be computed using the Feynman path integral.
If one assumes that the speckle potential is sufficiently weak, the so-called
thin phase grating approximation \cite{henkel_94} can be used. There, the
classical trajectories $\bi{r}_\mathrm{c}(t)$ are unaffected by the potential (\ie remain
straight lines), and the additional phase accumulated along each trajectory is
simply $\int{V(\bi{r}_\mathrm{c}(t)) dt/\hbar}$.
Along a path with length $\zeta$, the
typical accumulated phase will be  $\avV \zeta/\hbar {v} =\sqrt{\Delta/2}$. 
The weak scattering condition $\Delta\ll 1$ is thus equivalent to the requirement that the
accumulated phase be small, \ie to the applicability condition of the
thin phase grating approximation.  
The atomic wave is then only slightly distorted and scattered
after traveling a distance $\zeta,$ which also implies that the scattering mean free path
is larger than $\zeta.$

Equivalently, the weak scattering condition
\nref{Delta.def.eq} corresponds to a small
quantum reflection probability for a particle  
that is scattered by a 1D potential barrier with
height $\avV$ and linear size $\zeta$. 
To see this, two cases may be distinguished: 

(1) When the potential fluctuations exceed the correlation energy ($\avV
>E_\zeta$ or equivalently $\eta >1$) 
the weak scattering condition $\Delta\ll 1$ implies 
$E \gg \avV$ or equivalently $\varepsilon\gg \eta$: the atom flies well above
the potential fluctuations. The standard quantum reflection coefficient 
\cite{GalindoPascual}, 
in our reduced notations
\begin{equation}
\label{reflection}
R=
\left[ 1+ \frac{4(\varepsilon -\eta)}{\Delta\sin^2(2\sqrt{\varepsilon-\eta})}\right]^{-1}
\end{equation}
then is indeed small, $R\ll 1$, since the oscillating term
is bounded, $|\sin(x)/x|\le 1$ with $x=2\sqrt{\varepsilon-\eta}$. 
This case describes the regime of classical atomic motion. 

(2) In the opposite regime of small potential fluctuations ($\avV < 
E_\zeta$ or equivalently $\eta< 1$) the weak scattering condition can of course
also be realized with a fast atom, again without quantum corrections
to the classical transport. More interestingly, the weak
scattering condition can be met even in the case $E <\avV$, \ie even when the
atomic energy lies \emph{below} the average potential height. The weak
scattering regime then is realized if $E\ll E_\zeta$ where the large atomic de
Broglie wave length $\broglie{\lambda}\gg \zeta$ averages out the
short-scale fluctuations and makes the effective disorder weak. In
terms of the reflection coefficient (\ref{reflection}), this
corresponds to the case $\varepsilon < \eta$ and $|\sinh(x)/x|\approx 1$ with
$\Delta\ll 1$. We will see below that this is the regime where quantum
corrections to classical transport become important.

Finally, the weak scattering condition \nref{Delta.def.eq} can be rewritten as
\begin{equation}\label{eq:mobby}
E\gg\mobil{E}=\laser{V}^2/E_\zeta.
\end{equation}
Weak-scattering perturbation theory is
valid for atoms with a sufficiently high kinetic energy compared to the characteristic energy 
$\mobil{E}$. 
We will see in section \ref{sl3d.sec} that, in $d=3$ dimensions, $\mobil{E}$ 
essentially is the mobility edge \cite{PhysToday} 
which separates extended states with $E>\mobil{E}$ from 
localized states with $E <\mobil{E}$. 

\subsection{Scattering mean free path}
\label{ells.sec} 

Within the weak scattering approximation, the scattering mean free path, a central quantity
characterizing the disorder, can be
calculated analytically from the microscopic parameters as follows. 
The relevant self-energy contribution in the Born approximation
\nref{eq:bornapprox} is given by
the momentum convolution of the potential correlation
function $\Power_d$ with the free Green function $G_0$ (as detailed in \ref{app:self-energy}):
\begin{equation}\label{eq:fermirule}
\Sigma_2(\kappa,\varepsilon)= \eta^2\int \frac{\dd\kappa_1}{(2\pi)^{d}} \;
\Power_d(\bkappa-\bkappa_1) G_0(\kappa_1,\varepsilon), 
\end{equation}
To lowest order in $\Delta=\eta^2/\varepsilon$ the 
self-energy can be taken on-shell, 
$\Sigma_\kappa = \Sigma_2 (\kappa,\varepsilon_\kappa)$ with
$\varepsilon_\kappa = \kappa^2/2$. For
the same reason, the renormalization constant can be approximated by
$Z(\kappa) \approx 1$ such that 
$\scat{\gamma}= -2\, \Im \Sigma_\kappa$. 
Reverting to dimensionfull
quantities, the scattering rate defines the elastic
scattering mean free path $\ells/\zeta = {\kappa}/\scat{\gamma}$
for a quasi-monochromatic wave packet centered around the momentum ${k}={\kappa}/\zeta$:
\begin{equation} \label{eq:imf}
\frac{\zeta}{\ells}= -\,\frac{2\,\Im\Sigma_\kappa}{{k}\zeta} .
\end{equation}
The scattering mean free path defines the
distance over which a particle travels on average without being
scattered. The population of the wave packet 
decays over the distance $r$ by the
factor $e^{-r/\ells}$, analogously to Beer's law in optics \cite{bornwolf}.
Consistently with the weak disorder condition $|\Sigma_\kappa| \ll
\varepsilon$,
we have $\scat{\gamma}/2 \ll \varepsilon$ and
equivalently:
\begin{equation}
{k}\ells \gg 1. 
\label{weak_disorder.eq}
\end{equation}
Note that the
scattering mean free path in the Born approximation is inversely
proportional to $\Delta$ and thus to
$\laser{V}^2$ such that it does not depend on the sign of
the laser detuning $\laser{\delta}$. 

Taking the imaginary part of equation \nref{eq:fermirule} with the
help of $\Im G_0(\kappa,\varepsilon) =
-\pi\delta(\varepsilon-\kappa^2/2)$ gives the inverse scattering mean 
free path (\ref{eq:imf}) in the form
\begin{equation}\label{eq:mfp}
\frac{1}{{k}\ells} = \Delta \,
\Big(\frac{{k}\zeta}{2\pi}\Big)
^{d-2} \int \frac{\rmd\Omega_d}{4\pi} \;
\Power_d({k}\zeta,\theta). 
\end{equation}
Here, $\Power_d(\kappa,\theta) =
\Power_d(2\kappa|\sin\frac{\theta}{2}|) $ denotes the
angular correlation
function as a function of the scattering angle $\theta$ between $\bkappa$
and $\bkappa_1$
at fixed on-shell momenta $\kappa = \kappa_1 
= {k}\zeta$. The $d$-dimensional angular integration
measure is $\rmd\Omega_2 = \rmd \theta $ (integration
range from 0 to $2\pi$) and $\rmd\Omega_3 = 2\pi \sin\theta \, \rmd 
\theta$ (integration range from 0 to $\pi$).

\subsubsection{2D speckle}
\label{sec:2D_speckle}

\begin{figure}
\centering
\psfrag{tht}{\hspace{-2pt}\textcolor{Gray}{$\maxi{\theta}$}}
\psfrag{wavenum}{\quad \raisebox{-5pt}{${k}\zeta$}}
\psfrag{kells}{\hspace{5pt}${k}\ells$}
\psfrag{high}{\hspace{1pt}\textcolor{red}{\hspace{-4pt}$\Delta<1$}}
\psfrag{loww}{\hspace{-15pt}\textcolor{red}{$\;\Delta>1$}}
\subfigure
{(a)\includegraphics[width=.45\linewidth]{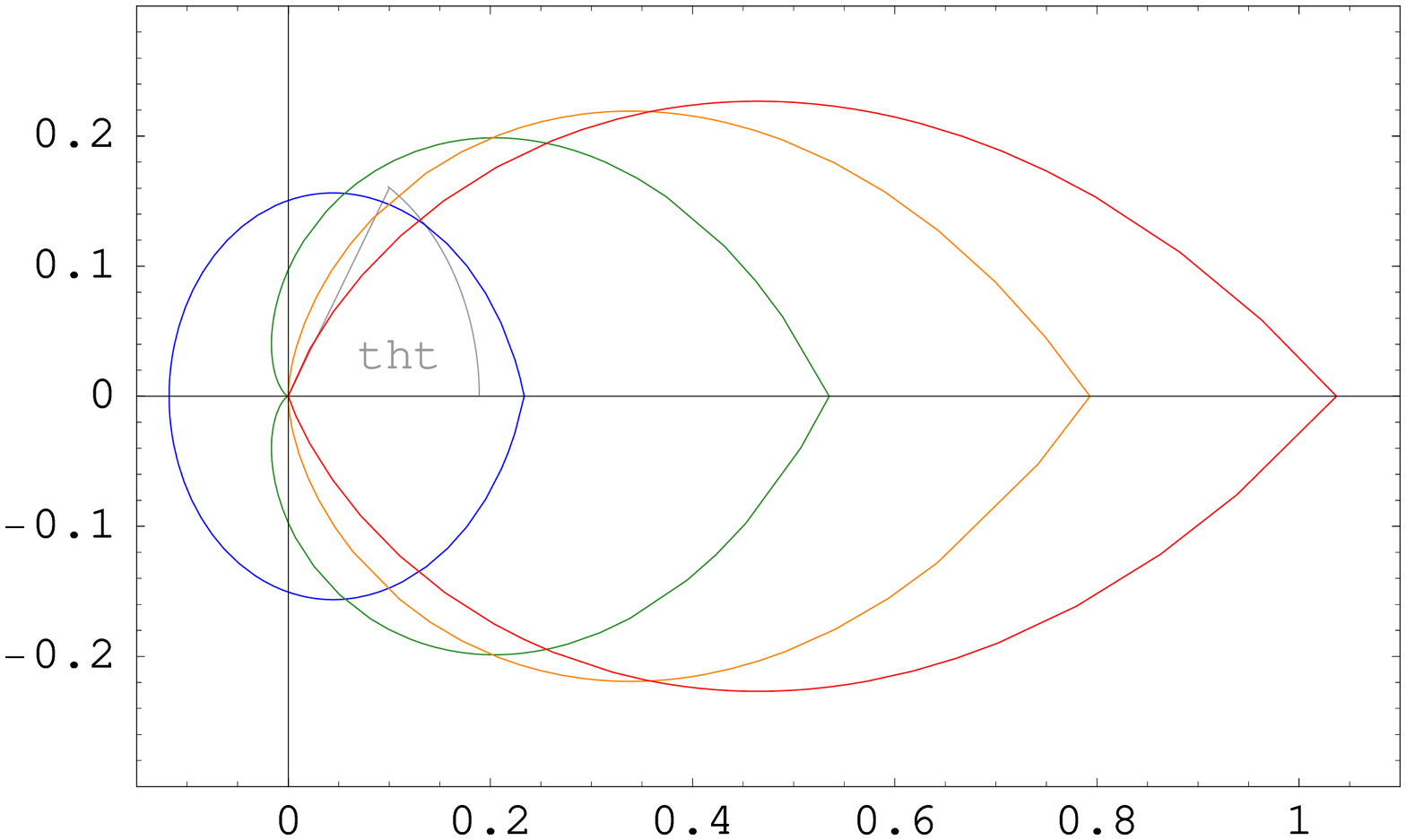}\label{fig:good}}
\hfill
\subfigure
{(b)\includegraphics[width=.475\linewidth]{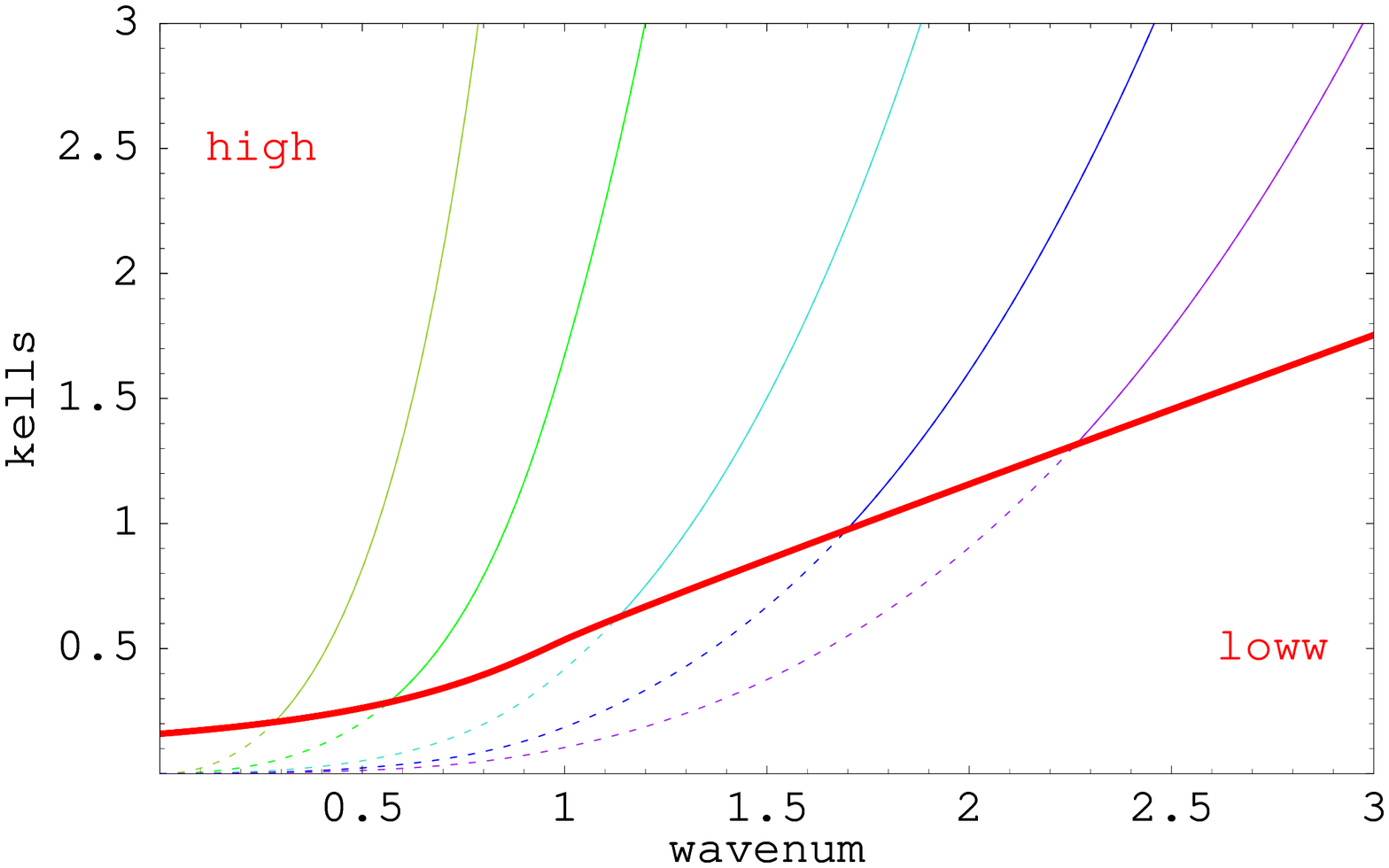}\label{fig:smfp2d}}
\caption{
2D: (a) Phase function (\ref{PF}) for atomic wave vectors
${k}\zeta = 0.4,\,1.0,\,1.4,\,1.8$ (from left to right: blue,
green, orange, red) in units of the speckle correlation length
$\zeta=1/\alpha\laser{k}$. Scattering is nearly isotropic for slow
atoms ${k}\zeta \ll 1$. For fast atoms ${k}\zeta \gg 1$, 
scattering is strongly peaked in the
forward direction, with a maximal deviation $\maxi{\theta} \approx 2/{k}\zeta$.\newline
(b) Disorder parameter ${k}\ells$ as a function of the reduced
matter wave number ${k}\zeta$ for different disorder strengths
$\eta=\avV/E_\zeta=0.2,\,0.4,\,0.8,\,1.2,\,1.6$ (thin curves
from left to right). The thick red line, connecting points of
${k}\ells$ where $\Delta=1$,
indicates the limit of validity of the weak scattering condition.}
\end{figure}

The 2D Fourier transform of 
the potential fluctuation correlation \nref{eq:corl2d} is a convolution of two
identical disks:
\begin{equation}
\label{eq:pow}
\Power_2(\bkappa) = 8
\Big[\arccos\tfrac{\kappa}{2}-\tfrac{\kappa}{2}\sqrt{1-\Big(\tfrac{\kappa}{2}\Big)^2}\,\Big]
\, \Theta(2-\kappa). 
\end{equation}
In the angular correlation function
$\Power_2(2{\kappa}\abs{\sin\frac{\theta}{2}})$, 
the Heaviside function $\Theta(1-{\kappa}\abs{\sin\frac{\theta}{2}})$ restricts the scattering angle to 
$\abs{\sin\frac{\theta}{2}}< 1/{\kappa}$. When ${\kappa} \le 1$, this condition is always
fulfilled, and all angles are possible.
When ${\kappa} > 1$, the scattering
direction is restricted to a maximum scattering angle $|\theta|\le
\maxi{\theta} = 2\arcsin (1/{\kappa})$: 
as ${\kappa}$ grows, the differential
scattering cross section is increasingly 
peaked in the forward
direction. This is illustrated by \fref{fig:good}, a polar
plot of the 2D phase function
\begin{equation}\label{PF}
f_2({\kappa},\theta) = \frac{\Power_2({\kappa},\theta)}
{\int \rmd\theta
\, \Power_2({\kappa},\theta)}. 
\end{equation}
For fast atoms ${\kappa} \gg 1$, the differential
scattering cross section is strongly peaked in the forward
direction which clearly reveals the anisotropic nature of the
scattering process. In this case, $\maxi{\theta} \approx
2/{\kappa} \ll 1$. For slow atoms ${\kappa} \ll 1$, the 
differential scattering cross section becomes
isotropic. In this case, the correlation function
$\Power_2(\brho)$ can be approximated by a delta function,
\ie a constant Fourier transform
$\Power_2(\bkappa)\approx 4\pi$. Hence, the angular dependence
is lost for ${\kappa} \ll 1$ and already the first scattering event
randomizes the direction of scattering. This isotropic scattering
limit, obtained for $\broglie{\lambda} \gg \zeta$, corresponds to the
$s$-wave scattering limit in scattering theory.

Using the angular correlation
$\Power_2(2{\kappa}\abs{\sin\frac{\theta}{2}})$ of 
\nref{eq:pow} in \eref{eq:mfp}, 
it is possible to calculate the scattering mean free path. 
For ${\kappa}=1$, we can do the integral exactly and obtain 
\begin{equation}
{k}\ells = \frac{\pi}{2(\pi^2-4)\eta^{2}}, \qquad
{k}\zeta = 1.
\label{smfp2Db}
\end{equation} 
We recall that $\zeta=(\alpha \laser{k})^{-1}$ is the 2D speckle correlation
length and $\eta=\laser{V}/E_\zeta$ the reduced potential strength. It is
also possible to obtain analytic results in the limiting cases ${\kappa}\gg 1$ and
${\kappa}\ll 1$ where the approximations $\sin{x} \approx x$ and
$\arccos{x}-x\sqrt{1-x^2}\approx \frac{\pi}{2}$, respectively, 
can be made:
\numparts
\begin{eqnarray}
\label{smfp2D}
{k}\ells &\approx 
\frac{({k}\zeta)^2}{4\pi\eta^{2}},
\quad \qquad {k}\zeta \ll 1, 
\label{smfp2Da}\\ 
k\ells &\approx 
\frac{3\pi({k}\zeta)^{3}}{32\eta^{2}} ,
\qquad {k}\zeta \gg 1. 
\label{smfp2Dc}
\end{eqnarray}
\endnumparts

The condition $\Delta \le 1$ implies the bound 
${k}\ells\ge 1/(2\pi),$ such that weak scattering 
$\Delta \ll 1$ indeed describes weak disorder ${k}\ells\gg1$,
even at very low momenta. At higher momenta (cf. \nref{smfp2Dc}), weak scattering 
$\Delta\le 1$ implies the bound $\ells \ge \frac{3\pi}{16}{\zeta},$ which agrees with the intuitive
expectation that the scattering mean free path cannot be 
considerably shorter than the 2D speckle correlation length $\zeta$ itself. 
\Fref{fig:smfp2d} shows a plot of $k\ells$ as a function of ${k}\zeta$ obtained
by numerical integration of (\ref{eq:mfp}). The boundary $\Delta=1$ 
indicates the limit of validity of the weak scattering approximation.

\subsubsection{3D speckle}
\label{sec:3D_speckle}

\begin{figure}
\centering
\psfrag{tht}{\hspace{2pt}\textcolor{Gray}{$\maxi{\theta}$}}
\psfrag{wavenum}{\quad \raisebox{-5pt}{${k}\zeta$}}
\psfrag{kells}{\hspace{5pt}${k}\ells$}
\psfrag{high}{\raisebox{5pt}{\textcolor{red}{\hspace{-4pt}$\Delta<1$}}}
\psfrag{loww}{\hspace{-8pt}\raisebox{-6pt}{\textcolor{red}{$\;\Delta>1$}}}
\subfigure
{(a)\includegraphics[width=.45\linewidth]{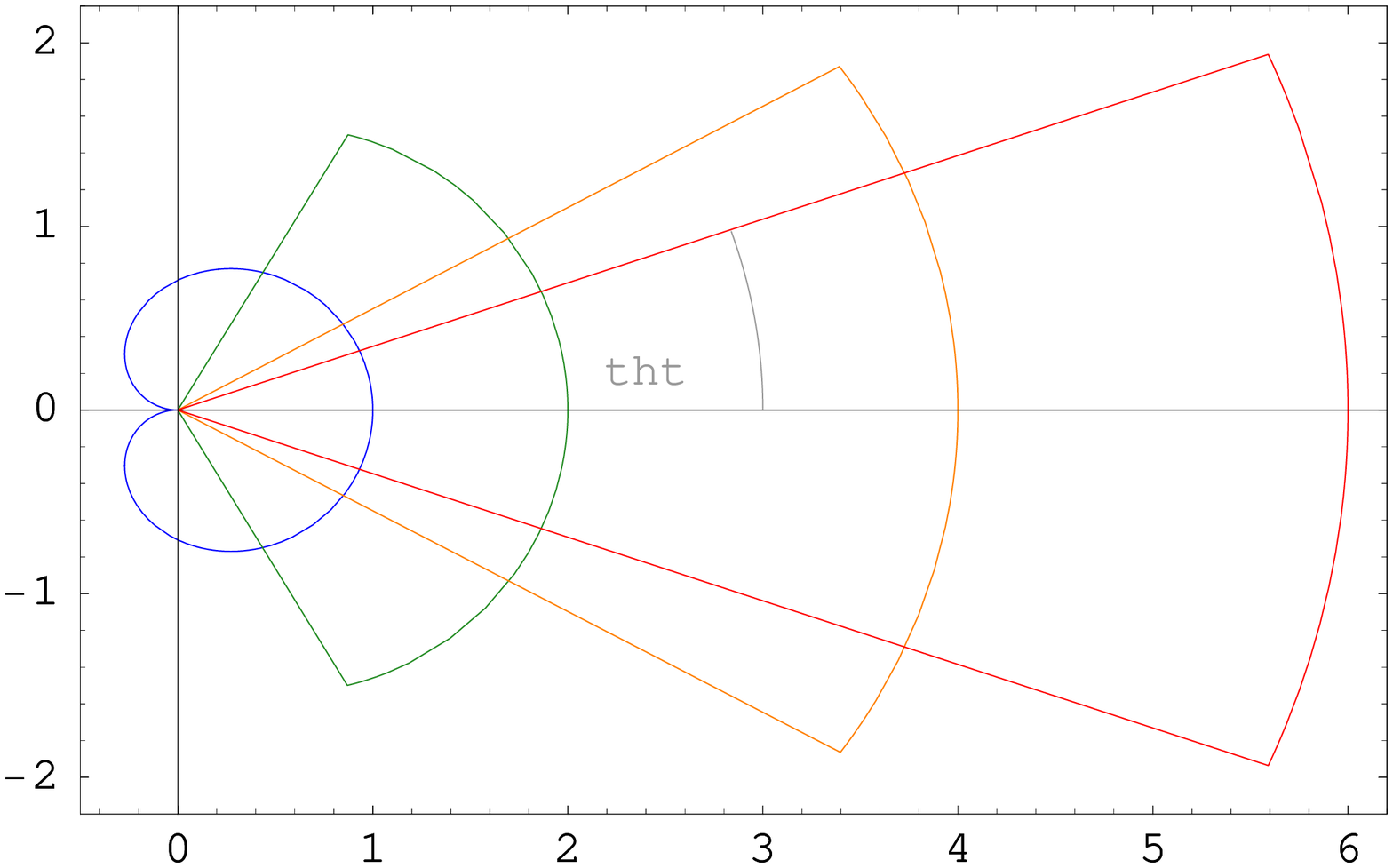}\label{fig:berry}}
\hfill
\subfigure
{(b)\includegraphics[width=.475\linewidth]{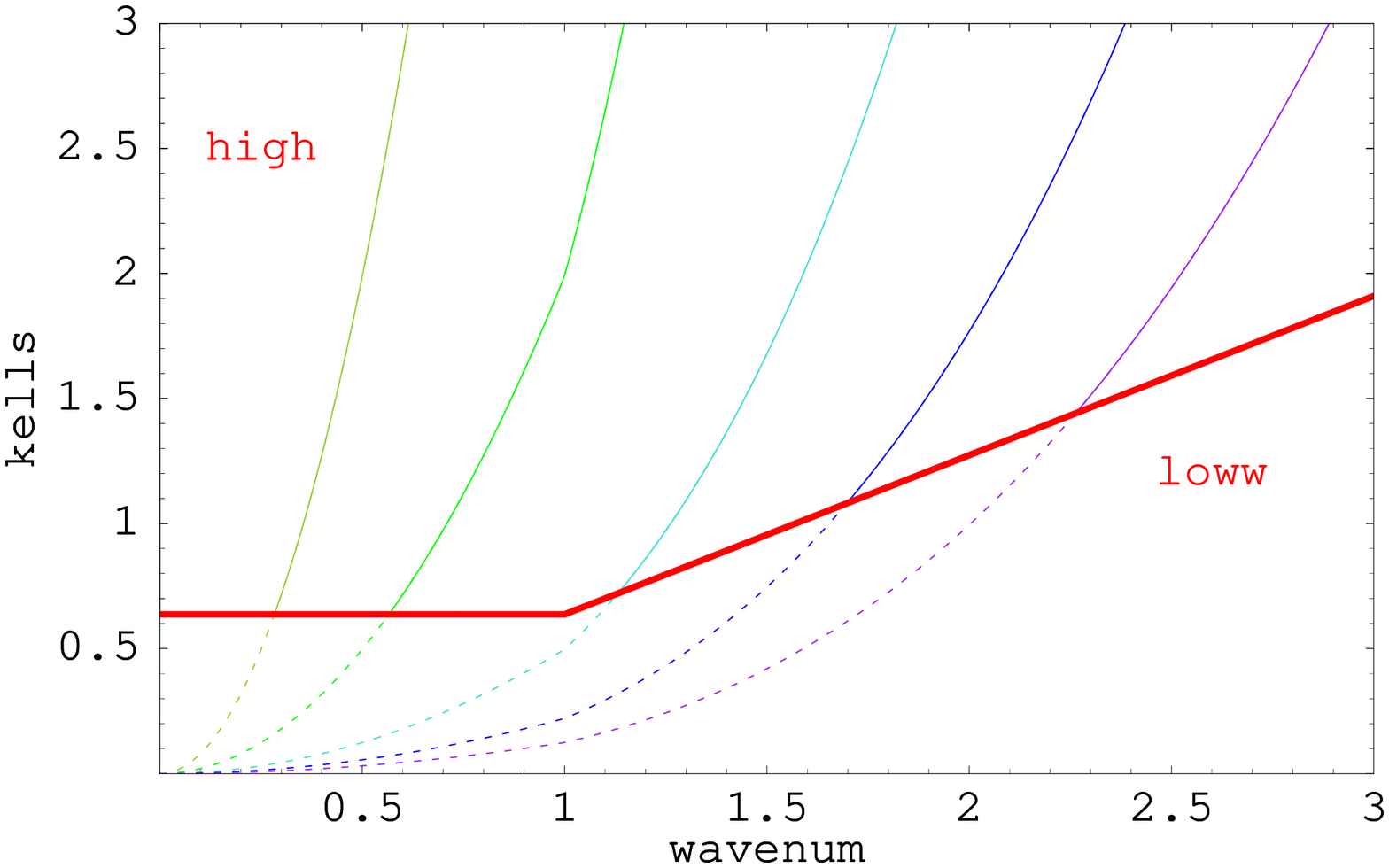}\label{fig:smfp3d}}
\caption{
3D: (a) Effective phase function
\nref{phase3D.eq} 
for atomic wave vectors ${k}\zeta = 1.0,\, 2.0,\, 4.0,\, 6.0$
(from left to right) in units of the speckle
correlation length $\zeta=1/\laser{k}$, equivalent to
velocity in units of recoil velocity, 
${k}\zeta \equiv {v}/\recoil{v}$. 
All plots with ${k}\zeta \leq 1$ are identical.
When ${k}\zeta \gg 1$, the phase function is strongly anisotropic and
displays the maximum scattering angle $\maxi{\theta}
\approx 2/{k}\zeta$.\newline
(b) Disorder parameter ${k}\ells$ as a function of the reduced
matter wave number ${k}\zeta$ for disorder strengths
$\eta=\avV/E_\zeta=0.2,\,0.4,\,0.8,\,1.2,\,1.6$ (thin curves
from left to right). The thick red line connects all points of
$k\ells$ where $\Delta=1$, 
indicating the limit of validity of the weak scattering condition.}
\end{figure}

In three dimensions, we consider a speckle pattern inside an ergodic cavity.
In this case the 3D Fourier transform of the fluctuation correlation
function \nref{eq:corl3d} is a convolution of two
identical spherical shells such that 
\begin{equation}
\label{eq:pow3d}
\Power_3(\bkappa) = \frac{\pi^2}{\kappa} \, \Theta(2-\kappa).
\end{equation}
Since $\Power_3({\kappa},\theta) =
\Power_3(2{\kappa}\sin\frac{\theta}{2})$ 
diverges in the forward direction $\theta \to 0$,
we plot in \fref{fig:berry} the effective phase function including the
angular Jacobian, 
\begin{equation}
f_3({\kappa},\theta) = \frac{\sin\theta \;
\Power_3({\kappa},\theta)}
{\int \rmd\Omega_3 \,\Power_3({\kappa},\theta)}.
\label{phase3D.eq}
\end{equation}
As for the 2D case, the plot shows bounded scattering 
$\abs{\theta}\le \maxi{\theta}$ for
fast atoms ${\kappa} > 1$ and unrestricted scattering for slow
atoms ${\kappa} \le 1$; exact backscattering $\theta=\pi$
appears suppressed due to the angular Jacobian. 
The inverse elastic mean free path (\ref{eq:mfp}) for $d=3$ is given by
\begin{equation}\label{eq:berrymean} 
\frac{1}{{k}\ells} = \pi \, \eta^2 \,[({k}\zeta)^{-3}\ \Theta({k}\zeta-1)
+ ({k}\zeta)^{-2} \ \Theta(1-{k}\zeta)],
\end{equation}
in terms of correlation length $\zeta=\laser{k}^{-1}$ and speckle strength $\eta=\laser{V}/E_\zeta$.

The condition $\Delta\le 1 $ implies the bound ${k}\ells \ge
\frac{2}{\pi},$ such that weak scattering $\Delta\ll 1$ indeed describes weak
disorder ${k}\ells\gg1$, even at low momenta. At high momenta, 
weak scattering $\Delta \le 1$ implies 
that $k\ells \ge \frac{2}{\pi}\,{k}\zeta$, \ie the lowest
achievable scattering mean free path is of the order of the 3D
speckle correlation length $\zeta$ itself. 
\Fref{fig:smfp3d} shows a plot of $k\ells$ as a function of $k \zeta$ as obtained
by \eref{eq:berrymean}.


In summary,  we have derived the elastic scattering mean free path
analytically as a function of all relevant physical parameters. We
have taken into account the correlations present in the speckle
pattern, using a 
diagrammatic perturbation theory up to order
$\eta^2$ in the reduced speckle strength.
The validity of the perturbation theory is limited to the regime
$\Delta = \eta^2/\varepsilon = \laser{V}^2/EE_\zeta\ll 1 $ of
not-too-large speckle fluctuations (at fixed atomic energy) and
not-too-slow atoms (at fixed speckle strength). 
Not surprisingly, fast atoms are only weakly deviated by the
disordered potential (${k}\ells\gg 1$). On the other hand, we
find that a strongly scattering disordered medium (${k}\ells
\to 1$) can be obtained for far-detuned speckle fields of moderate
strength and sub-recoil cooled atoms.

\section{Diffusive transport}
\label{diff.trans.sec}

In the present section, we show that the matter-wave dynamics is
diffusive on long time scales, and determine the corresponding diffusion constant by using a variant of 
Vollhardt and W\"olfle's diagrammatic perturbation theory  
\cite{vollhardt}.  

\subsection{Density kernels and continuity equation}
\label{IntKer.sec}

In the course of its propagation in a disordered potential, the initial matter
wave is rapidly turned into a diffuse matter wave invading the entire
scattering region. The dynamics of this process 
is described by 
the disorder-averaged and forward-propagated probability density to find an atom at position
$\brho$ at time $\tau \ge 0$: 
\begin{equation}
\label{prhotau.def}
p(\brho,\tau)=\Theta(\tau)\av{n(\brho,\tau)} 
= \Theta(\tau) \mathrm{Tr}\{ \av{\varrho(\tau)} \hat n(\brho)\} 
= \Theta(\tau)\langle \brho|\av{\varrho
(\tau)}|\brho\rangle
\end{equation}
 with $\hat n(\brho) = |\brho\rangle
\langle \brho|$ the local density operator. 
Propagating the initial density matrix
$\varrho_0$ in time, one finds  
$p(\brho,\tau) = \Theta(\tau) \bra{\brho}\av{U(\tau) \varrho_0
U(\tau)^\dagger}\ket{\brho}$. 
Its Fourier transform $p(\bi{q},\omega)=
\int \dd\!\rho \,\rmd \tau \exp[i(\omega \tau-\bi{q}\cdot\brho)]\, p(\brho,\tau)$ is
given by 
\begin{equation} \label{eq:probability.k}
p(\bi{q},\omega) = \int \frac{\dd \kappa}{(2\pi)^d}\;\varrho_0(\bkappa,\bi{q}) 
\Ker(\bkappa,\bi{q},\omega),
\end{equation}
where all information about the initial density distribution is
contained in $\varrho_0(\bkappa,\bi{q}) = \bra{{\bkappa_+}}\varrho_0 \ket{\bkappa_-}$ with
$\bkappa_\pm= \bkappa\pm\sfrac{\bi{q}}{2}$.
The subsequent propagation is determined by
the density propagation kernel 
\begin{equation} \label{relaxation_kernel.eq}
\Ker (\bkappa,\bi{q},\omega) = \int \frac{\dd \kappa'}{(2\pi)^d} \int\frac{\rmd \varepsilon}{2\pi} \;
\Phi(\bkappa,\bkappa',\bi{q},\varepsilon,\omega)
\end{equation}
defined in terms of the intensity propagator   
\begin{equation}
\label{intensity_kernel.eq}
\Phi(\bkappa,\bkappa',\bi{q},\varepsilon,\omega)
= 
\av{\bra{\bkappa'_+}G(\varepsilon_+)\ket{{\bkappa_+}}
\bra{\bkappa_-}G^\dagger(\varepsilon_-)\ket{\bkappa'_-}
}. 
\end{equation}
Here, $G(\varepsilon_+)$ is the full retarded propagator
\nref{eq:LSop} at energy $\varepsilon_+=
\varepsilon+\sfrac{\omega}{2}$, and $G^{\dag}(\varepsilon_-)$ the full
advanced propagator at energy $\varepsilon_-=
\varepsilon-\sfrac{\omega}{2}$. Note that the ensemble average is done
after taking their 
product which means that all correlations between different amplitudes
are included.  In solid-state physics, this kernel is also 
known as the ``particle-hole'' propagator because both a retarded
and an advanced Green's function appear. 

The average local current density is the expectation value 
$\bi{j}(\brho,\tau) =  \Theta(\tau)\mathrm{Tr}\{\av{\varrho(\tau)} \hat{\bi{j}}(\brho)\}$ of the
usual current density operator \cite{rammer}
$\hat{\bi{j}}(\brho)= \frac{1}{2} \left[\hat\bi{p} \hat n(\brho) +  \hat n(\brho)\hat\bi{p}\right]$.
Again going into Fourier components, it reads 
\begin{equation} \label{eq:current.k}
\bi{j}  (\bi{q},\omega) =
\int \frac{\dd \kappa}{(2\pi)^d}\varrho_0(\bkappa,\bi{q})
\;\Cur(\bkappa,\bi{q},\omega),  
\end{equation}
which is the analog of \nref{relaxation_kernel.eq}, i.e. 
the convolution of the initial distribution with the current kernel 
\begin{equation} \label{eq:current.kernel}
\Cur (\bkappa,\bi{q},\omega) =
 \int \frac{\dd \kappa'}{(2\pi)^d} 
\int \frac{\rmd \varepsilon}{2\pi} \; \bkappa' \; 
\Phi(\bkappa,\bkappa',\bi{q},\varepsilon,\omega)
\end{equation}
that transforms indeed like a vector. 

Any Hamiltonian of the form $H=\hat\bi{p}^2/2m + V(\hat\bi{r})$ 
leads to a so-called \emph{continuity equation}
that describes the local conservation of the probability density
\cite{rammer}. 
The only difference with the standard continuity equation in our case
comes from the ensemble average that requires to impose forward-time
propagation with the Heaviside distribution $\Theta(t)$. The
corresponding continuity equation  
\begin{equation}
  \label{eq:continuity}
\partial_\tau p(\brho,\tau) + \nabla\cdot\bi{j}(\brho,\tau) =  \delta (\tau) p(\brho,0)
 \end{equation}
features the initial condition on the right side.  
This conservation equation is of course valid for any initial
distribution; in terms of the kernels \nref{relaxation_kernel.eq} and \nref{eq:current.kernel}, 
it takes a very simple form: 
\begin{equation}
  \label{eq:cont:kernel}
  -i\omega \Ker(\bkappa,\bi{q},\omega) + i \bi{q}\cdot \Cur
  (\bkappa,\bi{q},\omega) = 1 . 
\end{equation}
This exact relation between the intensity and current kernel 
does not rely on an expansion for small $\bi{q}$ or
$\omega$ (in contrast to the approximate equation used by Vollhardt and W\"olfle
and the quantum transport literature.)
This is as far as general kinematic definitions can take us. We now have to specify the dynamics, i.e. the equation of motion obeyed by the intensity propagator \nref{intensity_kernel.eq}.  

\subsection{Bethe-Salpeter equation, linear response and diffusion equation}

The momentum matrix elements \nref{intensity_kernel.eq} define a four-point operator 
$\Phi(\varepsilon,\omega) = \av{G(\varepsilon_+) \otimes G^{\dag}(\varepsilon_-) }$  
that obeys a Bethe-Salpeter equation:
\begin{equation}\label{BS}
\Phi(\varepsilon,\omega) = [\av{G}(\varepsilon_+) \otimes \av{G^{\dag}}(\varepsilon_-)] +
[\av{G}(\varepsilon_+)\otimes\av{G^{\dag}}(\varepsilon_-)] \,U(\varepsilon,\omega)\,\Phi(\varepsilon,\omega). 
\end{equation}
The first term on the right-hand side represents the 
intensity propagation in the effective medium with uncorrelated average propagators \nref{eq:ave}. 
All correlated scattering events are described by the
irreducible intensity vertex $U$. 
The Bethe-Salpeter equation actually defines 
$U$, much in the same way that the Dyson equation \nref{eq:dysong}
defines the self-energy $\Sigma$. 

Starting from this Bethe-Salpeter equation, the quantum kinetic theory
described in \ref{app:QKE}   
permits to calculate the current kernel \nref{eq:current.kernel} as function of the
density kernel \nref{relaxation_kernel.eq}: 
\begin{equation}
\label{linearresponse.eq}
i\bi{q} \cdot\Cur(\bkappa,\bi{q},\omega) = \frac{q^2\kappa^2 \tautr(\kappa)}{d} \Ker(\bkappa,\bi{q},\omega) 
+i \tautr(\kappa) \bi{q}\cdot\bkappa.  
\end{equation}
This expression is valid in the linear-response regime and for large
distances and long times, $q\ells \ll1$ and $\omega \taus\ll1$. It
features the transport time $\tautr(\kappa)$ given by 
\begin{equation}\label{eq:tautr}
\fl\frac{1}{\tautr(\kappa)} = \int\frac{\rmd \varepsilon}{2\pi} 
\frac{A(\kappa,\varepsilon)}{2\pi\avdos(\varepsilon)}
\int\frac{\dd\kappa'\,\dd{\kappa''}}{(2\pi)^{2d}}
A(\kappa',\varepsilon) A({\kappa''},\varepsilon)\,
(1-\hat{\bkappa}'\cdot\hat{\bkappa}'')
U(\bkappa'',\bkappa',\varepsilon).
\end{equation}
The spectral function $A(\kappa,\varepsilon)$ has been defined in \nref{spectral_function.eq}, and
 $\avdos(\varepsilon)$ is the corresponding density of states.
The two coupled equations \nref{eq:cont:kernel} and \nref{linearresponse.eq} are easily solved for 
the density relaxation kernel \nref{relaxation_kernel.eq} in the diffusive regime, i.e. to leading order in $q$ and $\omega$:
\begin{equation}
\label{phikq.eq}
\Ker (\bkappa,\bi{q},\omega) = \frac{1}{-i\omega + \diko(\kappa) q^2}.
\end{equation}
Here, the \textit{momentum-dependent} diffusion constant is:
\begin{equation}\label{diko.def}
\diko(\kappa) = \frac{\kappa^2\tautr(\kappa)}{d}.
\end{equation}
Transformed back to time and position variables, the density relaxation kernel 
\begin{equation}
\Ker(\bkappa,\brho,\tau)=\frac{1}{(4\pi\diko(\kappa)\tau)^{\sfrac{d}{2}}}\,
\exp\Big[-\frac{\brho^2}{4\diko(\kappa)\tau}\Big]
\end{equation} 
takes indeed the well-known Gaussian form that obeys the diffusion equation 
for a unit source term: 
\begin{equation}
[\partial_\tau - \diko(\kappa)\nabla^2] \Ker(\bkappa,\brho,\tau) = \delta(\brho)\delta(\tau).
\end{equation}

\subsection{Boltzmann transport of matter waves}

The irreducible vertex operator $U$ entering
equation \nref{eq:tautr} describes the
average scattering of the local probability density.
$U$ can be expanded in a power series just 
like the self-energy \nref{self}: 
\begin{equation} \label{useries}
U = \sum_{n \ge 2} U_n,
\end{equation}
$U_n$ is proportional to the speckle strength $\eta$ to the power $n$, 
and contains all irreducible contributions with $n$
field correlations and at least one 
correlation between the retarded and advanced amplitude.
Due to the non-Gaussian character of intensity correlations, 
its diagrammatic representation (shown in \ref{app:intensity}) differs from
the standard form for Gaussian potentials. 

Generally, $U$ cannot be calculated exactly since correlations of arbitrary 
order are involved, and one has to resort to an approximation. 
In the Boltzmann approximation (also known as the independent scattering
approximation), the infinite series \nref{useries} is truncated
after the lowest-order contribution $U_2$ such that 
\begin{equation}
\label{eq:boltz}
U(\bkappa,\bkappa',\varepsilon) 
\approx \boltz{U}(\bkappa,\bkappa',\varepsilon) =  
\eta^2\,\Power_d(\bkappa-\bkappa'). 
\end{equation}
The corresponding Boltzmann intensity 
\nref{phikq.eq} describes multiple scattering as a 
sequence of scattering events
where both retarded and advanced amplitudes travel along the same
path. In other words, 
all interference effects have been discarded. 
This approximation thus provides a microscopic justification of 
the classical Boltzmann-Lorentz transport theory for non-interacting particles in
the presence of quenched disorder that has been
successfully applied to a large number of physical systems, ranging 
from the Drude
transport theory of electrons in metals
\cite{ashcroft}
to the radiative transfer equation in optics \cite{chandra, ishimaru}.

The Boltzmann approximation \nref{eq:boltz} for the intensity vertex
is very similar to the weak scattering
approximation \nref{eq:bornapprox} for the self-energy. In 
fact, these two approximations are intimately linked by the local
conservation of the probability density as expressed by the continuity
equation \nref{eq:cont:kernel} (which is guaranteed by a 
Ward identity \nref{eq:ward} in the diagrammatic perturbation
theory, see \ref{app:ward}). 
 
In the present case, inserting the vertex 
\nref{eq:boltz} into the general expressions \nref{eq:tautr} and \nref{diko.def}
defines a dimensionfull Boltzmann transport mean free path $\elltrb$
according to
\begin{equation}
\label{eq:BoltzD}
\boltz{\diko} = \frac{\kappa^2\boltz{\tautr}(\kappa)}{d} = 
\frac{m\boltz{D}}{\hbar} = \frac{{k}\elltrb}{d} . 
\end{equation}
To zeroth order in $\Delta=\eta^2/\varepsilon$, the
spectral functions in \nref{eq:tautr} reduce to the on-shell projector
$A_0(\kappa,\varepsilon) = 2\pi \delta(\varepsilon-\kappa^2/2)$ such that
\begin{equation}\label{eq:tmfp}
\frac{1}{{k}\elltrb} = \Delta \,
\Big(\frac{k \zeta}{2\pi}\Big)^{d-2}
\int \frac{\rmd\Omega_d}{4\pi} \; (1-\cos\theta) \;
\Power_d({k}\zeta,\theta),
\end{equation}
where
$\Power_d(\kappa,\theta)=\Power_d(2\kappa|\sin\frac{\theta}{2}|)$ 
is the angular potential correlation
function with $\theta$ the angle between $\bkappa$ and
$\bkappa'$ at on-shell momenta $\kappa=\kappa'=k \zeta$.
The transport mean-free path is the average distance required
to completely erase the memory of the initial direction of
propagation. 
It is related to
the scattering mean free path (\ref{eq:mfp}) through
\begin{equation}\label{eq:anis}
\frac{\ells}{\elltrb} = 1-\langle\cos\theta\rangle = 1- \int \rmd\Omega_d 
\; \cos\theta \;f_d({k}\zeta,\theta),
\end{equation}
where the cosine of the scattering angle is averaged over the phase
function \nref{PF} or \nref{phase3D.eq}. 
This term is known as
the \emph{anisotropy factor} of the scattering process. 
For fully isotropic scattering, it is of course zero, and
$\elltrb=\ells$. 
But for strongly anisotropic scattering, $\langle\cos\theta\rangle$
can take a value close to $1$. In this case, a large number of
scattering events is necessary to deviate the particle completely, 
$\elltrb \gg \ells$.
In the language of diagrammatic quantum transport theory, the
anisotropy factor \nref{eq:anis} is due to \emph{vertex corrections}
\cite{Mahan} that are obtained by summing the entire series
known as \emph{ladder diagrams}; see  \ref{app:intensity} for details.

\subsubsection{2D-speckle}

\begin{figure}
\begin{center}
\psfrag{cosinus}{\hspace{5pt}\raisebox{3pt}{$\langle \cos \theta\rangle$}}
\psfrag{kxi}{\hspace{10pt}\raisebox{-4pt}{${k}\zeta$}}
\psfrag{hallo}{(a) 2D}
\label{fig:tropy2D}
\includegraphics[width=.475\linewidth]{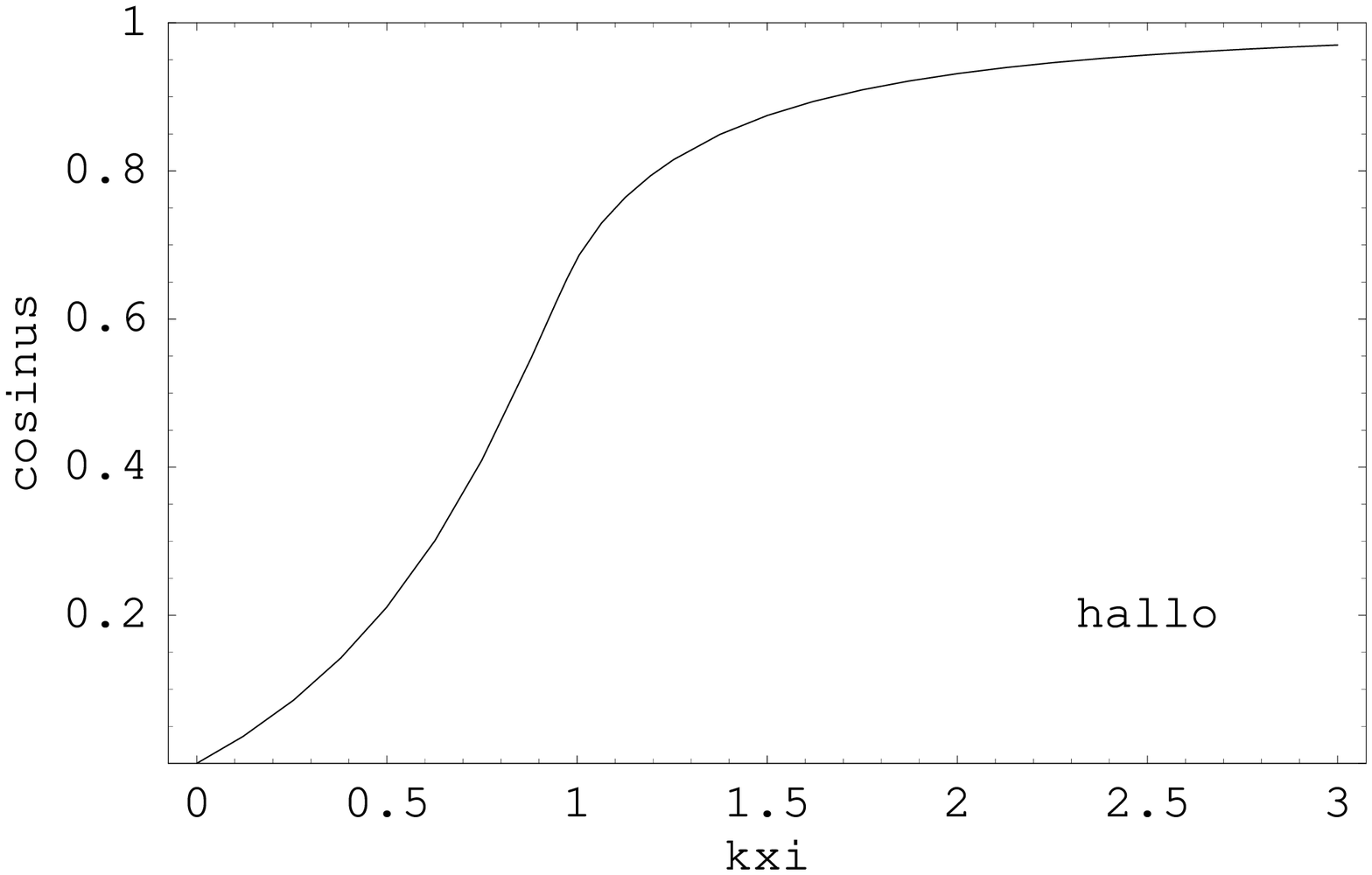}
\hfill
\label{fig:tropy3D}
\psfrag{hello}{(b) 3D}
\includegraphics[width=.475\linewidth]{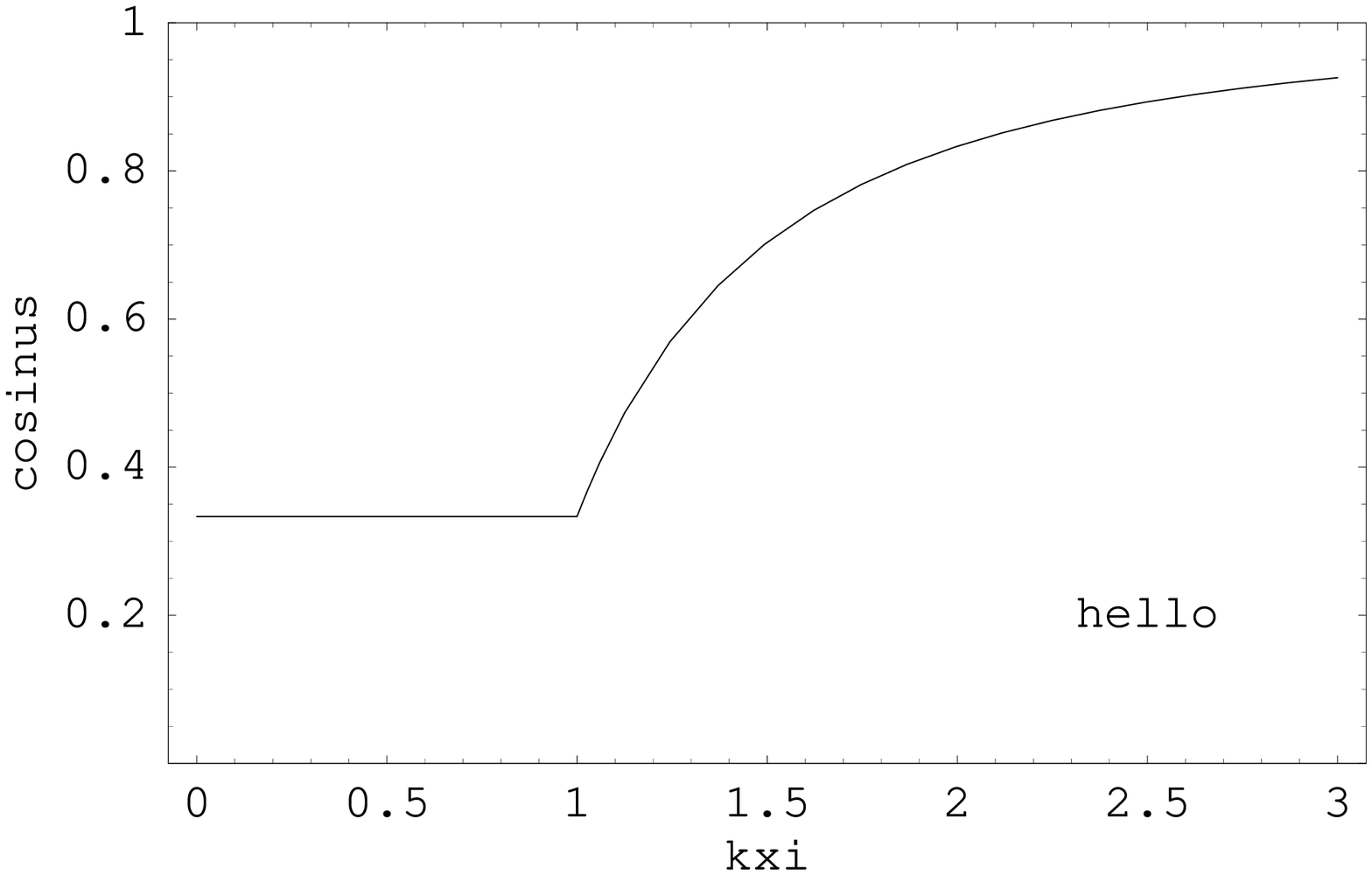}
\caption{Plot of the anisotropy factor 
as a function of the reduced atomic wave number
${k}\zeta$. 
\newline 
(a) 2D ($\zeta=1/\alpha\laser{k}$): For small wave numbers 
(${k}\zeta\to 0$) the anisotropy decreases ($\langle
\cos\theta\rangle\to 0$) and the scattering becomes isotropic
($\elltrb\to\ells)$. 
For large wave numbers (${k}\zeta \gg 1$), the anisotropy
becomes more pronounced ($\langle \cos\theta\rangle\to 1$). 
\newline 
(b) 3D ($\zeta=1/\laser{k}$): Large wave numbers show strong scattering
anisotropy. But even for small wave numbers, scattering is never really
isotropic, $\langle \cos\theta \rangle = \frac{1}{3}$ for all ${k}\zeta\le 1$, as already
indicated by the anisotropic phase function plotted in \fref{fig:berry}.
\label{fig:tropy}}
\end{center}
\end{figure}

In the limiting cases where analytical solutions are found (cf. \nref{smfp2Db}
and \nref{smfp2D}),
\numparts
\begin{eqnarray}\label{tmfp2D}
\elltrb &\approx \ells,\qquad\qquad\quad\;\;
{k}\zeta \ll 1, \label{tmfp2Da}\\
\elltrb &= \frac{\pi^2-4}{\pi^2-8}\,\ells,
\qquad\;\;\, {k}\zeta = \,1 \label{tmfp2Db}\\
\elltrb &\approx \frac{15}{4} \,({k}\zeta)^{2} \,\ells,
\qquad {k}\zeta \gg 1. \label{tmfp2Dc}
\end{eqnarray}
\endnumparts
A plot of the 2D anisotropy factor $\langle\cos\theta\rangle$ (computed numerically) as a function of ${k}\zeta$ is shown in \fref{fig:tropy2D}.
For small wave numbers ${k}\zeta\to 0$, one has $\langle
\cos\theta\rangle\to 0$ and $\elltrb \to \ells$: the scattering is
isotropic. When ${k}\zeta \gg 1$, the ratio $\elltrb/\ells$ scales 
as $({k}\zeta)^2$. This can be easily understood because the
phase function limits the angular integration to 
$|\theta|\le\maxi{\theta} \sim 1/{k}\zeta$ such that $1-\langle\cos\theta\rangle \approx \frac{1}{2}
\langle\theta^2\rangle \propto \maxi{\theta}^{2}$.
Thus roughly $({k}\zeta)^2$ independent scattering events are needed to
fully erase the memory of the initial direction. In other words,
the monochromatic laser photons with limited projected wave vectors
$\alpha\laser{k}$ which are present in the speckle field can only weakly deviate atoms with large momentum
$k\gg \alpha\laser{k}$.

\subsubsection{3D-speckle}

For the 3D case, the integration of \eref{eq:tmfp} yields the exact result:
\begin{equation}
\frac{1}{{k}\elltrb} 
= \frac{2\pi}{3} \, \eta^2\, [({k}\zeta)^{-5}\ \Theta({k}\zeta-1) + ({k}\zeta)^{-2}\ \Theta(1-{k}\zeta)].
\label{eq:berrytrans} 
\end{equation}
In terms of the scattering mean free path \nref{eq:berrymean} we have
\begin{equation}\label{3Dparam}
\elltrb = \frac{3\ells}{2} \;
[({k}\zeta)^2\ \Theta({k}\zeta-1) + \Theta(1-{k}\zeta)].
\end{equation}
As one can see in \fref{fig:tropy3D}, a slight anisotropy $\elltrb =
\frac{3}{2} \ells$ remains for all ${k}\zeta \leq 1$, implying $\langle \cos\theta \rangle = \frac{1}{3}$.
This is explained by the absence of scattering around the
backscattering direction after multiplying the finite differential
cross section by the vanishing angular Jacobian, as already evidenced 
by the anisotropic phase function plotted in \fref{fig:berry}. 
This residual anisotropy is due to the long range correlations in the
optical potential, as exemplified by the divergence of
$\Power_3(\bkappa),$ see eq.~(\ref{eq:pow3d}), near $\kappa=0.$
At higher momenta, ${k}\zeta \geq 1$, the ratio
$\elltrb/\ells$ scales as $({k}\zeta)^2$, for the same reason as in 2D. 

This closes our study of the classical transport properties of
monochromatic matter waves in correlated speckle potentials. Essentially, we
have found diffusive dynamics, as expected for particles in a
conservative random potential, with a transport mean free path
displaying a strong anisotropy for fast atoms, and becoming
approximately isotropic for cold enough atoms. In the following, we
investigate quantum corrections to classical transport.


\section{Coherent multiple scattering}
\label{coscat}

\subsection{Quantum corrections to classical transport}
\label{quantcorr.sec}

Within the Boltzmann approximation, all quantum interference
effects are discarded.
At first sight this seems reasonable
since any such effects could be expected to be suppressed
by the ensemble
average over all possible realizations of the random potential.
This means that the disorder average singles out
products of amplitudes and conjugate amplitudes traveling along
the same paths in the same direction
where no phase
differences are present. In the language of electronic quantum
transport, these are the ``particle-hole contributions''. They 
are insensitive to dephasing processes and therefore correspond
to classical propagation. 

It was realized however, that this argument is too
simplistic
\cite{langer}
for phase-coherent systems, where
interference between amplitudes of different scattering paths can occur. 
This can be easily understood by considering the return
probability to a given point 
in which case all scattering paths are closed
loops. Two waves propagating in
opposite directions around any such loop have zero phase difference
and interfere constructively 
(unless a magnetic field for
charged particles is applied, or dephasing
processes are at work). This constructive two-wave
interference enhances the return probability 
to twice the
classically expected value. 
An enhanced return probability in turn
implies a reduced diffusion constant for the onward propagation, 
an effect known as \emph{weak localization}.
 
Quantum corrections to the Boltzmann transport picture are
accounted for by including 
the sum of all counter-propagating amplitudes in the irreducible vertex $U$
used in \nref{eq:tautr}: $U = \boltz{U} + C$
\cite{vollhardt,pingsheng}. 
The so-called Cooperon contribution $C$ can be expressed in terms of
the diffusive intensity kernel \nref{phikq.eq} using a time-reversal
argument for the advanced amplitude as
\begin{equation} \label{defcooperon}
C(\bkappa,\bkappa',\varepsilon,\omega) 
= \frac{1-\langle\cos\theta\rangle}{4\pi \avdos(\varepsilon) \tau_s^2} \,
\frac{1}{-i\omega + \diko_\mathrm{B}(\kappa)
Q^2}. 
\end{equation}
It is a strongly peaked function around the backscattering direction
$\bi{Q}=\bkappa+\bkappa' =0$. The anisotropy
factor $(1-\langle\cos\theta\rangle)$ can be justified by considering dressed Hikami
boxes as explained in  \ref{app:WL}. Inserting 
\nref{defcooperon} into the general definition \nref{eq:tautr},  one
can replace the double integral over $\bkappa'$
and $\bkappa''$  by a single integral over
$\boldsymbol{Q}$ (with suitable cut-offs for very small and very large
momenta, see discussion below) such that the corrected inverse diffusion
constant \nref{diko.def} reads 
\begin{equation} 
\label{eq:weakloc:perturbative}
 \frac{1}{\wl\diko(\kappa,\omega)} = \frac{1}{\diko_\mathrm{B}(\kappa)} 
\left(1 + \frac{1}{\pi \dos(\varepsilon_\kappa)}\int \frac{\dd Q}{(2\pi)^d} \frac{1}{-i\omega + \diko_\mathrm{B}(\kappa)
Q^2}   \right ). 
\end{equation}
The interference correction inside the parenthesis is the leading term
in an expansion in powers of $1/k\elltrb \le 1/k\ells \ll 1$.
For this reason the denominator contains the free density of states
$\dos(\varepsilon)$ instead of the disorder-averaged density of
states $\avdos(\varepsilon)$. 

This perturbation theory
for the \emph{inverse} diffusion constant 
$\wl\diko(\kappa) ^{-1} = \diko_\mathrm{B}(\kappa) ^{-1} (1+\epsilon)$
is only meaningful for a small correction 
$\epsilon <1$ such that the resulting diffusion constant is bounded
from below as $ \wl\diko(\kappa) > \boltz\diko/2$. 
Vollhardt and W\"olfe \cite{vollhardt} have devised a self-consistent resummation of the
perturbation series that allows in principle to reach the strong localization
threshold  $\wl\diko(\kappa)\to 0$: one simply has to replace the
Boltzmann diffusion constant in the denominator of
\nref{eq:weakloc:perturbative} by the corrected diffusion constant
$\wl\diko(\kappa,\omega)$ itself. This prescription amounts to summing
iterated loops of counterpropagating amplitudes.    
As a net result, weak localization reduces the stationary diffusion
constant according to 
$\wl{\diko}=\boltz{\diko}-\delta \diko$,
with the quantum correction (see  \ref{app:WL} for more details): 
\begin{equation}\label{eq:weakloc}
\delta \diko  = \frac{1}{\pi
\dos(\varepsilon_\kappa)} \int \frac{\dd Q}{(2\pi)^d} \frac{1}{Q^2+ 1/\lambda_\ast^2}.
\end{equation} 
Here, the small-$Q$ divergence of the soft mode under the integral is cut off by the 
real quantity $\lambda_\ast := \lim_{\omega\to
0}[-i\omega/\wl\diko(\omega)]^{-1/2}$.
The characteristic length $L_\ast = \zeta \lambda_\ast$ can
encapsulate several effects that limit the interference of amplitudes
traveling around large loops of a characteristic size: 
$(i)$ limited system size $L$, 
$(ii)$ finite phase-coherence length $L_\phi$, and $(iii)$ strong
localization on a scale $\loc\xi$ such that in general 
\begin{equation}\label{eq:clim}
\frac{1}{L_*^2}=\frac{1}{L^2}+\frac{1}{L_\phi^2}+\frac{1}{\loc{\xi}^2}.
\end{equation}
This length has been found to monitor correctly the
behavior of the diffusion constant in bulk media ($L\to \infty$) in the presence of phase-breaking
mechanisms close to the strong localization threshold \cite{minkov}. 
The interesting physics associated with these effects in the weak
localization regime and at the threshold to strong localization 
will be discussed in the following subsections. 

The integral \nref{eq:weakloc} also 
diverges in the UV limit $Q\to\infty$ because we took the
spectral functions inside the double integral \nref{eq:tautr} at
$Q=0$. This divergence can thus be remedied by introducing an
ultraviolet cutoff $Q_\mathrm{max}=\zeta/\ell_c$ given a priori by the overlap of the
disorder-broadened spectral functions, \ie
$\ell_c=\ells$, as discussed by van Tiggelen \cite{bvtloc}.
We have chosen this cut-off in our previous publication \cite{letter}. 
However, in the present case of correlated scattering, loops
of counterpropagating amplitudes can only be closed on the larger
length scale $\boltz{\elltr} \ge\ells$. Therefore $\ell_c=\boltz{\ell}$ 
seems more adequate. In particular, this choice is consistent with the diffusion
approximation and provides particularly simple expressions in the
following. 
There is no need to
renormalize this microscopic cutoff length self-consistently as the
diffusion constant, \ie to replace $\elltrb$ by $\wl{\elltr}$, since
on the short time scale $\omega^{-1} \approx \boltz{\tautr}$ during
which the matter wave is scattered around a small closed loop, 
the weak
localization corrections \nref{eq:weakloc:perturbative} to classical
scattering are negligible. 
The precise choice for the small-scale cutoff $\ell_c$ in any case can only shift the
non-universal, perturbative
prediction of the onset, but 
does not affect universal predictions like critical exponenents at the transition to the
strongly localized regime; see section \ref{sl3d.sec} 
 below. 
 
\subsection{Weak localization}
\label{wl.sec}

\begin{figure}
\begin{minipage}[t]{0.475\textwidth}
\begin{center}
\psfrag{intensity}{\raisebox{-5pt}{$\laser{I}/\sat{I}$}}
\psfrag{weakloc}{$\delta D/\boltz{D}$}
\psfrag{loww}{ }
\psfrag{high}{\hspace{-4pt} (a) 2D}
\label{fig:wl2}\includegraphics[width=\linewidth]{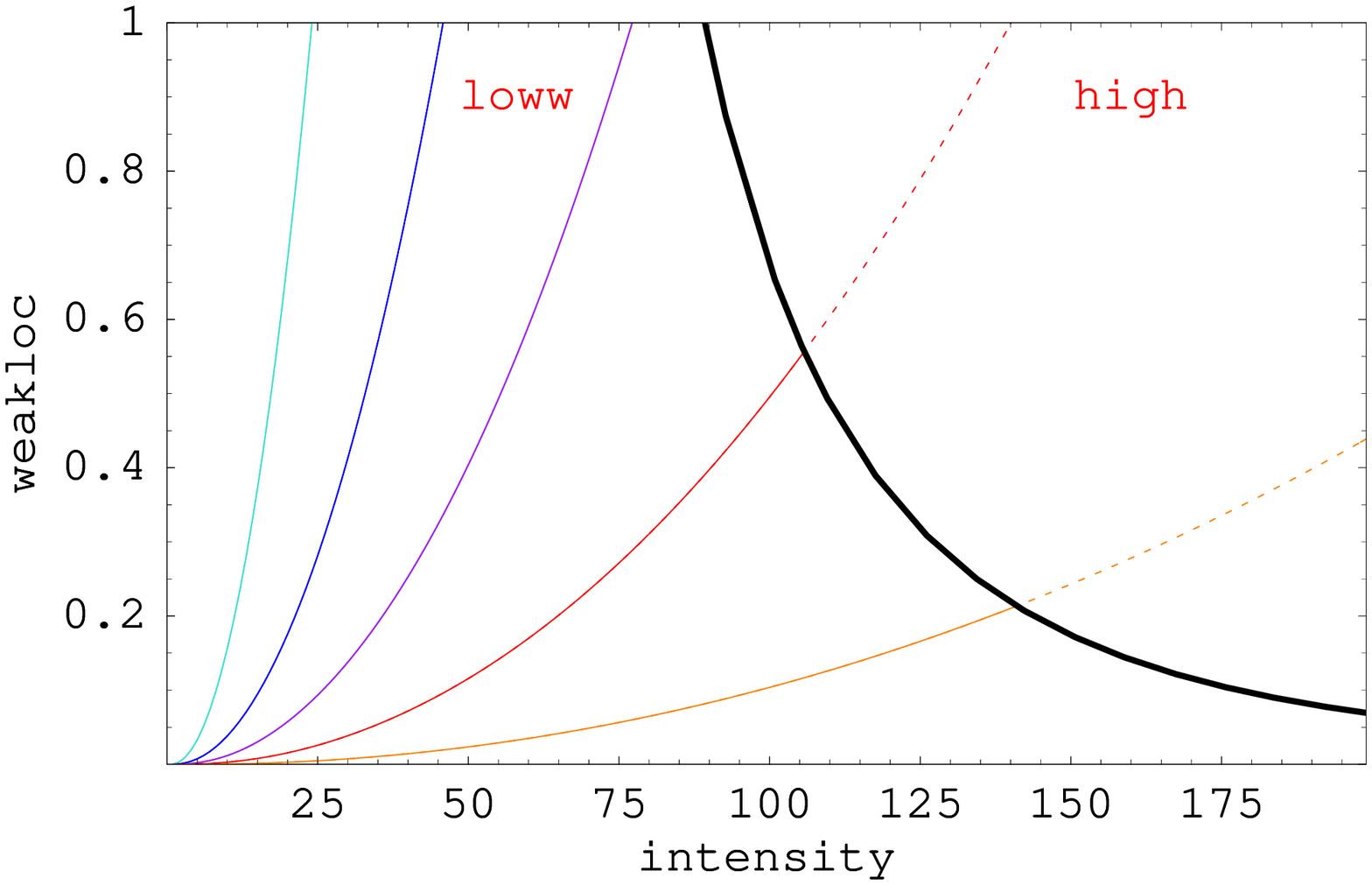}
\end{center}
\end{minipage}
\hfill
\begin{minipage}[t]{0.475\textwidth}
\begin{center}
\psfrag{power}{\raisebox{-5pt}{$\laser{I}/\sat{I}$}}
\psfrag{weakloc}{$\delta D/\boltz{D}$}
\psfrag{high}{ }
\psfrag{loww}{\hspace{-5pt} (b) 3D}
\label{fig:wl3p}\includegraphics[width=\linewidth]{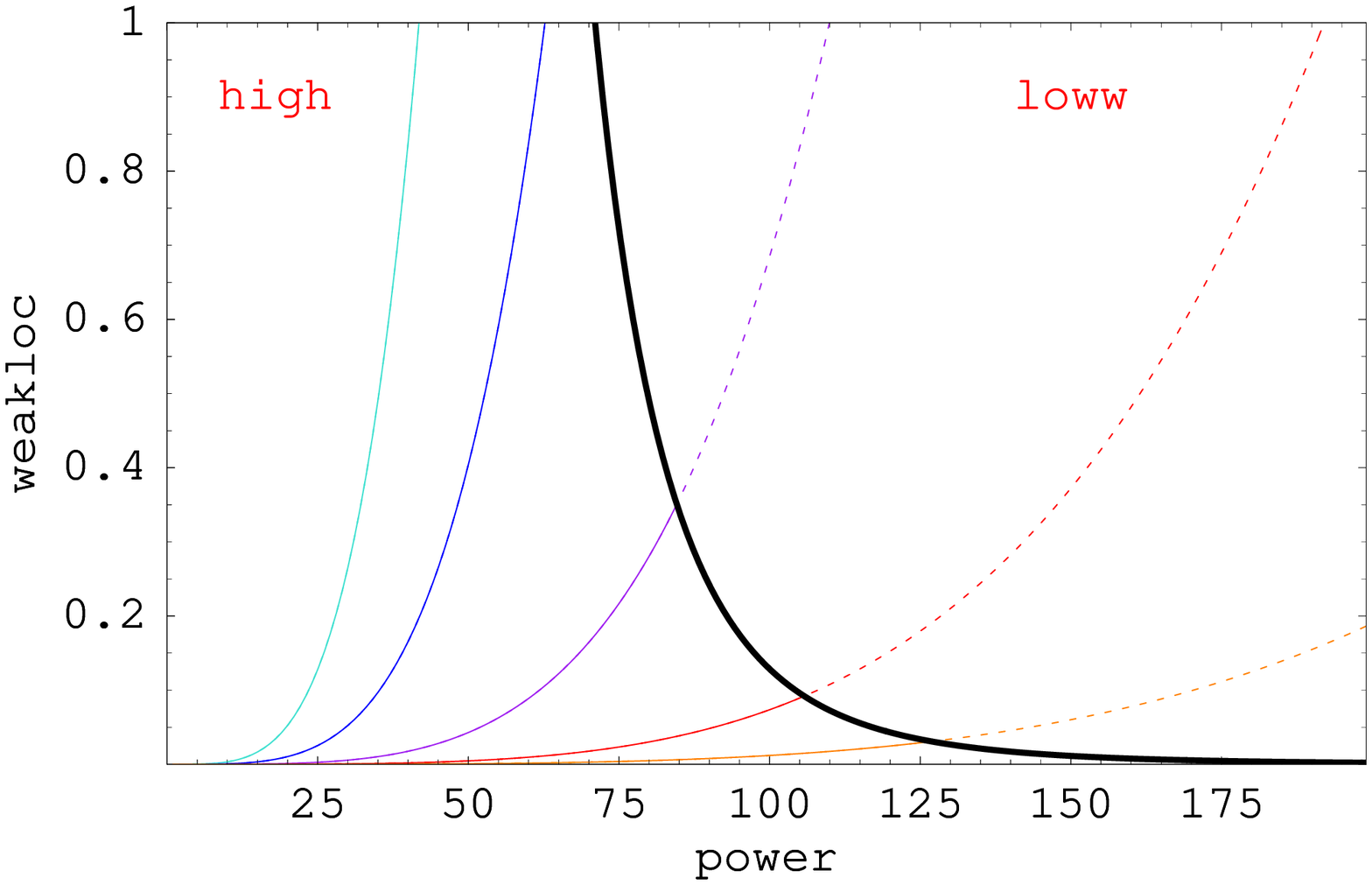}
\end{center}
\end{minipage}
\caption{Weak localization correction $\delta D/\boltz{D}$ 
as a function of the reduced intensity $\laser{I}/\sat{I}$ in a 
speckle field of size $L=2\,$cm. 
\newline 
(a) 2D: matter wave numbers
${k}\zeta \in \{0.8,\,1.0,\,1.2,\,1.5,\,2.0\}$ (from left to
right) with laser detuning $\laser{\delta}= 10^6\,\Gamma$.
\newline 
(b) 3D: matter wave numbers
${k}\zeta \in \{0.6,\,0.9,\,1.2,\,1.5,\,1.8\}$ (from left to
right) with laser detuning $\laser{\delta}=
10^4\,\Gamma$.
\newline
The weak scattering condition $\Delta<1$ is 
valid to the left of the thick black line.}
\label{fig:wl2and3}
\end{figure}

First we will study weak localization for a monochromatic diffusing 
matter wave with wave number 
${k}$ (determined by the cooling technology at hand) 
inside a speckle field of fixed size $L$, as a function of the laser 
intensity $\laser{I}$ and the detuning
$\laser{\delta}$ as
externally controllable parameters in the experiment.

For the observation of the weak localization correction a 
 \textit{coherent} diffusive process must be established inside a large enough
scattering region. This means that 
the total size of the scattering medium must be large enough  
 to admit diffusion in the first place, $L > \boltz\ell$. 
And phase-breaking events must occur at a small enough rate
$\gamma_\phi=1/\tau_\phi < \gamma_s $. 
Otherwise, if the phase coherence is immediately destroyed between two
consecutive scattering events, 
the propagation remains entirely classical, and the Boltzmann
transport theory of section \ref{diff.trans.sec} applies. 
The \emph{phase coherence length} 
$L_\phi= \sqrt{\boltz{D}\tau_\phi}$ is the scale beyond which phase breaking
mechanisms destroy weak localization. 
Both finite sample size and finite phase coherence  are taken into account by using
the characteristic infrared length \nref{eq:clim} in the weak localization
regime ($\loc\xi\to \infty$) as 
\begin{equation} 
\frac{1}{L_\ast^2}=\frac{1}{L^2}+\frac{1}{L_\phi^2}.
\label{lastwl}
\end{equation}
To ensure that interference corrections can be observed experimentally, one has to satisfy
both the diffusive and coherent transport condition, $L_\ast \gg \elltrb$.

For atoms experiencing the light shift \nref{eq:pot} inside the speckle
field, the phase-breaking mechanism
at work is inelastic photon scattering, \ie absorption and
spontaneous reemission of a photon into a different field mode 
accompanied by a recoil momentum kick for the scattering atom.
Thus $\tau_\phi = \tau_i = \gamma_i^{-1}$, 
with the inelastic scattering rate $\gamma_i \propto \laser{V}/\laser{\delta}$
given by \eref{eq:inel}. 
The coherence time can be made arbitrarily large 
by increasing the detuning $\laser{\delta}$ at a fixed potential
strength $\laser{V} \propto \laser{I}/\laser{\delta}$. 
We will show in the following that the phase coherence requirements can be met
with reasonable values for relevant experimental parameters.

The 2D weak localization correction \nref{eq:weakloc} relative to the
Boltzmann diffusion constant \nref{eq:BoltzD} reads in
dimensionfull units and in the
regime $\ell_c= \elltrb \ll L_\ast$ given by (\ref{lastwl}): 
\begin{equation}
\label{delta2D.eq}
\frac{\delta D}{\boltz{D}} = \frac{2}{\pi} \, \frac{\ln
(L_\ast/\elltrb)}{{k}\elltrb} . 
\end{equation}
Noticeable
corrections can be expected for strong disorder where
$k \elltrb$ is not too large. 

Figure \ref{fig:wl2and3}(a) shows the
relative weak localization correction $\delta
D/\boltz{D}$ as a function of the laser intensity 
$\laser{I}$ for different initial atomic velocities, at fixed laser detuning $\laser{\delta} =
10^6\,\Gamma$ and speckle size $L=2\,$cm. 
At its limit of validity  $\Delta=1$,
our theory predicts that the weak localization correction $\delta D$ reaches already 20\%
of the Boltzmann diffusion constant $\boltz{D}$
itself for ${k}\zeta = 2.0$. For a smaller wave
number ${k}\zeta = 1.5$, the value of $\delta D/\boltz{D}$ rises to 55\%.
Since $\zeta = 1/\alpha\laser{k} \gg 1/\laser{k}$, 
experimental evidence of 
2D weak localization requires initial temperatures for the atomic
sample well below the recoil temperature. In turn, this means using a
Bose-Einstein condensate as the initial atomic 
matter wave.
As a general rule, the colder the atoms,
the larger are the interference corrections.

In 3D, the weak localization correction \nref{eq:weakloc} relative to the
Boltzmann diffusion constant \nref{eq:BoltzD} reads for
$L_\ast\gg \ell_c=\elltrb$ 
\begin{equation}
\frac{\delta D}{\boltz{D}} =
\frac{3}{\pi} \,
\frac{1}{(k\elltrb)^2}.
\label{3dwl.eq}
\end{equation}
The 3D interference corrections are small in the weak
disorder regime ${k}\ells \gg 1$.
The relative weak localization correction 
$\delta D/\boltz{D}$ as a function of the laser intensity for a
speckle size $L=2\,\mathrm{cm}$ 
and detuning $\delta=10^4\,\Gamma$ are shown in
\fref{fig:wl2and3}(b). As expected, the largest interference corrections are obtained when
$k\zeta \leq 1$,
which means initial sub-recoil temperatures. 
Here again, an experimental
study may require a
Bose-Einstein condensate as the matter wave source.

\subsection{Strong localization: 2D}  
\label{sl2d.sec}

\begin{figure}
\begin{minipage}[t]{0.475\textwidth}
\begin{center}
\psfrag{detuning}{\raisebox{-5pt}{$\laser{\delta}/(10^6\,\Gamma)$}}
\psfrag{intensity}{$\laser{I}/\sat{I}$}
\psfrag{loww}{ }
\psfrag{high}{ }
\subfigure
{(a)\includegraphics[width=\linewidth]{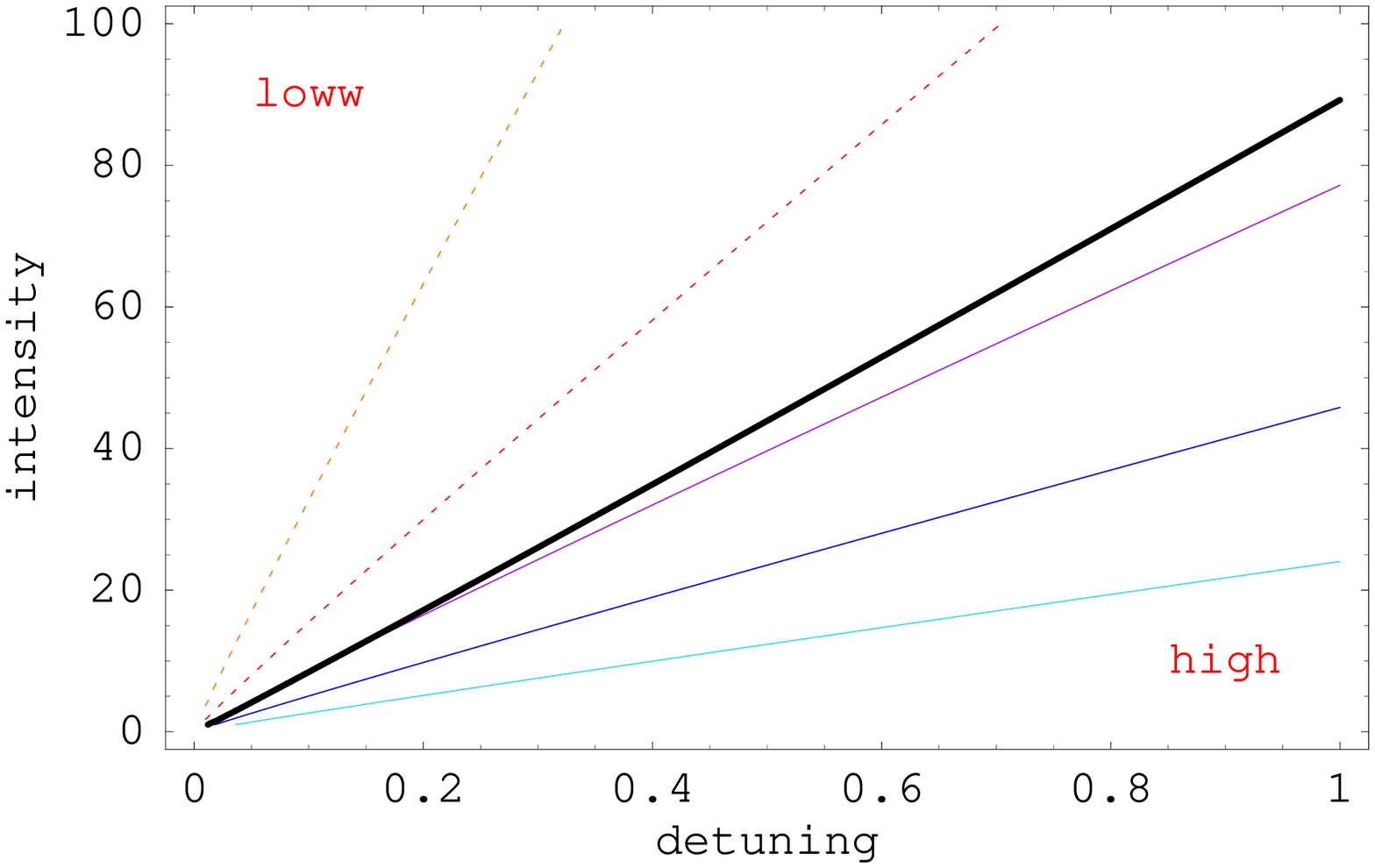}\label{fig:phpl2}}
\end{center}
\end{minipage}
\hfill
\begin{minipage}[t]{0.475\textwidth}
\begin{center}
\psfrag{intensity}{\raisebox{-5pt}{$\laser{I}/\sat{I}$}}
\psfrag{loc}{$\hspace{-22pt}\loc{\xi}$, $L_\phi$, $\elltr$ [m]}
\psfrag{sizt}{\raisebox{-7pt}{\textcolor{Cyan}{$\elltrb$}}}
\psfrag{sizc}{\hspace{-10pt}\raisebox{0pt}{\hspace{-4pt}\textcolor{Purple}{$\elltrwl$}}}
\psfrag{size}{\raisebox{-1pt}{\raisebox{-1pt}{$L$}}}
\psfrag{sizl}{\hspace{5pt}\raisebox{0pt}{\hspace{-3pt}\textcolor{red}{$\loc{\xi}$}}}
\psfrag{sizf}{\raisebox{-5pt}{\textcolor{blue}{$L_\phi$}}}
\subfigure
{(b)\includegraphics[width=\linewidth]{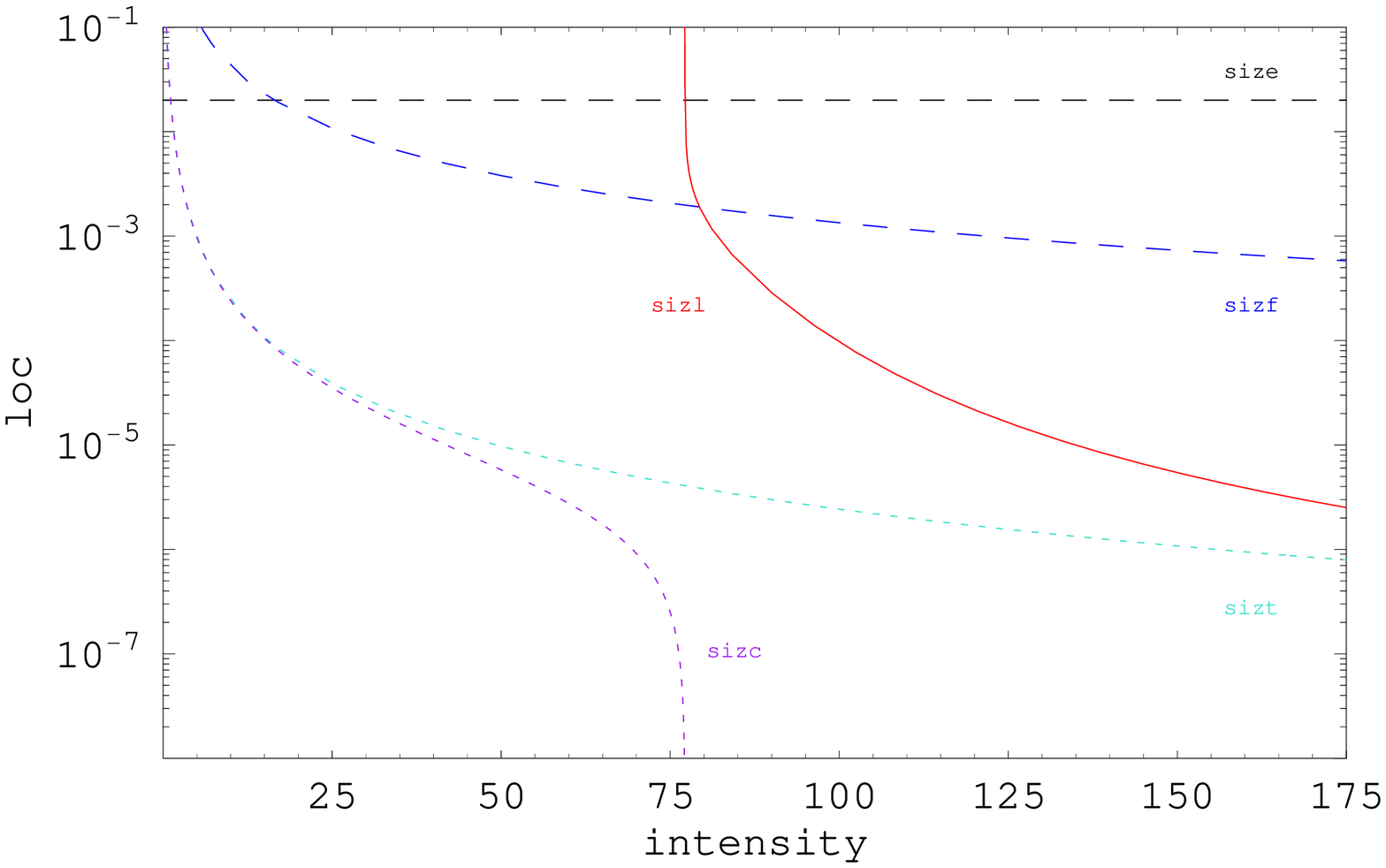}\label{fig:xiloc2d}}
\end{center}
\end{minipage}
\caption{
2D: (a)
Strong localization onset defined by $\delta D=\boltz{D}$
in ($\laser{\delta},\laser{I}$) phase space
for different atomic wavenumbers
$k\zeta \in \{0.8,\,1.0,\,1.2,\,1.5,\,2.0\}$
(from right to left) in a speckle field of size $L=2\,$cm.
For each value of $k\zeta$, the strong localization regime
lies above and to the left of the corresponding line.
The red line corresponds to the criterion $\Delta=1$. Solid curves
below can reach the strong localization onset within the weak
scattering regime; dotted curves are merely extrapolated. \newline
(b) Logarithmic plot of the
Boltzmann transport mean free path $\boltz\ell$,
the weak localization corrected transport mean free path $\elltrwl$, the phase
coherence length 
$L_\phi$, and the localization
length $\loc{\xi}$ as a function of laser intensity $\laser{I}$ in units of
saturation intensity $\sat{I}$ for a fixed atomic 
wave vector ${k}\zeta=1.2$ and detuning
$\laser{\delta}=10^6\,\Gamma$ inside a speckle field of $L=2\,$cm. 
At the strong localization threshold 
$\laser{I}= 77\,\sat{I}$, the corrected transport mean free path
$\elltrwl=2m\wl{D}/\hbar {k}$ vanishes. For higher intensities
the matter wave is localized with a finite localization length 
$\loc{\xi}$.}
\end{figure}

The 2D interference
correction \nref{delta2D.eq} to the diffusion constant diverges with $L_\ast \to
\infty$ which indicates 
that a perfectly phase-coherent wave in an infinite disordered 2D system is in fact always
localized, as predicted by the single-parameter scaling theory 
\cite{gang4}.  The
corrected diffusion constant vanishes, $\wl{D} \to 0$, at the threshold
determined for $k\boltz{\ell} \geq k\ells\gg 1$ by 
\begin{equation} \label{eq:loc2d}
  \ln
(L_\ast/\elltrb) = \frac{\pi}{2} {k}\elltrb . 
\end{equation}
This condition
defines a curve in ($\laser{\delta},\laser{I}$) parameter space. 
In \fref{fig:phpl2} we plot the corresponding ``phase diagram'' showing the
boundary between the weak and the strong localization regime for different
atomic velocities. The curves are almost 
straight lines because the Boltzmann 
mean free path $\boltz\ell$ in the denominator
of \eref{delta2D.eq} scales as $\eta^{2}
\propto \laser{V}^{2}$. 
As a consequence $\delta D/\boltz{D}$ scales as
$(\laser{I}/\laser{\delta})^2$, with small corrections from the 
logarithmic dependence on $L_*/\boltz\ell$. 
Consequently, the strong localization onset has a linear dependence in the ($\laser{\delta},\laser{I}$)
plane with a slope determined by the 
atomic wave number $k$.

For each point on these curves, one can deduce the corresponding values
 for the multiple scattering parameters. For example, for Rubidium 87 atoms 
($m=1.44\, 10^{-25} \, \mathrm{kg}$, $\atom{\lambda} = 780\,$nm,
$\sat{I}=1.67\,$mW/cm$^2$, $\Gamma /2\pi = 6
\,$MHz) at
$k\zeta=1.2$,
$L=2\,\mathrm{cm}$, 
$\laser{I}=77\,\sat{I}$,
and $\laser{\delta}=10^6\,
\Gamma$, we find a scattering mean free path $\ells=0.8\,\mu$m,
a transport mean free path $\elltrb=4.1\,\mu$m and a phase coherence
length $L_\phi\approx 2\,\mathrm{mm}$. 
This places the strong localization threshold at ${k} \ells \approx 0.81$.
These numbers are of course to be taken with a grain of salt since
they are obtained by applying the weak scattering approximation quite
close to the limit of its validity (at the transition point $\Delta
\approx 0.83$).

In the strong localization regime, extended atomic wavefunctions become
exponentially localized as a function of the distance, and the
corresponding localization length $\loc{\xi}$ 
enters as a new length scale.  
In a bulk system $L\to\infty$ the characteristic
length \nref{eq:clim} then reads $1/L_*^2 =
1/L_\phi^2+1/\loc{\xi}^2$ \cite{minkov}. 
Together with equation \nref{eq:loc2d}, this determines the 2D localization
length $\loc{\xi}$ as function of the atomic wave vector and the other
experimentally relevant parameters. 

We study in \fref{fig:xiloc2d} the characteristic length scales on
both sides of the strong localization threshold 
as a function of the laser intensity $\laser{I}$ for 
${k}\zeta =1.2$, $L=2\,$cm, $\alpha=0.1$, and fixed laser
detuning $\laser{\delta}=10^6\,\Gamma$.
With increasing laser intensity, the phase coherence length $L_\phi$
(blue dashed curve) decreases since the probability of spontaneous photon scattering
increases. 
The Boltzmann transport mean free path $\elltrb$
(turquoise dashed curve), a purely local quantity, shows no particular
singularity, but the corrected mean-free path $\elltrwl$
(violet dashed line) plunges to zero
together with the corrected diffusion constant $\wl{D} =
\hbar{k} \elltrwl/2m$ at 
the threshold 
value $\laser{I}=77\,\sat{I}$. This threshold value corresponds
precisely to the transition point on the phase boundary in
\fref{fig:phpl2} reached  with $L_\ast = L_\phi =2\,$mm.  
At the same point, the localization length
$\loc{\xi}$ (red full curve) comes down from infinity and tends to the
expression derived from \nref{eq:loc2d} in the
limit $\loc\xi \ll L_\phi$ of strong localization: 
$\loc\xi = \boltz\ell \exp \left[\frac{\pi}{2}k\boltz\ell\right]$.

\subsection{Strong localization: 3D}
\label{sl3d.sec}

Contrary to the 2D case, the 3D correction \nref{3dwl.eq} remains finite as $L_\ast \to \infty$.
Therefore, a 
transition to the strongly localized regime cannot be driven solely by
large-scale coherence, but requires strong
enough local disorder. This essential difference between the 2D and the 3D case can be
traced back to the fact that a random walk in 2D returns to the
origin with unit probability, whereas in 3D it escapes with finite
probability to infinity 
\cite{randomwalk, domb}.

As shown by Vollhardt and W\"olfle \cite{real}, the self-consistent
weak localization prediction \nref{eq:weakloc} allows to calculate the critical exponents of
the localization length and the diffusion constant at the
Anderson phase transition in an arbitrary dimension $d>2$. We shall
use their findings for $d=3$. 
The integral in \nref{eq:weakloc} can be evaluated with an arbitrary
UV-cutoff $\zeta/\ell_c$ using the identity
$\sfrac{x^2}{(1+x^2)} = 1-\sfrac{1}{(1+x^2)}$. At the threshold
$\delta\diko = \boltz\diko$, it yields a transcendental 
equation for $L_*$ \cite{rammer, real}: 
\begin{equation}\label{eq:condeq}
\frac{L_*}{\cut{\ell}} \bigg[1-\Big(\frac{\gamma_0}{\gamma}\Big)^{\!2}\bigg] = \arctan \frac{L_*}{\cut{\ell}}.
\end{equation}
Here the disorder parameter $\gamma$ and its critical
value $\gamma_0$ are defined as 
\begin{equation}\label{eq:dip}
\gamma = \frac{1}{k\sqrt{\cut{\ell}\boltz\ell}}, 
\qquad\qquad\gamma_0 = \sqrt{\pi/3}. 
\end{equation}
The ratio $\gamma/\gamma_0$ determines whether \eref{eq:condeq}
admits a solution $L_\ast>0$ 
and thus a finite localization
length. The usual graphical solution 
of $b x =\arctan x$ shows that there is no finite solution for
$L_\ast$  in the diffusive
regime of small disorder $\gamma<\gamma_0$. In this case, 
the matter
wave shows truly diffusive dynamics on all scales larger than
$\elltrb$ with finite diffusion constant $\wl D >0$ weakly localized by
the correction \nref{3dwl.eq}.  

In the localized regime $\gamma>\gamma_0$,  \eref{eq:condeq}
admits a finite solution 
$L_\ast>0$ that tends critically towards infinity as
$\gamma\to\gamma_0$. 
Using $\arctan(\sfrac{L_*}{\cut{\ell}}) \approx 
\sfrac{\pi}{2}$, one finds  
$L_* \sim \abs{\gamma-\gamma_0}^{-\nu}$ with the critical
exponent $\nu=1$. Both the diagrammatic perturbation theory and the
scaling theory \cite{kramer} thus lead to the same critical
exponent $\nu=1$. 

\begin{figure}
\begin{minipage}[t]{0.475\textwidth}
\begin{center}
\psfrag{detuning}{\raisebox{-5pt}{$\laser{\delta}/(10^4\,\Gamma)$}}
\psfrag{intensity}{$\laser{I}/\sat{I}$}
\psfrag{high}{ }
\psfrag{loww}{ }
\subfigure
{(a)\includegraphics[width=\linewidth]{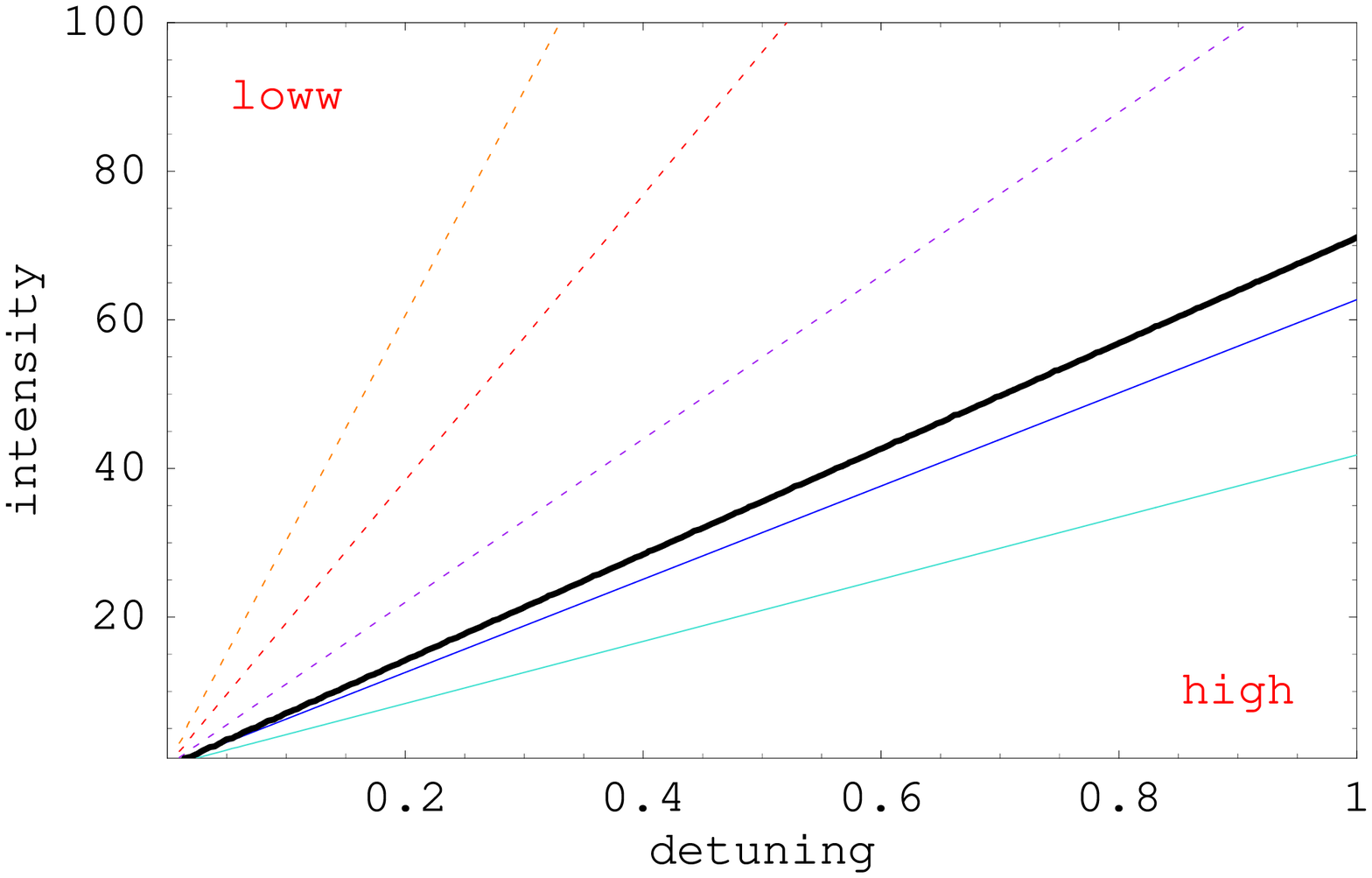}\label{fig:phpl3}}
\end{center}
\end{minipage}
\hfill
\begin{minipage}[t]{0.475\textwidth}
\begin{center}
\psfrag{power}{\raisebox{-5pt}{$\laser{I}/\sat{I}$}}
\psfrag{loc}{$\hspace{-22pt}\loc{\xi}$, $L_\phi$, $\elltr$ [m]}
\psfrag{sizt}{\raisebox{0pt}{\textcolor{Cyan}{$\elltrb$}}}
\psfrag{sizc}{\raisebox{0pt}{\hspace{-3pt}\textcolor{Purple}{$\elltrwl$}}}
\psfrag{sizl}{\raisebox{2pt}{\hspace{12pt}\textcolor{red}{$\loc{\xi}$}}}
\psfrag{sizf}{\raisebox{-1pt}{\textcolor{blue}{$L_\phi$}}}
\subfigure
{(b)\includegraphics[width=\linewidth]{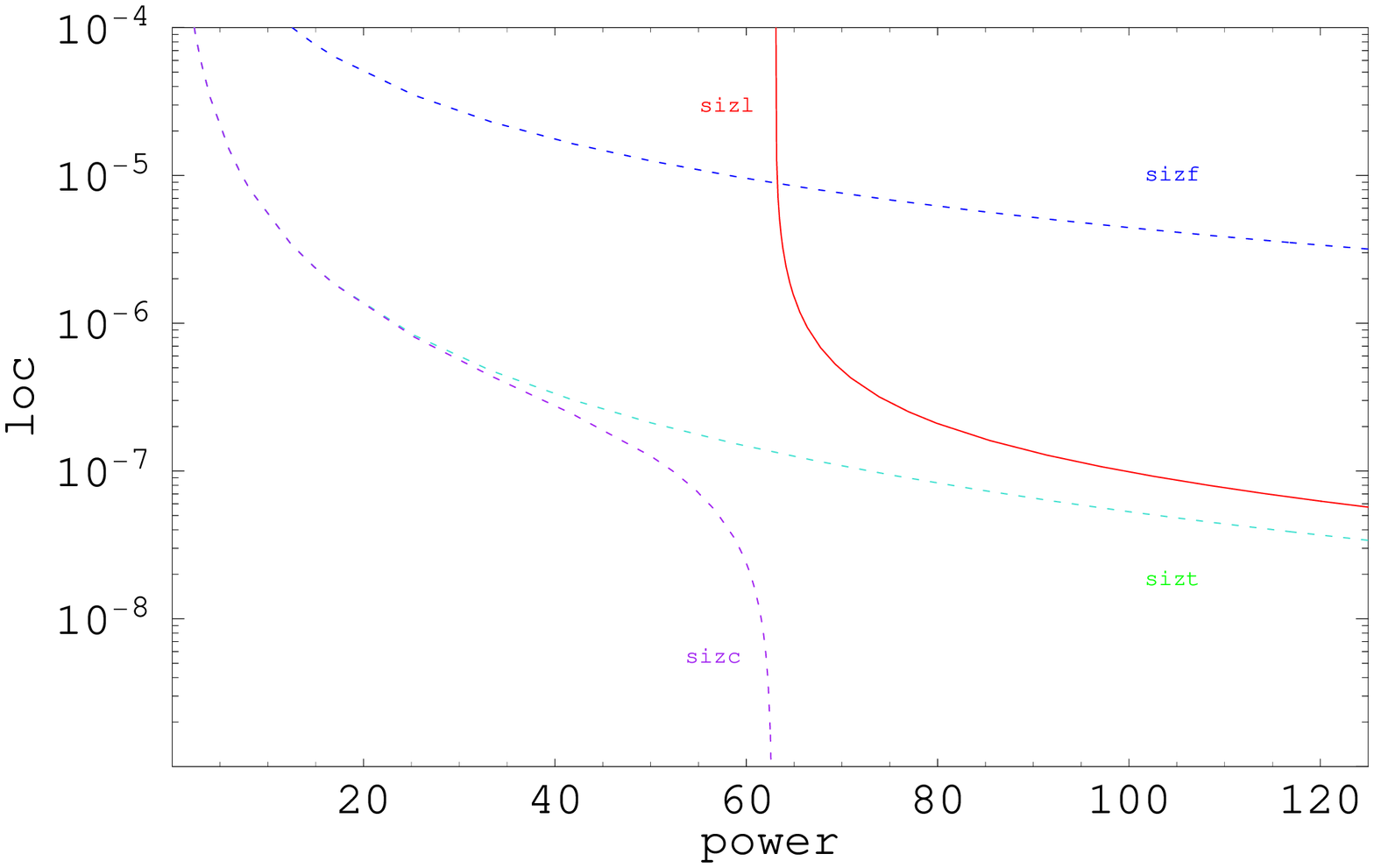}
\label{fig:xiloc3p}}
\end{center}
\end{minipage}
\caption{
3D: (a)
Strong localization onset defined by $\delta D=\boltz{D}$
in ($\laser{\delta},\laser{I}$) phase space
for different atomic wavenumbers
$k\zeta \in \{0.6,\,0.9,\,1.2,\,1.5,\,1.8\}$
(from right to left) in a speckle field of size $L=2\,$cm.
For each value of $k\zeta$, the strong localization regime
lies above and to the left of the corresponding line.
The red line corresponds to the criterion $\Delta=1$. Solid curves
below can reach the strong localization onset within the weak
scattering regime; dotted curves are merely extrapolated. \newline
(b) Logarithmic plot of  the Boltzmann transport mean free path $\ell_B$,
the weak localization corrected transport mean free path $\elltrwl$, the phase
coherence length 
$L_\phi$, and the  
localization length $\loc{\xi}$ as a function of laser intensity $\laser{I}$ in units of
saturation intensity $\sat{I}$ for 
$k\zeta=0.9$ and $\delta=10^4\,\Gamma$.
At the strong localization threshold 
$\laser{I}= 63\,\sat{I}$, the corrected transport mean free path
$\elltrwl=3m\wl{D}/\hbar {k}$ vanishes. For higher intensities
the matter wave is localized with a finite localization length 
$\loc{\xi}$.}
\end{figure}

Universal quantities like critical exponents are independent of the cut-off $\ell_c$,  in
contrast to the disorder parameter and the critical threshold as
defined in \eref{eq:dip}.  
The parameter $\gamma$ simplifies considerably if we identify the
cut-off length $\cut{\ell}$ with the Boltzmann transport mean free
path as justified in section \ref{quantcorr.sec} above. With this
choice $\gamma=1/k\boltz\ell$, 
we locate the strong localisation threshold $\gamma_\mathrm{c}=\gamma_0$
or $\delta
D/\boltz{D}=1$ at the critical disorder strength or critical Ioffe--Regel parameter \cite{bvtloc}
\begin{equation}\label{eq:ioffe}
k\boltz{\ell}^\mathrm{c}=\sqrt{\frac{3}{\pi}} \approx 0.95.
\end{equation}
In order to estimate whether this strong localization threshold
can be reached with current experimental techniques, we use the 
perturbative results of section \ref{AvPro} and
\ref{diff.trans.sec} 
to calculate the laser intensity and detuning for different atomic
momenta. For sufficiently large and phase-coherent systems, $L_\ast \gg \boltz\ell$, we have
$\sfrac{\delta D}{\boltz{D}}\propto 
(\sfrac{\laser{I}}{\laser{\delta}})^{4}$.
In the ($\laser{I}, \laser{\delta}$) parameter plane, the
strong localization threshold $\delta D/\boltz{D}=1$
is thus characterized by the simple linear scaling
$\laser{I} \propto \laser{\delta}$.
This can be seen in the phase diagram of the 3D strong localization onset
for different atomic wave vectors in \fref{fig:phpl3}. 

For an atom prepared at recoil momentum ${k}=0.9\,\laser{k}$, and a
realistic detuning $\laser{\delta} = 10^4\Gamma$, we
locate the strong localization threshold at a very reasonable laser intensity 
$\laser{I}=63\,\sat{I}$.
At this point
$\ells=0.09\,\mu$m, $\elltrb=0.13\,\mu$m and the phase coherence length
$L_\phi=9\,\mu$m.
Precisely at the transition point, the corrected transport mean-free
path vanishes, $\elltrwl=0$.
Since $L_\ast \gg \elltrb$ the coherent 
transport condition is well fulfilled.
Again these numbers have to be taken with a grain of salt
since the underlying perturbative description is bound to break down
in the strongly disordered regime.

The self-consistent equation \nref{eq:condeq} permits to describe the
localized regime. 
Close to the
threshold, it reads 
\begin{equation}\label{eq:loc3d}
\frac{L_*}{\boltz{\ell}} = \frac{\sfrac{3}{2}}{\sfrac{3}{\pi}-(k\boltz{\ell})^2}.
\end{equation}
Together with \eref{eq:clim} this expression determines the 3D
localization length $\loc\xi$.

\Fref{fig:xiloc3p} shows a logarithmic plot of the relevant length
scales on both sides of the localization threshold for a fixed atomic
momentum of $k=0.9\,\laser{k}$ as function of the intensity
$\laser{I}$  at $\delta=10^4\,\Gamma$.
Just as in 2D, the phase coherence length decreases monotonically
because spontaneous photon emission is enhanced as the power
increases. The strong localization threshold is reached for
$\laser{I}=63\,\sat{I}$, where the corrected transport mean free path
$\elltrwl$ vanishes together with the diffusion constant. 
As expected the localization length $\loc{\xi}$ comes down from
infinity at the threshold and tends toward its bulk value 
$\loc\xi = \frac{3}{2}\boltz{\ell}/(\frac{3}{\pi}-k^2\boltz{\ell}^2)$ 
given by \nref{eq:loc3d} in the regime $\loc\xi \ll L_\phi,L$.

Often the critical behavior of $L_*$ is stated as a function of the
energy which enters in our case the Boltzmann transport mean free path
$\boltz\ell(k)$. 
For sub-recoil atoms $k\zeta\le 1$, the disorder parameter 
can be expressed in terms of the energy  
making use of
\eref{eq:mobby} and \eref{eq:berrytrans}: 
\begin{equation}\label{eq:mooo}
\frac{\gamma}{\gamma_0}=\frac{\mobby{E}}{E}=\sqrt{\frac{\pi}{3}}\,\frac{\mobil{E}}{E}.
\end{equation}
where $\mobby{E}$ denotes the mobility edge. Up to the non-universal
factor $\gamma_0=\sqrt{\pi/3}$ very close to unity, it is given by
the weak-scattering energy scale $\mobil{E}$ defined in
\eref{eq:mobby}. 
The critical exponent as function of energy is still the same, 
$L_* \propto (\mobby{E}-E)^{-\nu}$ with $\nu=1$.  
For above-recoil atoms $k\zeta\ge 1$, the threshold condition
$\gamma=\gamma_0$ leads to an energy such that the weak-scattering
parameter $\Delta\ge 1$ lies formally beyond its limit of validity; in this regime,
our first-order perturbation theory cannot give trustworthy estimates
for the mobility edge.

Similarly, we can calculate the
critical exponent of the diffusion constant decreasing to zero on the
diffusive side $\gamma<\gamma_0$ of the transition, 
$\wl{D}\propto (\gamma_0-\gamma)^s\propto (E-\mobby{E})^s$
with the same critical exponent $s=1$ as predicted by the scaling theory \cite{kramer}.
This is particularly remarkable since the diagrammatic perturbation
theory is in principle only valid for weak disorder and a priori
cannot describe the 
strong localization onset. The accurate description of the critical
exponents within this theory however can be seen as a strong hint that
the quantum interference corrections, which are responsible for weak
localization, also remain the dominant contribution for strong
localization.

\section{Conclusion and Outlook}

We have formulated a quantum transport theory for 
non-interacting matter
waves in two- and three-dimensional disordered optical speckle 
potentials with a finite, and rather large, correlation length $\zeta$. 
Making use of diagrammatic perturbation theory for weak scattering, we have determined all relevant 
microscopic transport parameters (scattering mean free path $\ells$,
transport mean free path $\elltrb$ and Boltzmann diffusion constant
$\boltz{D}$) that are necessary to describe the average diffusion
process, as 
functions of experimentally controllable quantities such as the
initial atomic momentum, the laser detuning and the laser intensity. 

As a consequence of the speckle correlations we encounter an essential
difference compared to the common treatment for Gaussian white noise. 
The differential single scattering cross section for correlated
fluctuations is anisotropic and its degree of anisotropy depends
strongly on the initial energy of the matter wave. This is a general
property of correlated potentials which has important implications for
the small perturbation parameter that determines the limit of the weak
scattering regime. 
Together with the strength of the potential fluctuations $\laser{V}$,
the speckle correlation length $\zeta$ enters---in form of the
correlation energy $E_\zeta=\hbar^2/m\zeta^2$---the important energy
scale $\mobil{E}=\avV^2/E_\zeta$, which determines the minimum energy
for weak scattering. 
The validity of our results is restricted to the regime where the
atomic energy is larger than the weak scattering energy
$\mobil{E}$. In this regime the atoms are only weakly scattered by the
speckle fluctuations. 

Using a linear response theory, we calculate the weak localization
correction to classical transport.
These interference corrections are 
sensitive to phase-breaking mechanisms such as spontaneous photon emission.
These spontaneous dissipative processes
can be maintained at harmless rates for realistic 
values of the experimental parameters.
We have shown that in this case the weak localization
correction $\delta D$ can
reach a considerable fraction of the Boltzmann diffusion
constant $\boltz{D}$ within the weak scattering regime for atoms at
recoil or sub-recoil temperatures. Even the strong localization
threshold is within reach of current experimental techniques; we have
calculated the corresponding critical disorder values and localization
lengths. 
In order to measure the diffusion constant and its weak localization corrections in an actual
experiment, one could for instance release a confined atomic cloud and monitor its
lont-time spread inside the speckle field by time-of-flight or in-situ
imaging techniques. The crossover to the strongly localized regime would appear
as a freezing of the diffusive process as the cloud reaches the
localization length, with a final density distribution mapping out the
longest-lived localized eigenmodes of each realization that permit the corresponding localisation
length to be 
measured. 

The extension to non-monochromatic matter wave distributions
(such as the ones used in the recent publications
\cite{Shapiro07,LSP06b}) is straightforward 
and will be the subject of a subsequent publication. 

As the use of sub-recoil temperatures favors localization effects, the
most promising candidate for matter wave experiments seems to be a
Bose-Einstein condensate. In this respect, another extension
of this work would be to consider the dynamics of a Bose-Einstein
condensate \cite{LSP06b} in the 2D and 3D speckle field and to
study the role of interactions  (for instance using the approach
developed in \cite{LSP06a}) and quantum statistics on localization.
Another possible extension of this work would be to study the
influence of the internal spin structure of the atom. 

Finally, this work could be extended to disordered magnetic potentials.
One immediate advantage compared to the optical speckle potential
would be the Gaussian character of the magnetic potential fluctuations
\cite{Wang}. 
A first numerical study of the expansion of a Bose-Einstein condensate
in a one-dimensional disordered atomic waveguide has revealed
characteristic signatures of localization in the non-interacting
regime \cite{peter}. 
This could open the door to further experiments studying disorder on
atom chips \cite{schmiedmayer, wire}.

\ack 
We acknowledge financial support from DFG, BFHZ-CCUFB,
the PROCOPE and the Marie Curie program. R. Kuhn is particularly
indebted to the DAAD for financial support within the doctoral
fellowship program. R. Kuhn would like to thank G. Montambaux for
helpful technical discussions and the Institut Non Lin\'eaire de Nice,
Laboratoire Kastler Brossel and the Physics Department of the National
University of Singapore for their kind hospitality. The 
authors would like to thank  R. Kaiser, D. Wilkowski, G. Labeyrie,
T. Chaneli\`ere, T. Wellens, B. Gr\'emaud, R. Sapienza,
L. Sanchez-Palencia for their interest, and finally an anonymous
referee for helpful remarks. 

\appendix

\section{Self-energy $\Sigma$ and small perturbation parameter}
\label{app:self-energy}

Let us recall the self-energy series \nref{self},
\begin{equation}\label{app:self}
\Sigma = \sum_{n \ge 2} \, \Sigma_n,
\end{equation}
where $\Sigma_n$ represents all one-propagator irreducible diagrams of
order $n$ in field correlations and in powers of $\eta = \laser{V}/E_\zeta$, the speckle strength in units of correlation energy.
The first terms are:
\begin{eqnarray}
\fl\Sigma_2 =  \st\ffgff =: \st\vgv \label{sigma2.app} \\
\fl\Sigma_3 =  \st\fvgvgv \label{sigma3.app} \\
\fl\Sigma_4 =  \st\cross + \, \st\bows \nonumber \\
\fl \qquad + \, \st\fvieri \, + \, \st\fvierii + \, \st\fvieriii \label{sigma4}\\
\fl\Sigma_5 = \, \st\scinq \,
         + \, \st\scinqb \, + \, \ldots\label{sigma5}
\end{eqnarray}

In these diagrams, every straight line represents the free 
propagator $G_0$ defined in \eref{G0.eq}.
``Irreducible'' means that by cutting any such single
propagator line, the diagrams do not split into independent parts. A dotted line like $\diagram{\oleero}=\eta\gamma_d$ represents the
field correlation function \nref{correlE}. 
A double dotted line $\diagram{\ffleerff}=\eta^2\Power_d=\eta^2|\gamma_d|^2$
represents the intensity correlation function 
\nref{correl}, that we note $\eta^2\Power_d = \diagram{\vleerv}$ for
simplicity in the following. 
To evaluate a diagram, one writes down its expression
in terms of correlation functions and propagators and integrates over
all free internal variables. This integration is most conveniently
done in momentum space since the
free propagator $G_0(\kappa,\varepsilon)$ is then diagonal.

As an example, let us calculate the first diagram $\Sigma_2
=\diagram{\vgv}$ explicitly.
The potential fluctuation operator $\diagram{\otimes} =\eta\delta V$ is diagonal in
real space and therefore translation invariant in momentum space: 
$\langle \bkappa' |\fluct| \bkappa\rangle=
\fluct(\bkappa-\bkappa')$. 
The Fourier transform of the
potential fluctuations is defined by
\begin{equation}
\fl \fluct(\bkappa) = \int \dd\rho \; \fluct(\brho) \,
e^{-i\bkappa \cdot \brho}, \qquad 
\fluct(\brho) = \int \frac{\dd\kappa}{(2\pi)^{d}} \; \fluct(\bkappa) \,
e^{i\bkappa\cdot \brho}. 
\end{equation}
The 2-point correlation function of the potential fluctuations in momentum space reads
\begin{equation}\label{eq:khi}
\av{\fluct(\bkappa_1)\,\fluct(\bkappa_2)} =
(2\pi)^d \delta(\bkappa_1+\bkappa_2)\;\Power_d(\bkappa_1),
\end{equation}
where $\Power_d(\bkappa)$ is the Fourier transform of
$\Power_d(\brho)$ that was defined in (\ref{correl}). 
The matrix elements of $\Sigma_2(\varepsilon)$ are given by 
$\bra{\bkappa'}\Sigma_2(\varepsilon)\ket{\bkappa}
= (2\pi)^d\delta (\bkappa- \bkappa')\,
\Sigma_2(\kappa,\varepsilon) $; as expected, the self-energy operator is diagonal in momentum
space and isotropic. Its diagonal entries are
\begin{equation}\label{app:fermirule}
\Sigma_2(\kappa,\varepsilon)= \eta^2 \int \frac{\dd\kappa_1}{(2\pi)^{d}} \;
\Power_d(\bkappa-\bkappa_1) G_0(\kappa_1,\varepsilon), 
\end{equation}
which is simply the momentum convolution of the potential correlation
function $\Power_d$ with the free Green function $G_0$.

The appearance of odd terms $\Sigma_{2q+1}$ in the formal series \nref{app:self} reflects the
non-Gaussian character of the potential fluctuations.
We now wish to determine the effective small parameter and
corresponding range of validity of the self-energy series.
To this aim, we discuss separately the high-energy
and low-energy cases ${\kappa} = \sqrt{2\varepsilon} \gg 1$ and
${\kappa} \ll 1$, respectively. 

In the high-energy limit, it is possible to
determine the dependence of $\Sigma_n$ on the ultraviolet momentum
${\kappa}\gg 1$ by
simple power counting. 
Each irreducible diagram contributing to $\Sigma_n$ contains $n-1$
internal propagators $G_0(\varepsilon)$, $p$ field-correlation
functions ($0\le p\le n$), and $(n-p)/2$ intensity-correlation
functions. Taking into account all momentum conservation laws, this
leaves exactly $(n+p)/2$ independent variables $\bkappa_i$ that
can be chosen to be the arguments $\gamma_d(\bkappa_i)$ or
$\Power_d(\bkappa_i)$ of the correlation functions
\nref{eq:corl2d} and \nref{eq:corl3d}.
Because of the strict momentum cutoff, 
these correlation functions constrain the norms $\kappa_i$ to remain
of order unity or smaller. The only dependence on ${\kappa}$
comes from the Green functions $G_0(\kappa_m,\varepsilon)$ that are
evaluated at momenta $\bkappa_m= {\bkappa}-\sum_i\alpha_i\bkappa_i$
with coefficients $\alpha_i\in\{0,\pm1\}$ that describe the diagram's topology. 
Linearizing around the on-shell value
$\varepsilon={\kappa}^2/2$ for ${\kappa}\gg \kappa_i$,
each Green's function contributes 
a power ${\kappa}^{-1}$ such that 
$\Sigma_n \sim  a_n \eta^n {\kappa}^{1-n} = a_n \kappa (\eta/\kappa)^n$, where $a_n$ is
related to the number of $n$-point irreducible diagrams. 
We thus can identify the effective expansion parameter
$g=\eta/{\kappa}$ that should be small, which in turn justifies the choice of the weak
scattering condition $\Delta = \eta^2/\varepsilon= 2 g^2\ll 1 $ in
section \ref{weak} above. 
Of course, such an analysis of a superficial degree of divergence \cite{lebellac}
can only
determine how the small parameter depends on energy, but not fix
numerical factors of order unity. 
We can rewrite the ratio of the self-energy (\ref{app:self}) to the
kinetic energy $\varepsilon$ as 
\begin{equation}\label{eq:coup}
\Sigma/\varepsilon \sim \sum_{n \ge
2}\,\frac{a_n}{{\kappa}}\,g^n,\qquad ({\kappa} \gg 1),
\end{equation}
In this high-energy regime, non-Gaussian
terms $n=2q+1$ contribute to the series. In other words, fast atoms live in a potential that has definitely non-Gaussian statistics beyond the leading-order Born approximation $n=2$.    

The low-energy limit ${\kappa}\ll 1$ in 2D allows
for a similar analysis. Each of the $(n+p)/2$ correlation
functions in $\Sigma_n$ tends towards its finite limit as $\kappa\to 0$  
and thus becomes effectively $\delta$-correlated in real space. 
The contributions 
${\kappa}^{n+p}$ from the $(n+p)/2$ integration measures and
${\kappa}^{2-2n}$ from the $n-1$ Green's functions yield a scaling ${\kappa}^{2-n+p}$ for
$\Sigma_n$. If $n$ is even, $\Sigma_n$ is therefore 
dominated by the diagrams with $p=0$, containing only intensity-correlation 
functions and diverging like ${\kappa}^{2-n}$. 
If $n$ is odd, the dominant contribution comes from the
diagrams with the smallest number of field correlations, $p=3$, that
appear first in $\Sigma_3$ and reappear subsequently in higher non-Gaussian terms
$\Sigma_{2q+1}$. 
One can then rewrite the dominant contributions to (\ref{app:self}) in
the form 
\begin{equation}\label{selfp1}
\Sigma/\varepsilon \sim \sum_{q\geq
1}(a_{2q}g^{2q}+a_{2q+1}{\kappa}^{3}g^{2q+1}) ,\qquad
(\mathrm{2D\ and\  } {\kappa}\ll 1).
\end{equation}
Remarkably, $g=\eta/{\kappa}$ is still the expansion parameter, and
remains small if $\eta\ll{\kappa}\ll 1$. 
Terms which are negligible compared to both their neighbors in the
series can be omitted. This applies to all odd terms when
${\kappa}^3g^{2q+1} \ll g^{2q+2} $, \ie ${\kappa}^3\ll
g$ at fixed $g$. In this low-energy and weak-scattering regime, the self-energy does no longer depend on the
field-correlation functions, and only the Gaussian terms 
survive:
\begin{equation}\label{selfp1g}
\Sigma/\varepsilon \sim \sum_{q\geq 1} a_{2q}g^{2q},\qquad
(\mathrm{2D\ and\ } {\kappa}^3\ll g). 
\end{equation}

The 3D case requires a separate discussion. As far as the pure intensity
correlations ($p=0$) are concerned, actually the same reasoning as for
2D holds because the
low-${\kappa}$-divergence of the intensity correlation $\propto
{\kappa}^{-1}$ is compensated by a supplementary factor from the
integration measure. However, the pure field-correlation functions
$\gamma_3(\kappa)=\delta(1-\kappa)$ cannot
tend to a constant as $\kappa\to 0$, but project all integration 
momenta onto the unit sphere. Consequently, a
small-${\kappa}$-contribution 
of the integration measure and the correlation functions can only come
from the $(n-p)/2$ variables intervening in the intensity-correlation
functions: it reads ${\kappa}^{n-p}$. The contribution of a
propagator is either ${\kappa}^{-2}$ (as in the 2D case) if it does not depend on a 
 field-correlation momentum, or independent of
${\kappa}$ in the limit ${\kappa}\rightarrow 0$ because
field correlation momenta of order unity remain present. Thus, the
diagrams depending only on the field-correlations $(n=p)$ behave like
${\kappa}^0$. At fixed $n$ and $p$, the dominant diagrams mixing field-
and intensity-correlation functions 
are those with the largest number of propagator lines independent of
the $p$ field-correlation variables. This happens when field
correlations can be written as products of the largest number of independent
field-correlation sub-diagrams
that never cross the intensity-correlation lines. An
example for such a diagram is the first contribution 
to $\Sigma_5$ shown in \eref{sigma5} that displays a $\Sigma_3$-type
field correlation inside a $\Sigma_2$ intensity correlation. 
In all dominant cases, the field sub-diagrams will contain at most 3, 4 or 5
vertices (as those shown in $\Sigma_3$, $\Sigma_4$, and $\Sigma_5$), 
because higher-order field correlations could be factorized into these
elementary ones, thus yielding an additional independent propagator. 
Writing $p=3n_1+4n_2+5n_3$, where the $n_i$ are
non-negative integers, the largest possible number of sub-diagrams is
obtained by maximizing the sum $n_1+n_2+n_3$. The number
of propagator lines giving a ${\kappa}^{-2}$ contribution to the
diagram is then $n-p+n_1+n_2+n_3-1$, and the total contribution of the
diagram is ${\kappa}^{2-n+n_1+2n_2+3n_3}$. It turns out that
when $n$ is even, the dominant contribution ${\kappa}^{2-n}$ to
$\Sigma_n$ comes from the intensity-correlation diagrams $(p=0)$. When
$n$ is odd, the main contribution ${\kappa}^{3-n}$ is due to the
diagrams with $n_1=1$, \ie $p=3$. Similarly to the 2D case, the
dominant contribution to the self-energy (\ref{app:self}) is
\begin{equation} 
\label{app:self3d}
\Sigma/\varepsilon=\sum_{q\geq 1}(a_{2q}g^{2q}+a_{2q+1}{\kappa}g^{2q+1}).
\end{equation}
As in the 2D case, $g=\eta/{\kappa}$ is the expansion
parameter. Again, terms which are negligible compared to both their
neighbors can be omitted. This applies to the odd terms when
${\kappa}^2\ll\eta\ll{\kappa}\ll 1$ at fixed $g$, \ie in
the quantum regime discussed at the end of \sref{weak}. In this regime,
an effective $\delta$-correlated Gaussian potential is recovered. 

Since the number of diagrams $a_n$ grows factorially with $n$, we face the
well-known troublesome fact that even for $g\ll 1$, our global weak
scattering condition covering both the high- and low-energy regime, 
the series (\ref{app:self})
formally diverges. It can only be 
understood as an \emph{asymptotic} series that can be accurately approximated by just the
first few terms \cite{bender}. When the effective
coupling constant $g$ is sufficiently small, then a truncation to the first
term already gives a good approximation to the self-energy. The
weak scattering condition is thus given by $g \ll 1$, or
equivalently by $\eta \ll \kappa$, which amounts to $E \gg
\mobil{E} = \avV^2/E_\zeta$.

Self-energy diagrams like
\diagram{\vgv} and \diagram{\bowso} and similarly nested
higher-order diagrams with an outer correlation function can be
self-consistently summed up to give the diagram \diagram{\vggv},
where the thick line represents the average Green function.
The same procedure can be applied to all
diagrams like \diagram{\crosso} with two outer correlation
functions yielding the self-consistent diagram \diagram{\croggo},
etc.

\section{Diffusion from quantum kinetic equation}
\label{app:QKE}
\label{app:ward}

This appendix is devoted to the derivation of the diffusion constant
using a quantum linear response theory that we believe to be a
slightly improved version of Vollhardt and W\"olfle's approach \cite{vollhardt}.  

\subsection{Quantum kinetic equation and continuity equation}

We first rewrite the Bethe-Salpeter equation \nref{BS} for the intensity propagator
$\Phi(\bkappa,\bkappa',\bi{q},\varepsilon,\omega)$ in a more useful way.  
For the scalar quantities the product of the average propagators $\av{G^{\dag}}(\varepsilon_-) \otimes
\av{G}(\varepsilon_+)$   
can be reformulated in momentum space by using the identity $\av{G^{*}}\,\av{G} = (\av{G^{*}}-\av{G})/(\av{G}^{-1}-\av{G^{*}}^{-1})$
as
\begin{equation} \label{eq:idi}
\av{G^{*}}(\bkappa'_-,\varepsilon_-)
\av{G}(\bkappa'_+,\varepsilon_+) 
= \frac{- \Delta G(\bkappa',\bi{q},\varepsilon,\omega)}
{\omega - \bi{q}\cdot\bkappa' -\Delta \Sigma(\bkappa',\bi{q},\varepsilon,\omega)},
\end{equation}
where $\bkappa'_\pm= \bkappa'\pm\sfrac{\bi{q}}{2}$  
and 
$\Delta G(\bkappa,\bi{q},\varepsilon,\omega) 
= \av{G}(\bkappa_+,\varepsilon_+) -
\av{G^*}(\bkappa_-,\varepsilon_-)$ as well as 
$\Delta \Sigma(\bkappa,\bi{q},\varepsilon,\omega) =
\Sigma(\bkappa_+,\varepsilon_+)-\Sigma^*(\bkappa_-,\varepsilon_-).$
Multiplying the denominator to the other side  
leads to the quantum kinetic equation \cite{rammer}
\begin{equation}\label{eq:qke}
\fl \left[ \omega - \bi{q}\cdot\bkappa' \right]
\Phi(\bkappa,\bkappa',\bi{q},\varepsilon,\omega)=
-(2\pi)^d\delta(\bkappa-\bkappa') \Delta
G(\bkappa',\bi{q},\varepsilon,\omega)
 + \calC[ \Phi].   
\end{equation} 
This is the standard form of a kinetic equation with a
Fourier-transformed drift derivative on the
left hand side, and on the right hand side first a source term and
then all scattering information contained in the linear collision functional 
\begin{eqnarray}
  \label{eq:def.collision}
\fl \calC[ \Phi]  = \Delta
\Sigma(\bkappa',\bi{q},\varepsilon,\omega)
\Phi(\bkappa,\bkappa',\bi{q},\varepsilon,\omega)  \nonumber \\
-  \Delta
G(\bkappa',\bi{q},\varepsilon,\omega)
 \int \frac{\dd{\kappa''}}{(2\pi)^d}\;
 \Phi(\bkappa,\bkappa'',\bi{q},\varepsilon,\omega)
U(\bkappa'',\bkappa',\bi{q},\varepsilon,\omega).
\end{eqnarray}

As shown by Vollhardt and W{\"o}lfle \cite{vollhardt}, 
the irreducible vertex $U$ and the self-energy $\Sigma$ are intimately linked 
through the Ward identity \cite{rammer,vollhardt} 
\begin{equation}\label{eq:ward}
\Delta \Sigma(\bkappa',\bi{q},\varepsilon,\omega)= \int
\dkpi{{\kappa''}} \,\Delta G(\bkappa'',\bi{q},\varepsilon,\omega)\,
U(\bkappa'',\bkappa',\bi{q},\varepsilon,\omega).
\end{equation}
Notably, the proof works also for correlated potentials and thus applies
without restriction to our case (in contrast to scattering by resonant
potentials where corrections to the Ward identity renormalize the transport speed of the propagating wave
\cite{resonant}). This Ward identity plays the role of an optical
theorem: everything that disappears from the forward propagating mode
is scattered into other modes. With its help, the scattering functional can be rewritten as  
\begin{eqnarray}
  \label{eq:def.collision2}
\fl \calC[ \Phi]  = 
\int \frac{\dd{\kappa''}}{(2\pi)^d}\;
 \big[\Delta G(\bkappa'',\bi{q},\varepsilon,\omega)
 \Phi(\bkappa,\bkappa',\bi{q},\varepsilon,\omega)
\nonumber \\ 
-  \Delta G(\bkappa',\bi{q},\varepsilon,\omega)
 \Phi(\bkappa,\bkappa'',\bi{q},\varepsilon,\omega)
 \big]
U(\bkappa'',\bkappa',\bi{q},\varepsilon,\omega).
\end{eqnarray}
By parity, it vanishes identically under integration over
$\bkappa'$, $   \int \dd \kappa' \calC[\Phi]= 0$. 
Moreover, one can verify that 
$  \int \rmd \varepsilon  \Delta
G(\bkappa,\bi{q},\varepsilon,\omega) = 2\pi i $
by going back to the definition of the
time-evolution propagator and observing that $U(t=0)=\mathbbm{1}$
while defining $\Theta(0)=\frac{1}{2}$.  
By integrating the quantum kinetic equation over $\bkappa'$ and
$\varepsilon$, we thus recover the 
continuity equation \nref{eq:cont:kernel} exactly to all orders in $\bi{q}$ and
$\omega$. This is slightly different from Vollhardt and
W\"olfle's approach, followed by the entire quantum transport
literature, which uses different kernels defined by sums 
over $\bkappa'$ and $\bkappa$ (instead of
$\bkappa'$ and $\varepsilon$) that only obey an approximate continuity
equation.

\subsection{Linear response} 

Starting with the continuity equation, we now need a second 
equation that expresses the current $\Cur(\bkappa,\bi{q},\omega)$ as 
function of a small gradient in density, thus defining a
linear-response coefficient.  
Multiplying the QKE \nref{eq:qke} by $\bi{q} \cdot\bkappa'$ and integrating over
$\varepsilon$ and $\bkappa'$ gives an equation for the current kernel
$\Cur$: 
\begin{eqnarray}\label{eq:qke.cur}
\fl \omega \bi{q} \cdot \Cur(\bkappa,\bi{q},\omega)
-  \int\frac{\rmd \varepsilon}{2\pi} 
  \int \frac{\dd\kappa'}{(2\pi)^d}
(\bi{q}\cdot\bkappa')^2
\Phi(\bkappa,\bkappa',\bi{q},\varepsilon,\omega) \nonumber \\
= - i \bi{q} \cdot\bkappa
 +\int\frac{\rmd \varepsilon}{2\pi} 
  \int \frac{\dd\kappa'}{(2\pi)^d} (\bi{q}\cdot\bkappa') \calC[ \Phi].
\end{eqnarray} 
This is not a closed equation for $\Ker$ and $\Cur$ because it
involves higher-order moments of the full distribution $\Phi$. If one
continues to write equations for the new unknowns, one generates a
hierarchy that must be truncated judiciously in order to 
extract the density kernel $\Ker$.     

In the long time and long-distance limit $q,\omega\to 0$, we can make
two assumptions that facilitate the derivation of the transport time
in the linear response regime.    
Firstly, we perform the usual expansion 
\begin{equation}\label{diffusion.expansion}
\fl \tilde\Phi (\bkappa,\bkappa',\bi{q},\varepsilon,\omega)  =   
 \frac{A(\kappa',\varepsilon)}{2\pi\avdos(\varepsilon)} 
 \int \frac{\dd{\kappa''}}{(2\pi)^d}
 \left[1+  \frac{d}{2\varepsilon}
(\bkappa'\cdot\hat{\bi{q}})(\hat{\bi{q}}\cdot\bkappa'') \right] \Phi (\bkappa,\bkappa'',\bi{q},\varepsilon,\omega)
\end{equation}
into the lowest-order terms in powers of
the angular dependence on $\hat\bi{q}\cdot\bkappa'$, isotropic and
vectorial. 
The factorized spectral function $A(\kappa',\varepsilon)$ encapsulates the
dependence on the modulus of $\kappa'$ which is constrained to the
disorder-broadened energy shell; indeed, 
the Bethe-Salpeter equation \nref{BS} shows that the intensity relaxation kernel
$\Phi$ is proportional to $\Delta G(\bkappa',\bi{q},\varepsilon,\omega)$ which is itself very well
approximated by the strongly peaked spectral function
$A(\kappa,\varepsilon)$ since $\Delta G(\bkappa,\varepsilon)= 2 i \Im
\av{G}(\kappa,\varepsilon)= 
-iA(\kappa,\varepsilon)$. 
Secondly, by symmetry in $\bkappa$ and $\bkappa'$, the intensity relaxation kernel
$\Phi$ is also proportional to $A(\kappa,\varepsilon)$. 
As a sharply peaked function of $\varepsilon$, the kernel can therefore be well approximated by  
$\Phi(\bkappa,\bkappa'',\bi{q},\varepsilon,\omega) \approx
A(\kappa,\varepsilon) \int\frac{\rmd \varepsilon'}{2\pi}
\Phi(\bkappa,\bkappa'',\bi{q},\varepsilon',\omega)$.
Using this in \eref{diffusion.expansion} leads to the ansatz
\begin{equation}\label{kernel.expansion}
\fl \tilde\Phi (\bkappa,\bkappa',\bi{q},\varepsilon,\omega) 
= \frac{A(\kappa,\varepsilon) A(\kappa',\varepsilon)}{2\pi\avdos(\varepsilon)}  
\left[ \Ker(\bkappa,\bi{q},\omega) + \frac{d}{2\varepsilon}
(\bkappa'\cdot\hat{\bi{q}}) \hat{\bi{q}}\cdot\Cur(\bkappa,\bi{q},\omega)
\right].
\end{equation}
This expression in terms of the relevant kernels $\Ker$ and $\Cur$  
is custom-tailored such that  
the consistency relations $\tilde\Phi_0 = \Ker$ and
$\tilde{\boldsymbol{\Phi}}_1=\Cur$ hold. 
We can therefore insert $\Phi \approx
\tilde\Phi $ into \eref{eq:qke.cur} and calculate the two
remaining terms. On the left-hand side, only the
$\Ker$-term gives a contribution since the $\Cur$-term vanishes by
parity.    
On the right-hand side, the collision  integral of the
isotropic $\Ker$-term vanishes by parity if  $\Delta 
\Sigma$, $\Delta G$, and $U$ are evaluated at $\bi{q}=0$ and $\omega=0$. 
This is allowed to leading order in $q,\omega$ because the collision
integral is already of order $1/\taus$.      
The remaining integral over the $\Cur$-term defines the
inverse transport time \nref{eq:tautr}. 
Rearranging the equation, we 
can solve for $\bi{q}\cdot\Cur$ and obtain \eref{linearresponse.eq}.

Together with the continuity equation \nref{eq:cont:kernel} this
provides a closed set of equations for $\Ker$
and $\Cur$ 
with the following solutions for $\omega\tautr(\kappa) \ll 1$:
\begin{equation}\label{PhiJ}
\fl \Ker(\bkappa,\bi{q},\omega)  = \frac{1- i \tautr(\kappa) \bi{q}\cdot\bkappa}{-i\omega + \diko (\kappa)q^2},\qquad
i \bi{q}\cdot \Cur(\bkappa,\bi{q},\omega) = \frac{\diko(\kappa)
q^2}{-i\omega +\diko(\kappa) q^2} 
\end{equation}
with the reduced diffusion constant $\diko(\kappa) = \kappa^2\tau(\kappa)/d$. 
Inserting these solutions into the expansion \nref{kernel.expansion} 
yields the corresponding diffusive intensity propagation kernel 
\begin{equation}\label{diffkernel}
\Phi (\bkappa,\bkappa',\bi{q},\varepsilon,\omega) 
= \frac{A(\kappa,\varepsilon) A(\kappa',\varepsilon)}{2\pi\avdos(\varepsilon)}
\,\frac{1 - i \tautr(\kappa) \bi{q}\cdot(\bkappa+\bkappa') }{-i\omega + \diko (\kappa) q^2}
.
\end{equation}

\section{Intensity diagrams and Boltzmann transport theory}
\label{app:intensity}

The Bethe-Salpeter equation \nref{BS} can be formally recast into the form 
\begin{equation}\label{eq:phi}
\Phi = [\av{G^{\dag}}\otimes\av{G}] +
[\av{G^{\dag}}\otimes\av{G}] \,R
\,[\av{G^{\dag}}\otimes\av{G}], 
\end{equation}
where the reducible vertex is 
$R = U + U\,[\av{G^{\dag}}\otimes\av{G}]\,R$.
The first terms of the power series $U = \sum_{n\ge 2} U_n $
for the irreducible intensity vertex are the following diagrams: 
\begin{eqnarray}\label{eq:U}
\fl  U_{2} & = & \stab \label{U2}\\[0.75em] 
\fl  U_{3} & = & \pik + \herz \label{U3}\\[0.75em] 
\fl   U_{4} & = & \icks + \baum + \blume + \blatt + \spade + \dots\label{U4}
\end{eqnarray}
As before, ``irreducible'' means that these diagrams 
do not fall apart into independent sub-diagrams by cutting one of the
thick lines that represent the average propagators
$\av{G}$. The upper lines belong to
retarded amplitudes 
$\av{G}$ while the lower entries belong to advanced amplitudes $\av{G^{\dag}}$. 
Dotted lines connecting two $\otimes$ (resp. $\bullet$) represent an intensity
(resp. field) correlation function. 
In addition to the familiar 
potential correlations as in $U_2$ and $U_4$, we find
field-correlation diagrams as in $U_3$ and all higher orders, 
since the potential
fluctuations do not obey Gaussian statistics.

Through the Ward identity \nref{eq:ward}, every self-consistent
diagram in the expansion of the self-energy 
corresponds to a set of diagrams in the expansion of $U$.
For example, the diagram \nref{U2} is linked to the
self-consistent diagram \diagram{\vggv}, exactly as in the electronic
case. This means in particular that the Boltzmann approximation $U
\approx \boltz{U} = U_2$ and the weak scattering approximation $\Sigma
\approx \Sigma_2$ are consistent, and the Ward identity \nref{eq:ward}
appears as the expression \nref{eq:mfp} of the scattering mean free path in
terms of the Boltzmann vertex \nref{eq:boltz}. 
The next two diagrams \nref{U3} are linked to the field correlations
\diagram{\fggs} 
defining $\Sigma_3$, and the following \nref{U4} are linked to
\diagram{\croggo}, etc. 

In the Boltzmann approximation $U=U_2$, the reducible vertex at the
core of the intensity propagator \nref{eq:phi} becomes 
$\boltz{R} = U_2 +U_2 [\av{G^{\dag}}\otimes\av{G}]U_2+ \dots   = U_2 + L$. 
Here $L$ denotes the so-called \emph{diffuson}, the letter $L$
referring to the ladder topology of diagrams: 
\begin{equation}\label{eq:LS.app}
L = \st\lad + \st\ladder + \dots =: \laddad
\end{equation}
We will need an analytical expression for the diffuson in the
diffusive regime. To this end, we consider the geometric series for $\boltz{\Phi}$ in
the Boltzmann approximation: 
\begin{equation}\label{eq:dikern}
\boltz{\Phi} = \st{\legs} + \st{\legs\stab\legs} + \st{\legs\laddad\legs}
\end{equation}
The first term on the right-hand side describes ballistic propagation, 
whereas the second term describes single scattering in the effective medium.
These two terms can be neglected compared to the remaining diffuson
contribution to the series which is given by \eref{diffkernel} with $\diko=\boltz\diko$.
Making use of the identity \nref{eq:idi} in the long time and large
distance limit, \ie for $\omega=0$ and $\bi{q}=0$, the spectral
functions in \nref{diffkernel} can be expressed as
$A(\kappa,\varepsilon)=\av{G^*}(\kappa,\varepsilon)\av{G}(\kappa,\varepsilon)/\taus$.
Both spectral functions $A(\kappa,\varepsilon)$ and
$A(\kappa',\varepsilon)$ each contribute a factor $1/\taus$. The
products of the retarded and the advanced Green functions as functions
of $\kappa$ and $\kappa'$ can be identified with the
entering and exiting propagators to the diffuson \nref{eq:LS.app} in
\eref{eq:dikern}. 
The diffuson is therefore given by
\begin{equation}
L(\bkappa,\bkappa',\bi{q},\varepsilon,\omega)
= \frac{1}{2\pi\avdos(\varepsilon)\taus^2}\,
\frac{1 - i \tautr(\kappa) \bi{q}\cdot(\bkappa+\bkappa') }{-i\omega + \boltz{\diko} (\kappa) q^2}.
\end{equation}
Only the prefactor differs form the diffusive intensity propagation kernel \nref{diffkernel}. In particular, the diffuson displays the same diffusive pole as the diffusive intensity propagation kernel.

\section{Weak localization correction for anisotropic scattering}
\label{app:WL}

The weak localization correction to the transport time \nref{eq:tautr}
is contained in the intensity scattering vertex 
$U=U_B+C$, where $C$ incorporates the full \emph{cooperon} contribution
\begin{equation}\label{eq:uabc}
C = \hikami{C}{A} + \hikami{C}{B} + \hikami{C}{C}
\end{equation}
The momentum diagrams corresponding to these three contributions are
\\
\begin{equation}\label{eq:momco}
\fl C = \st\icksix + \st\baumix + \st\blumix =: \st\icksplode
\end{equation}
\bigskip 

Special attention has to be paid to the cooperon contribution in the
case of anisotropic scattering. The simple substitution $U=U_B+C_A$
which is sometimes used in the literature (cf. \cite{pingsheng}) is
not sufficient in this case. 
For electron transport in highly anisotropic systems W\"olfle and 
Bhatt \cite{aniso} have performed the calculation of the
cooperon contribution for the conductivity tensor. In this calculation
the two additional conductivity diagrams $C_B$ and $C_C$ corresponding to dressed Hikami boxes \cite{akker} are taken into account.
This calculation for the anisotropic conductivity tensor is similar to
the calculation of the weak localization correction to the diffusion
constant that we wish to perform. We therefore need to include the
same additional Hikami diagrams $\hikami{C}{B}$ and $\hikami{C}{C}$
adapted to our case.  

The total transport time including the weak localization correction
takes the form: 
\begin{eqnarray}\label{eq:flix}\nonumber
\fl \frac{\boltz{\tau}}{\wl{\tau}(\kappa)}= 
\frac{\boltz{\tau}}{\wl{\hikami{\tau}{A}}(\kappa)}+\frac{\boltz{\tau}}{\taus^2}
\, \int\frac{\rmd
\varepsilon}{2\pi}\frac{A(\kappa,\varepsilon)}{2\pi\avdos
(\varepsilon)}\,\int \frac{\dd \kappa'}{(2\pi)^d} \frac{\dd
{\kappa''}}{(2\pi)^d}\; \\\nonumber
\times\Big[[\av{G}(\kappa')]^2\av{G^*}(\kappa')\;[\av{G}({\kappa''})]^2\av{G^*}({\kappa''})+\av{G}(\kappa')[\av{G^*}(\kappa')]^2\;\av{G}({\kappa''})[\av{G^*}({\kappa''})]^2\Big]
\\
\qquad\qquad\quad
\times(1+\hat{\bkappa}'\cdot\hat{\bkappa}'')\,\Power_d(\bkappa'-\bkappa'')\int
\frac{\dd Q}{(2\pi)^d} \;L_0(\kappa,Q,\omega) 
\end{eqnarray}
where the energy dependence of $\varepsilon$ of the Green functions
has been left out for better visibility and where 
\begin{equation}
L_0(\kappa,Q,\omega)=\frac{1}{2\pi\avdos(\varepsilon)\taus^2} \,\frac{1}{-i\omega+\boltz{\diko}(\kappa) Q^2}.
\end{equation}
In order to evaluate the integral over $\kappa'$ and ${\kappa''}$ in \eref{eq:flix}
one makes use of the peaked structure of the Green functions as a
function of $\kappa'$ and ${\kappa''}$ around the point
$\kappa_\varepsilon=\sqrt{2\varepsilon}$ for weak disorder. To leading
order in the small disorder parameter $1/\varepsilon\taus\ll 1$ 
the integral over $\kappa'$ and ${\kappa''}$ then yields:
\begin{eqnarray}\label{eq:totwl}
\fl
\frac{\boltz{\tau}}{\wl{\tau}(\kappa)}=1+\frac{\boltz{\tau}}{\taus^2}\int
\frac{\rmd\varepsilon}{2\pi}\frac{A(\kappa,\varepsilon)}{2\pi\avdos(\varepsilon)}\,\int
\frac{\dd Q}{(2\pi)^d}\;L_0(\kappa,Q,\omega)\nonumber\\ 
\times
\bigg[2\,[\fuchs^{2,2}]+\int\frac{\rmd \Omega'_d}{\Omega'_d}
(1+\hat{\bkappa}'\cdot\hat{\bkappa}'')\,\Power_d(\kappa_\varepsilon,\theta)\,
\Big[[\fuchs^{2,1}]^2+[\fuchs^{1,2}]^2\Big]\bigg].
\end{eqnarray}
The function $f^{n,m}(\taus)$ is defined by \cite{akker}:
\begin{equation}\label{eq:nice}
\fuchs^{n,m}(\taus)=2\pi\dos\,i^{\,n-m}\,\frac{(n+m-2)!}{(m-1)!(n-1)!}(\taus)^{ n+m-1}.
\end{equation}
Reinserting the functions $f^{n,m}(\taus)$ into \eref{eq:totwl}, the
term in brackets yields the factor $4\pi\dos\taus^3(1-\langle
\cos\theta \rangle)$. Evaluating the integral over $\varepsilon$ to
leading order in $1/\varepsilon\taus$, we obtain the final result 
\begin{equation}
\frac{\boltz{\tau}}{\wl{\tau}(\kappa)}=1+\frac{1}{\pi\dos(\varepsilon_\kappa)}
\int\frac{\dd Q}{(2\pi)^d}\;\frac{1}{-i\omega+\boltz{\diko}(\kappa) Q^2}
\end{equation}
that leads to the corrected inverse diffusion constant \nref{eq:weakloc:perturbative}.

\bibliographystyle{unsrt_mod}
\section*{References}

\end{document}